\documentclass{article}
\usepackage{bm,latexsym,amsmath,amssymb,amsfonts,fancyhdr,color,graphicx,multirow,slashed,cite,multirow}
\usepackage[a4paper,bottom=3cm,top=2.5cm,head=0mm,width=17cm,dvipdfm]{geometry}
\usepackage[usenames,dvipsnames,svgnames,table]{xcolor}
\usepackage[colorlinks=true,
            linkcolor=blue,
            urlcolor=blue,
            citecolor=green,          
						bookmarks=true,
						bookmarksnumbered=true,
						breaklinks=true,
						pdfpagemode=Fullscreen,
						pdfstartview=FitBH]{hyperref}
\usepackage[dotinlabels]{titletoc}
\usepackage{titlesec}
\usepackage{booktabs}
\usepackage{authblk,ulem}

\numberwithin{equation}{section}
\allowdisplaybreaks[4]

\titlelabel{\thetitle.\quad \hspace{-0.8em}}
\titlecontents{section}
              [1.5em]
              {\vspace{4mm} \large \bf}
              {\contentslabel{1em}}
              {\hspace*{-1em}}
              {\titlerule*[.5pc]{.}\contentspage}
\titlecontents{subsection}
              [3.5em]
              {\vspace{2mm}}
              {\contentslabel{1.8em}}
              {\hspace*{.3em}}
              {\titlerule*[.5pc]{.}\contentspage}
\titlecontents{subsubsection}
              [5.5em]
              {\vspace{2mm}}
              {\contentslabel{2.5em}}
              {\hspace*{.3em}}
              {\titlerule*[.5pc]{.}\contentspage}

\newcommand{\titledef}{Dark Matter Inelastic Scattering with Nuclei for Direct Detection} % Insert Title here!!!
\hypersetup{ pdfauthor = {Shao-Feng Ge},
	     pdftitle = {\titledef}, % Insert title here!!!
	     pdfsubject = {}, % Insert subject here!!!
             pdfkeywords = {}, % Insert keywords here!!!
	     pdfcreator = {LaTeX with hyperref package},
	     pdfproducer = {dvips + ps2pdf} }

\definecolor{gesfblack}{rgb}{0,0,0}

\definecolor{gesfblue}{rgb}{0.08,0.42,0.76}
\newcommand{\gblue}[1]{{\color{gesfblue} #1}}
\definecolor{gesfgreen}{rgb}{0,1,0}

\definecolor{gesfgrey}{rgb}{0.5,0.5,0.5}

\definecolor{gesflanse}{rgb}{0.00,0.50,0.50}

\definecolor{gesfpurple}{rgb}{0.47,0.19,0.42}

\definecolor{gesfred}{rgb}{1,0,0}
\newcommand{\gred}[1]{{\color{gesfred} #1}}
\definecolor{gesfwhite}{rgb}{1,1,1}

\definecolor{gesfyellow}{rgb}{0.7,0.4,0.3}

\newcommand{\gsec}[1]{{\hypersetup{linkcolor=red}Sec.\,\ref{#1}\hypersetup{linkcolor=blue}}}

\newcommand{\geqn}[1]{\hypersetup{linkcolor=blue}Eq.\,(\ref{#1})\hypersetup{linkcolor=blue}}
\newcommand{\gfig}[1]{{\hypersetup{linkcolor=violet}Fig.\,\ref{#1}\hypersetup{linkcolor=blue}}}
\newcommand{\gtab}[1]{{\hypersetup{linkcolor=gesflanse}Table~\ref{#1}\hypersetup{linkcolor=blue}}}

\definecolor{Orange}{cmyk}{0,0.61,0.87,0}
\definecolor{JungleGreen}{cmyk}{0.99,0,0.52,0}
\definecolor{OliveGreen}{cmyk}{0.64,0,0.95,0.40}
\definecolor{Brown}{cmyk}{0,0.81,1,0.60}
\definecolor{RoyalBlue}{cmyk}{0.71,0.53,0,0.12}
\definecolor{Gray}{cmyk}{0,0,0,0.40}
\definecolor{LightPink}{cmyk}{0.0,0.25,0,0}
\definecolor{LLightPink}{cmyk}{0.0,0.10,0,0}
\definecolor{LightBlue}{cmyk}{0.25,0,0,0}
\definecolor{LightGray}{cmyk}{0,0,0,0.2}

\usepackage{leftindex}
\newcommand{\bra}[1]{\left \langle #1 \right\vert}
\newcommand{\ket}[1]{\left\vert #1 \right\rangle}
\newcommand{\braket}[2]{\left\langle #1 \middle\vert #2 \right\rangle} % overlap
\newcommand{\me}[3]{\left \langle #1  \middle\vert  #2 \middle\vert #3\right \rangle} % matrix element
\newcommand{\rme}[3]{\left \langle #1  \middle\Vert  #2 \middle\Vert #3\right \rangle} % reduced matrix element
\newcommand{\CG}[3]{C_{#1\,#2}^{#3}}% CG coefficients

\graphicspath{{figs/}}

\begin{document}
\fontsize{12pt}{14pt}\selectfont

\title{%\begin{flushright}
       %\mbox{\normalsize IPMU18-xxxx}
       %\end{flushright}
			 %\vskip 20pt
       \textbf{\huge \titledef}} % Insert title here!!!
\author[1,2]{{\large Shao-Feng Ge} \footnote{\href{mailto:gesf@sjtu.edu.cn}{gesf@sjtu.edu.cn}}}
\affil[1]{State Key Laboratory of Dark Matter Physics, Tsung-Dao Lee Institute \& School of Physics and Astronomy, Shanghai Jiao Tong University, Shanghai 200240, China} 
\affil[2]{Key Laboratory for Particle Astrophysics and Cosmology (MOE) \& Shanghai Key Laboratory for Particle Physics and Cosmology, Shanghai Jiao Tong University, Shanghai 200240, China}

\author[1,2]{{\large Oleg Titov} \footnote{\href{mailto:titov_o@sjtu.edu.cn}{titov\_o@sjtu.edu.cn}}}

\author[3]{{\large Yakun Wang} \footnote{\href{mailto:wangyk@buaa.edu.cn}{wangyk@buaa.edu.cn}}}
\affil[3]{School of Physics, Beihang University, Beijing 102206, China}

\date{\today}

\maketitle

\begin{abstract}
\fontsize{12pt}{14pt}\selectfont
We investigate the nuclear responses for the WIMP-nucleus
scattering in the dark matter direct detection with particular
emphasis on the inelastic channel for the $^{129}$Xe and
$^{131}$Xe isotopes. Our generalization
incorporates both the elastic and inelastic scattering channels. 
With multipole expansion of the effective operators, the angular
momentum, parity and time-reversal selection rules can effectively
determine the allowed transitions. Instead of the nuclear shell
model, we use the state-of-the-art relativistic configuration-interaction
density functional theory, which is more suitable for heavy nuclei
such as xenon isotopes, to calculate the nuclear response functions.
For certain interaction operators, the inelastic contribution can
be comparable as its elastic counterpart and some can even dominate
by up to three orders of magnitude. Additionally, the higher
excited nuclear states can have comparable signal rate as the
first excited states. We compare our results with the nuclear
shell model calculation. The differences would have significant
effects for interpreting the dark matter direct detection searches.
\end{abstract}

\section{Introduction}

Dark matter (DM) constitutes more than $80\%$ of the matter in our Universe~\cite{Arbey:2021gdg,Young:2016ala}, yet its nature remains one of the biggest unsolved mysteries in physics.
One of the strongest-motivated DM candidates are the weakly interacting massive particles (WIMPs)~\cite{Arcadi:2024ukq}.
The search for WIMPs in direct detection experiments is mainly
using the elastic scattering on nuclei ${}^A_Z X$
\cite{MarrodanUndagoitia:2015veg,Cooley:2021rws}.
In this case, both the initial and final nuclei are
in the ground state. All the transferred energy deposits
as the kinetic energy of the final nucleus. The interest
to the elastic channel is primarily motivated by the fact
that the cross sections are enhanced by a factor of $A^2$,
where $A$ is the nucleon number,
for the simplest case of spin-independent (SI) interactions.
This is exactly the original idea proposed by Goodman and
Witten \cite{Goodman:1984dc,Wasserman:1986hh} using the so-called coherent
scattering \cite{Freedman:1973yd,Drukier:1984vhf}.

Nevertheless, it was noted early on~\cite{Goodman:1984dc,Ellis:1988nb}
that the kinetic energies of sufficiently heavy WIMPs are
comparable to the energies of low-lying excited states
in some nuclei. So the WIMP scattering can induce not
just the elastic coherent but also inelastic nuclear
transitions. In this scenario, one can search for the
nuclear recoil signal accompanied by a nuclear de-excitation.
The authors of \cite{Goodman:1984dc} have pointed out
that the inelastic channel is particularly interesting
for studying the spin-dependent (SD) interactions. 

More detailed theoretical studies of the inelastic
channel followed in 2000s, motivated by both new
developments in the calculations of nuclear structure
and advancement in the detection techniques
\cite{Engel:1999kv,Vergados:2003st,Toivanen:2008zz,Toivanen:2009zza,Baudis:2013bba,Vietze:2014vsa,Vergados:2015tua,McCabe:2015eia,Vergados:2016ytt,Arcadi:2019hrw}.
In particular, Refs.\,\cite{Toivanen:2008zz,Toivanen:2009zza,Baudis:2013bba,Vietze:2014vsa,McCabe:2015eia} considered the potential of xenon targets. Specifically, transitions to the lowest excited states of $^{129}_{54}$Xe (with excitation energy $E^* = 39.6$\,keV) and $^{131}_{54}$Xe ($E = 80.2$\,keV) were studied. It was shown that for SD interactions, the contribution of inelastic channel in these isotopes can be comparable to the elastic one. 
More recently, in Ref.~\cite{Arcadi:2019hrw}, inelastic
scattering in xenon was studied using the non-relativistic
effective field theory (NR EFT) \cite{Fitzpatrick:2012ix, Anand:2013yka}.
The authors showed that inelastic signal may even become
dominant for some DM-nucleon interactions.
Besides, the inelastic channel could be very important
for the fermionic absorption DM beyond the WIMP scenario,
especially with SD interactions \cite{Ge:2024euk}.

Accurate predictions of both elastic and inelastic DM-nucleus scattering rates require a reliable description of the nuclear many-body structure. 
Most existing calculations of nuclear responses to DM scattering have been performed within the nuclear shell model (NSM) \cite{Fitzpatrick:2012ix, Anand:2013yka,Menendez2012PRD,AbdelKhaleq:2023ipt,Klos:2013rwa}. 
In this framework, nuclear many-body wave functions are obtained by diagonalizing an effective Hamiltonian in a configuration space constructed from Slater determinants of active nucleons occupying a selected set of single-particle orbitals, conventionally referred to as the model space. 
As the number of active single-particle orbitals and valence nucleons increases, however, the dimension of the resulting configuration space grows combinatorially. 
Consequently, the practical shell-model calculations are necessarily restricted to a limited model space and, for sufficiently large configuration spaces, may require additional truncations.
For xenon isotopes $^{128,129,131}$Xe, more details can be found in
\cite{Menendez2012PRD,AbdelKhaleq:2023ipt,Klos:2013rwa}.
Such truncations inevitably raise questions about the convergence of the calculated nuclear responses and motivate calculations with reduced restrictions on both the model space and the configuration space.

A further limitation of the existing studies is related
to the nuclear final states considered for the inelastic
DM scattering. So far, the NSM calculations have mainly
focused on the elastic scattering and the inelastic transitions
to the first excited state of the target nucleus.  The
nuclear responses associated with transitions to higher
excited states have not yet been explored. A theoretical
framework capable of treating xenon isotopes in a sufficiently
large configuration space while simultaneously describing
transitions to multiple excited states is therefore highly
desirable.

The relativistic configuration-interaction density
functional (ReCD) theory is a configuration-interaction
approach rooted in the relativistic nuclear density
functional theory (DFT) \cite{Meng2016book}. 
It inherits an important advantage of the nuclear DFT
that all nucleons are treated as active particles and
a large single-particle space can be employed without
imposing severe model-space truncations that are
typically required in the conventional NSM calculations. 
At the same time, the many-body correlations beyond
the pure DFT can be incorporated through configuration
mixing to enable a microscopic description of both
the nuclear ground and excited states. 
The ReCD theory has so far been successfully applied
to a variety of nuclear phenomena, including the
nuclear charge radii \cite{Qu:2026vsb}, high-spin
spectroscopy \cite{Wang:2022wxl,Zhao:2016umr},
chiral rotation \cite{Wang:2023qll}, wobbling
motion \cite{Qu:2025qvy}, as well as the two-neutrino
and neutrinoless double beta decays
\cite{Wang:2023hkc,Wang:2024zkl,Wang:2026xd9n}.
These features make the ReCD theory particularly
suitable for studying the DM scattering off
medium-mass and heavy nuclei such as xenon isotopes. 
Its large single-particle space and capability of
describing multiple nuclear states within a unified
microscopic framework provide an opportunity to
reduce the model- and configuration-space restrictions
encountered in previous calculations. More importantly,
it can systematically investigate not only the elastic
DM–nucleus scattering but also the inelastic transitions
to higher-lying excited states.

In this paper, we present new evaluation of elastic and inelastic scattering rates in a natural xenon target for all NR EFT operators, using the state-of-the-art ReCD theory. To the best of our knowledge, the existing literature only considered the contributions of the first excited state to the inelastic channel, implicitly assuming the contributions of higher excited states to be subdominant. In this work, we quantitatively study the contributions of all the kinematically accessible levels in xenon. 
We also highlight the impact of nuclear-structure modeling on DM-nucleus scattering by comparing the ReCD-based nuclear responses with the available NSM results. 
The differences between the two approaches can lead to substantially different recoil spectra and consequently affect the interpretation of DM direct-detection searches. 
This is particularly relevant in light of the recent LUX-ZEPLIN (LZ) observation of an event consistent with a DM-induced nuclear recoil at $E_R \approx 250\,$keV \cite{Akerib:2026jyz}, with a local statistical significance of up to $3.4\sigma$, whose interpretation relies on NSM nuclear response functions. 
We find that replacing the NSM responses with the ReCD results can substantially reduce the predicted event rate around the observed recoil energy for certain interactions, thereby making the DM-signal interpretation less favorable.

This paper is organized as follows. We first discuss
the kinematics in \gsec{sec:kinematics} to show the
allowed mass range for the the WIMP-nucleus inelastic
scattering. The following \gsec{sec:EFT&responses}
gives a review of the NR EFT formalism and presents
its generalization for the inelastic scattering.
\gsec{sec:nuclear} provides a pedagogical introduction
to the ReCD theory and elaborates the evaluation of
nuclear density matrix elements.
In~\gsec{sec:results}, we present our results for
the total and differential event rates in a xenon
target and discuss the impact of the first and higher
excited states on the expected signals. Especially,
we comment on the recent LZ event in \gsec{sec:LZ}.
Our summary and conclusions can be found in \gsec{sec:conclusion}.

\section{Kinematics and Inelastic Scattering}
\label{sec:kinematics}

For the DM scattering with nuclei, it is possible to
excite the target nuclei if the DM kinetic energy
is large enough. Since both the halo DM and nucleus
are non-relativistic, the energy and momentum
conservation in the lab frame take the form
\cite{Engel:1999kv,Baudis:2013bba},
\begin{align}
  \frac{\bm p_\chi^2}{2 m_\chi}
=
  \frac{\bm p_\chi'^2}{2 m_\chi}
+ \frac {\bm q^2}{2 m_A}
+ E^*,
\quad \mbox{with} \quad
 \bm p_\chi
=
 \bm p_\chi' - \bm q,
\label{eq:E-mom-conservation}
\end{align}
where $\bm p_\chi$ and $\bm p'_\chi$ are the
initial- and final-state DM momenta while
$\bm q \equiv \bm p'_\chi - \bm p_\chi$ is
the momentum transfer. The initial DM kinetic
energy $\bm p^2_\chi / 2 m_\chi$, where $m_\chi$
is the DM mass, splits into three parts:
the final-state DM kinetic energy
$\bm p'^2 / 2 m_\chi$, the nuclear recoil
energy $E_R \equiv \bm q^2 / 2 m_A$ with
$m_A$ being the nuclear mass, and
the excitation energy $E^* \geq 0$.
Since the excitation energy $E^*$ is much
smaller than $m_A$, we
neglect the mass difference between the
nuclear ground and excited states in the
denominator of \geqn{eq:E-mom-conservation}.

Replacing the final-state DM momentum
$\bm p'_\chi$ with the sum of
its initial-state counterpart $\bm p_\chi$
and the momentum transfer $\bm q$, the
energy conservation summarized in the
first equation of \geqn{eq:E-mom-conservation}
becomes 
%,
\begin{align}
  \bm q^2
+ 2 \mu v \cos \theta |\bm q|
+ 2 \mu E^*
=
  0,
\label{eq:q_quad_eq}
\end{align}
where $\mu \equiv m_A m_\chi/(m_A + m_\chi)$
is the reduced DM-nucleus mass and
$\bm v = \bm p_\chi /m_\chi$ is the initial
DM velocity with $v \equiv |\bm v|$.
The scattering angle $\theta$
is the angle
between the initial DM momentum $\bm p_\chi$
and the momentum transfer $\bm q$.

For the inelastic scattering, with $E^* > 0$,
the following requirements must be satisfied.
Since the first and last terms in \geqn{eq:q_quad_eq}
are always positive, the equation can have solutions
only when the second term is negative and has a
sufficiently large magnitude \cite{Engel:1999kv,Baudis:2013bba},
\begin{align}
  \cos \theta < 0,
\quad
  v
= \frac 1 {|\cos \theta|}
\left(
  \frac{|\bm q|}{2 \mu}
+  \frac{E^*}{|\bm q|}
\right)
\geq
  \frac{|\bm q|}{2 \mu}
+ \frac{E^*}{|\bm q|}
\equiv
  v_\text{min}.
\label{eq:vmin}
\end{align}
Thus, the inelastic scattering is possible for DM velocities
in a narrower range than the elastic case ($E^* = 0$).

From \geqn{eq:q_quad_eq}, we can find
the magnitude of
the momentum transfer as \cite{Baudis:2013bba}
%,
\begin{align}
  |\bm q|
=
  -\mu v \cos \theta 
  \left( 1 \pm \sqrt{1 - \frac{2 E^*}{\mu v^2 \cos^2 \theta}} \right).
\label{eq:NR_scatter_exc_nucleus_q_sols}
\end{align}
For a given DM initial velocity $v$, these solutions are physical
only if the argument of the square root is non-negative \cite{Engel:1999kv,Baudis:2013bba},
\begin{align}
  E^*
\leq
  \frac  1 2 \mu v^2 \cos^2 \theta
\leq
  \frac  1 2 \mu v^2
\leq
  \frac  1 2 \mu v_\text{max}^2.
\label{eq:nucl_exc_en_NR_no_split}
\end{align}
Here $v_\text{max} \equiv v_\text{esc} + v_E$ with
$v_\text{esc} = 533$\,km/s \cite{Piffl:2013mla} is
%being
the Galactic escape velocity and $v_E = 232$\,km/s
is the average speed of the Earth in the Galaxy.
\geqn{eq:nucl_exc_en_NR_no_split} imposes a requirement
on the minimal mass $m_\chi^\text{min}$ for the DM particle~\cite{Vergados:2015tua},
\begin{align}
   m_\chi 
\geq 
  m_\chi^\text{min} 
\equiv 
    \frac{m_A E^*}{m_A v_\text{max}^2 /2 - E^*}
\geq 0.
\label{eq:min_mchi_inel}
\end{align}
For a xenon target, the denominator of \geqn{eq:min_mchi_inel}
is positive for $E^* \leq m_A v_\text{max}^2 /2 \approx 400\,$keV.
For natural xenon isotopes presented in \gtab{tab:xenon_isotopes},
the first excited levels of $^{129}$Xe ($E^* = 39.6$\,keV)
and $^{131}$Xe ($E^* = 80.2$\,keV) have small enough energies
to satisfy \geqn{eq:min_mchi_inel}, with the minimal
detectable DM masses 
$m_\chi^\text{min} \approx 14$\,GeV and
$m_\chi^\text{min} \approx 31$\,GeV, respectively.
Moreover, higher excited states of these two isotopes
are also accessible for sufficiently large DM masses,
as shown in \gtab{tab:xenon_isotopes}.
For other xenon isotopes, even the first excited states
have very high excitation energies $E^* > 400\,$keV,
for which no DM mass can satisfy the kinematics condition in
\geqn{eq:min_mchi_inel}. Therefore, halo DM can scatter
on these nuclei only elastically.

\begin{table}[h]
\centering
\begin{tabular}{ccccc}
  Isotope & Abundance ($\%$) & $J^\pi$ & $E^*$ (keV) & $m_\chi^\text{min}$ (GeV)
\\
\hline
     $^{128}_{54}$Xe & 
     $1.91$  & 
     $0^+$ & 
     0
     &
     0
\\
     & 
     & 
     $2^+$ &
     442.911
     &
     -
\\
\hline
     $^{129}_{54}$Xe & 
     $26.4$   &
     $1/2^+$ &
     0
     &
     0
\\
     & 
     & 
     $3/2^+$ &
     39.5774
     & 14
\\
     & 
     & 
     $11/2^-$ &
     236.14
     & 182
\\
     & 
     & 
     $9/2^-$ &
     274.29
     & 279
\\
     & 
     & 
     $3/2^+$ &
     318.1787
     & 512
\\
     & 
     & 
     $5/2^+$ &
     321.711
     & 543
\\
\hline
     $^{130}_{54}$Xe &  $4.07$ & $0^+$  &  0  & 0
\\
     &  & $2^+$ & 536.068 & -
\\
\hline
     $^{131}_{54}$Xe &  $21.2$ & $3/2^+$  &  0 & 0
\\
     &  &  $1/2^+$ & 80.1854 & 31
\\
     &  & $11/2^-$ & 163.93 &  86
\\
     &  & $9/2^-$ & 341.144 &  718
\\
     &  & $5/2^+$ & 364.49 &  1279
\\
\hline
     $^{132}_{54}$Xe &  $26.9$  & $0^+$  &  0 & 0
\\
     &  & $2^+$ &  667.715& -
\\
\hline
     $^{134}_{54}$Xe &  $10.4$  & $0^+$  &  0 & 0
\\
     &  & $2^+$ &  847.041& -
\\
\hline
     $^{136}_{54}$Xe &  $8.86$  & $0^+$  &  0 & 0
\\
     &  & $2^+$ & 1313.06 & -
\end{tabular}
\caption{Parameters of natural xenon isotopes: abundance,
spin and parities $J^\pi$,
the excitation energies $E^*$ of lower-lying nuclear levels,
as well as the minimal masses for exciting to the corresponding
excited energy levels as derived in
Eq.\,\eqref{eq:min_mchi_inel}.
The nuclear data are taken from \cite{IAEA}.
}
\label{tab:xenon_isotopes}
\end{table}

From \geqn{eq:NR_scatter_exc_nucleus_q_sols},
one can find the minimal and maximal recoil energies
\begin{align}
    E_R^{\text{(min)}}
&=
    \frac{\mu^2 v^2}{2 m_A} \left(1 - \sqrt{1 - \frac{2 E^*}{\mu v^2}} \right)^2,
\quad
    E_R^{\text{(max)}}
=
    \frac{\mu^2 v^2}{2 m_A} \left(1 + \sqrt{1 - \frac{2 E^*}{\mu v^2}} \right)^2.
\label{eq:inel_recoil_energy_range}
\end{align}
Thus, the range of possible recoil energies is maximal for
the case of elastic scattering, $E^* = 0$, reducing to
$E_R^{\text{(min)}} = 0$ and
$E_R^{\text{(max)}} = 2 \mu^2 v^2 / m_A$.

\section{Effective Operators and Nuclear Responses}
\label{sec:EFT&responses}

The differential event rate for inelastic scattering has the same form as in the elastic case, 
\begin{align}
    \frac{dR}{dE_R}
&=
    \frac{1}{32 \pi m_A^2 m_\chi^2} \frac{\rho_\chi}{m_\chi} 
    \int_{v \geq v_\text{min}} d^3 \bm v\,
    \frac{f_E (\bm v) }{v} \, \overline{|\mathcal{M}|^2}.
\label{eq:diff_event_rate}
\end{align}
Here $\rho_\chi \approx 0.4\,$GeV/cm$^{3}$ \cite{Arbey:2021gdg} 
is the local DM energy density, $f_E (\bm v)$ is the DM
velocity distribution at Earth's location in the rest frame
of the detector, $\overline{|\mathcal{M}|^2}$ is the averaged
squared scattering amplitude. The difference between the
elastic and inelastic regimes appears in the lower integration
limit of \geqn{eq:vmin} and at the amplitude level.
With the kinematics already discussed in \gsec{sec:kinematics},
we further explore the latter in this section. While
\gsec{sec:EFT_operators} defines all the four-fermi operators
between the DM and nucleon spinor fields, the DM and nuclear
response functions are derived in \gsec{sec:DM_nuclear_responses}.

\subsection{Effective Operators for Dark Matter-Nucleon Interaction}
\label{sec:EFT_operators}

The DM direct detection experiments with nuclear
targets aim to observe the nuclear recoil
caused by the DM-nucleus scattering. Conventionally,
only the DM kinetic energy would convert to the
nuclear recoil energy to deposit in the detector.
Then for the halo DM, both DM and nucleons are non-relativistic.
Although nucleons can have Fermi motion with momentum
up to around 200\,MeV, the non-relativistic approximation
is good enough. In addition, the involved momentum transfer
is much lower than the possible new physics scale. It is
a good approximation to use the effective operators between
the DM and nucleon fields \cite{Fitzpatrick:2012ix,Anand:2013yka}.
In this subsection, we first
summarize all the effective operators that satisfy
the Lorentz invariance in \gsec{sec:operators_Lorentz}
and then their non-relativistic expansions
in \gsec{sec:NR_EFT_operators}.

\subsubsection{Spin-Independent and Spin-Dependent Operators}
\label{sec:operators_Lorentz}

As pointed out above, the DM-nucleus scattering can be
described by the effective interactions at the nucleon level.
For a fermionic DM, the effective Lagrangian is essentially
four-fermion operators,
\begin{align}
  \mathcal L (x)
=
  g_{\chi N}
  \overline \Psi_\chi (x) \Gamma_\chi \Psi_\chi (x)
  \overline \Psi_N (x) \Gamma_N \Psi_N (x),
\label{eq:effective_relativistic_Lg}
\end{align}
where $g_{\chi N}$ is the effective coupling.
Sandwiched between the DM and nucleon fields
$\Psi_\chi$ and $\Psi_N$, the operators $\Gamma_\chi$,
$\Gamma_N = 1$, $\gamma_5$, $\gamma^\mu$, $\gamma^\mu \gamma_5$,
and $\sigma^{\mu \nu}$ act in the four-component spinor space.

The field operators satisfy the Dirac equation
\begin{align}\label{eq:Dirac_eq}
  (i \gamma^\mu \partial_\mu - m) \Psi (x)
=
  (i \gamma^\mu \partial_\mu - m)
\left\lgroup
\begin{matrix}
  \psi (x)\\
  \xi (x)
\end{matrix}
\right\rgroup
=
  0,
\end{align}
where $\psi (x)$ and $\xi (x)$ are two-component fields
in the Dirac representation.
Both the initial- and final-state DM are energy and
momentum eigenstates. On the other hand, the nucleons
confined inside nucleus are in energy eigenstates.
Therefore, both DM and nucleons can be described by
field operators as 
$\psi(x) = e^{iHt} \psi (\bm x) e^{-iHt}$
and $\xi(x) = e^{iHt} \xi (\bm x) e^{-iHt}$.
Then for the stationary states of energy $E$,
the small component has the form
\begin{align}
  \xi (\bm x)
=
- \frac{i}{E+m}
  \bm \sigma \cdot \bm \nabla \psi (\bm x),
\label{eq:2spinors_relation}
\end{align}
and is typically highly suppressed for non-relativistic
particles.

Thus, we can express both the DM and nucleon bilinears
involved in the Lagrangian \geqn{eq:effective_relativistic_Lg}
in terms of spinor fields $\psi (\bm x)$,
\begin{subequations}
\begin{align}
    \overline \Psi_\chi (\bm x) \Gamma_\chi \Psi_\chi (\bm x)
& =
    \psi_\chi^\dagger (\bm x) \mathcal O_\chi
\left(
  \frac{-i \overleftarrow{\bm \nabla}_\chi}{E_\chi + m_\chi},
  \frac{-i \overrightarrow{\bm \nabla}_\chi}{E_\chi + m_\chi}
\right)
  \psi_\chi (\bm x),
\\
    \overline \Psi_N (\bm x) \Gamma_N \Psi_N (\bm x)
& =
    \psi_N^\dagger (\bm x) \mathcal O_N \left(\frac{-i \overleftarrow{\bm \nabla}_N}{E_N + m_N}, \frac{-i \overrightarrow{\bm \nabla}_N}{E_N + m_N} \right) \psi_N (\bm x),
\end{align}
\end{subequations}
with the operators $\mathcal O_\chi$ and $\mathcal O_N$
acting in the DM and nucleon two-component spinor spaces,
respectively. The dependence of $\mathcal O_\chi$ and
$\mathcal O_N$ on the nabla operator arises from
\geqn{eq:2spinors_relation} and consequently always
appear with a factor $1/(E+m)$. In general, the nabla
operator can act from both left and right as indicated
by arrows.

To extract the non-relativistic form of these effective
interactions as detailed in the next \gsec{sec:NR_EFT_operators},
we take the DM fields as plane waves for illustration.
With definite initial and final momenta $\bm p_\chi$
and $\bm p'_\chi$, we can effectively replace the nabla
operators in the bilinear via $-i\bm \nabla \to \bm p$,
\begin{align}
    \overline \Psi_\chi (\bm x) \Gamma_\chi \Psi_\chi (\bm x)
&=
    \psi_\chi^\dagger (\bm x) \mathcal O_\chi \left(\frac{\bm p'_\chi}{E_\chi + m_\chi}, \frac{\bm p_\chi}{E_\chi + m_\chi} \right) \psi_\chi (\bm x).
\end{align}
However, the nucleons confined inside nuclei do not
have definite momentum and hence we cannot simply replace
the nabla operator with some momentum.
The Lagrangian $\mathcal L (\bm x)$ in
\geqn{eq:effective_relativistic_Lg} then takes the form 
\begin{align}
    \mathcal L (\bm x)
&=
    g_{\chi N}
    \psi_\chi^\dagger (\bm x)\psi_N^\dagger (\bm x)
    \mathcal O \left(\frac{\bm p'_\chi}{E_\chi + m_\chi}, \frac{\bm p_\chi}{E_\chi + m_\chi}, \frac{-i \overleftarrow{\bm \nabla}_N}{E_N + m_N}, \frac{-i \overrightarrow{\bm \nabla}_N}{E_N + m_N} \right) \psi_\chi (\bm x) \psi_N (\bm x),
\label{eq:eff_Lg_2comp}
\end{align}
with $\mathcal O \equiv \mathcal O_\chi \otimes \mathcal O_N$
bieng the direct product of $\mathcal O_\chi$ and $\mathcal O_N$.

For practical computations, it is convenient to take the non-relativistic limit of the operator $\mathcal O$. As shown in \gsec{sec:kinematics}, the typical values of the arguments of operator $\mathcal O_\chi$ are $|\bm p'_\chi|/(E_\chi + m_\chi) \sim |\bm p_\chi|/(E_\chi + m_\chi)\sim |\bm p_\chi|/2m_\chi\sim v \sim 10^{-3}$, so one can expand in powers of these parameters.
The nucleons inside the nucleus are not in momentum eigenstates. 
However, the magnitude of the expectation value of the operator $-i \bm \nabla_N /(E_N + m_N)$ can be estimated as $\lesssim p_F/2m_N \sim  0.1$, where $p_F \approx 250$\,MeV is the Fermi momentum of the nucleus.
Therefore, the full operator $\mathcal O$ can be expanded in its arguments up to some power.

As an example, in the case of a scalar mediator, if we keep only zeroth-order terms in DM and nucleon momenta, the Lagrangian~\geqn{eq:effective_relativistic_Lg} becomes 
\begin{align}
    \overline \Psi_\chi (\bm x) \Psi_\chi (\bm x)
    \overline \Psi_N (\bm x) \Psi_N (\bm x)
\approx
    \psi_\chi^\dagger (\bm x)\psi_N^\dagger (\bm x)
    (1_\chi \otimes 1_N)
    \psi_\chi (\bm x) \psi_N (\bm x)
\label{eq:SI_Lg},
\end{align}
with $1_\chi$ and $1_N$ being the identity operators acting in the two-component DM and nucleon spin spaces. The corresponding non-relativistic (NR) operator $(1_\chi \otimes 1_N)$ 
was dubbed ``spin-independent'' (SI).
Another example is a vector mediator with a pure axial coupling,
\begin{align}
  \overline \Psi_\chi (\bm x) \gamma_\mu \gamma_5 \Psi_\chi (\bm x)
  \overline \Psi_N (\bm x) \gamma^\mu \gamma_5 \Psi_N (\bm x)
\approx
    4
    \psi_\chi^\dagger (\bm x)\psi_N^\dagger (\bm x)
    (\bm S_\chi \otimes \bm S_N)
    \psi_\chi (\bm x) \psi_N (\bm x),
\label{eq:SD_Lg}
\end{align}
where $\bm S_\chi$ and $\bm S_N$ are the DM and
nucleon spin operators. The operator $(\bm S_\chi \otimes \bm S_N)$
is usually called
``spin-dependent'' (SD) with dependence
on the nucleon spin $\bm S_N$.

It turns out that for any Dirac bilinears,
the lowest-order expansion of \geqn{eq:effective_relativistic_Lg}
can only lead to NR operators \geqn{eq:SI_Lg}
or~\geqn{eq:SD_Lg} or vanish.
For instance, a pure vector mediator results in the SI operator~\geqn{eq:SI_Lg}.
For this reason, the early works \cite{Goodman:1984dc, Lewin+_1996, Jungman+_1996}
on DM direct detection focus on SI and SD interactions.
In addition, the results of direct detection experiments
are typically reported in terms of these interactions.

\subsubsection{Non-Relativistic Effective Operators}
\label{sec:NR_EFT_operators}

In principle, one can keep those higher-order terms
in the non-relativistic expansion of
$\mathcal O_\chi \otimes \mathcal O_N$ and induce
other momentum-dependent effective non-relativistic
operators. For example, in the case of a pseudoscalar
mediator, the first non-vanishing contribution turns
out to be quadratic in the momentum transfer $\bm q$,
\begin{align}
    \mathcal O_P
& \equiv
    \frac{\bm q \cdot \bm S_\chi}{m_\chi}
    \frac{\bm q \cdot \bm S_N}{m_N},
\label{eq:pseudoscalar_operator_NR}
\end{align}
where $m_\chi$ and $m_N$ are the DM and nucleon masses,
respectively. However, different relativistic effective
Lagrangians may induce the same effective non-relativistic operators.
Moreover, some relativistic operators may reduce to
a linear combination of standard effective non-relativistic
operators. For example, the tensor interaction induces
operators in \geqn{eq:SD_Lg} and \geqn{eq:pseudoscalar_operator_NR}.
It is then convenient to adopt the EFT point
of view and assign effective coupling coefficients
to all possible non-relativistic operators separately,
without involving too much model details.

This approach has been systematically studied in
Refs.\,\cite{Fitzpatrick:2012ix,Anand:2013yka}
(see also earlier works \cite{Dobrescu:2006au,Fan+_2010}).
Being Galilean-invariant and Hermitian, all the
non-relativistic operators can be built from the
identity matrices $1_\chi$ and $1_N$, the spin
operators $\bm{S}_\chi$ and $\bm{S}_N$, the
Hermitized momentum transfer vector $i\bm{q}$
and the transverse relative DM-nucleon velocity
$\bm{v}^\perp$,
\begin{align}
  \bm{v}^\perp
& \equiv 
  \frac 1 2
\left(
  \frac{\bm p'_\chi + \bm p_\chi}{m_\chi}
- \frac{i \overleftarrow{\bm \nabla}_N - i \overrightarrow{\bm \nabla}_N}{m_N}
\right),
\quad
  \bm v^\perp \cdot \bm q = 0.
\label{eq:v_perp_definition}
\end{align}
The full set includes 15 operators \cite{Anand:2013yka},
\begin{subequations}
\begin{align}
    \mathcal{O}_1 & \equiv 1_\chi 1_N,\quad \mathcal{O}_2 \equiv (\bm \upsilon^\perp)^2,\quad \mathcal{O}_3 \equiv i\bm{S}_N\cdotp\left(\frac{\bm{q}}{m_N}\times \bm{\upsilon}^{\perp}\right),
\\
    \mathcal{O}_4 & \equiv \bm{S}_\chi\cdotp\bm{S}_N,\quad \mathcal{O}_5 \equiv i\bm{S}_\chi\cdotp\left(\frac{\bm{q}}{m_N}\times\bm{\upsilon}^\perp\right),\quad \mathcal{O}_6 \equiv \left(\bm{S}_\chi\cdotp\frac{\bm{q}}{m_N}\right)\left(\bm{S}_N\cdotp\frac{\bm{q}}{m_N}\right),
\\
    \mathcal{O}_7 & \equiv \bm{S}_N\cdotp\bm{\upsilon}^\perp,
    \quad
    \mathcal{O}_8 \equiv \bm{S}_\chi\cdotp\bm{\upsilon}^\perp,\quad \mathcal{O}_9 \equiv i\bm{S}_\chi\cdotp\left(\bm{S}_N\times\frac{\bm{q}}{m_N}\right),
\\
     \mathcal{O}_{10} & \equiv i\bm{S}_N\cdotp\frac{\bm{q}}{m_N},
\quad
     \mathcal{O}_{11} \equiv i\bm{S}_\chi\cdotp\frac{\bm{q}}{m_N},
\quad
    \mathcal{O}_{12} \equiv \bm{S}_\chi\cdotp(\bm{S}_N\times\bm{\upsilon}^\perp),
\\
    \mathcal{O}_{13}& \equiv i(\bm{S}_\chi\cdotp\bm{\upsilon}^\perp)\left(\bm{S}_N\cdotp\frac{\bm{q}}{m_N}\right), 
\quad
    \mathcal{O}_{14} \equiv i\left(\bm{S}_\chi\cdotp\frac{\bm{q}}{m_N}\right)(\bm{S}_N\cdotp\bm{\upsilon}^\perp),
\\
    \mathcal{O}_{15}& \equiv -\left(\bm{S}_\chi\cdotp\frac{\bm{q}}{m_N}\right)\left[(\bm{S}_N\times\bm{\upsilon}^\perp)\cdotp\frac{\bm{q}}{m_N}\right].
\end{align}
\label{eq:NR_EFT_operators}
\end{subequations}
We can readily see that this set includes the standard
SI ($\mathcal O_1$) and SD ($\mathcal O_4$) operators
in \geqn{eq:SI_Lg} and \geqn{eq:SD_Lg}. The
pseudo-scalar operator \geqn{eq:pseudoscalar_operator_NR}
also enters this set, with $\mathcal O_6 \equiv -\frac{m_\chi}{m_N}\mathcal O_P$.

With the only exceptions of $\mathcal{O}_1$ and $\mathcal{O}_2$,
all operators here are spin-dependent in the sense that
they involve either the nucleon ($\bm S_N$) or DM
($\bm S_\chi$) spin operators (or both).
In particular,
not only $\mathcal O_4$, but also $\mathcal O_3$, $\mathcal O_6$,
$\mathcal O_7$, $\mathcal O_9$, $\mathcal O_{10}$, and
from $\mathcal O_{12}$ to $\mathcal O_{15}$ involve
the nucleon spin $\bm S_N$.
However, these extra operators need not to have the same
property as $\mathcal O_4$ whose subsequent elastic-scattering cross section vanishes with the nuclear spin. 
Besides $\mathcal O_4$, one may expect the spin
cancellation rule to also apply for $\mathcal O_6$, $\mathcal O_9$,
and $\mathcal O_{10}$ which involves only the nucleon
spin $\bm S_N$ on the nuclear side. The other operators
depend on not just the nucleon spin but also the nucleon
gradient through $\bm v^\perp$ and hence may not
obey the naive spin cancellation expectation.
As we will prove explicitly in \gsec{sec:selection_rules}, the operators $\mathcal{O}_{3},\,\mathcal{O}_{12},$ and $\mathcal{O}_{15}$ may cause a signal even for the elastic scattering on spinless nuclei.
Note also that, since the transverse velocity $\bm{v}^\perp$
defined in \geqn{eq:v_perp_definition} acts on both the DM
and nucleon fields, the operators in
\geqn{eq:NR_EFT_operators} involving $\bm{v}^\perp$
are not simple direct products of the form
$\mathcal O_\chi \otimes \mathcal O_N$.

The authors of \cite{Fitzpatrick:2012ix,Anand:2013yka}
also matched various relativistic operators of the form
in \geqn{eq:effective_relativistic_Lg} to the non-relativistic
EFT operators in \geqn{eq:NR_EFT_operators}, making the
connection between the relativistic coefficients $g_{\chi N}$ 
%$a$ 
and non-relativistic coefficients $c$. It turns out
that the operator $\mathcal O_2$ does not arise in the
leading-order non-relativistic expansion of any relativistic
structure. So it is typically discarded from the
non-relativistic EFT analysis.

Within the non-relativistic EFT framework,
one distinguishes between the DM-proton and
DM-neutron couplings $c_i^p$ and $c_i^n$ with
$i=1,\dots,15$ for the operators in \geqn{eq:NR_EFT_operators}.
In practical nuclear-physics computations,
it is often convenient to work in the isospin-basis
$c_i^\tau$ ($\tau = 0,1$):
\begin{align}
    c_i^0
&\equiv
    \frac{1}{2}
    (c_i^p + c_i^n),
\quad
    c_i^1
\equiv
    \frac{1}{2}
    (c_i^p - c_i^n).
\end{align}
Then the effective DM-nucleon interaction Hamiltonian density can be written as
\begin{align}
  \mathcal H
\equiv
  \psi_\chi^\dagger (\bm x)\psi_N^\dagger (\bm x)
\left[
  \sum_{\tau =0,1}
  \sum_{i=1}^{15}
  c_i^\tau
  \mathcal{O}_i
  t^\tau
\right]
  \psi_\chi (\bm x) \psi_N (\bm x).
\label{eq:NR_EFT_Hamiltonian}
\end{align}
The operators $t^\tau$ in~\geqn{eq:NR_EFT_Hamiltonian} act in the isospin space,
\begin{align}
    t^0
&\equiv
\left\lgroup
\begin{matrix}
  1 & 0 \\
  0 & 1
\end{matrix}
\right\rgroup,
\quad
    t^1
\equiv
\left\lgroup
\begin{matrix}
  1 & 0 \\
  0 & -1
\end{matrix}
\right\rgroup,
\end{align}
Following the above comment about the operator
$\mathcal O_2$, in the following we set $c_2^\alpha = c_2^\tau = 0$.

\subsection{Dark Matter and Nuclear Responses}
\label{sec:DM_nuclear_responses}

The spin-independent and spin-dependent scatterings would have
quite different scaling behaviors with the number of nucleons.
First, the SI interaction $\mathcal O_{1}$ in \geqn{eq:SI_Lg} is
proportional to the nucleon number density operator
$\psi_N^\dagger \psi_N$.
Then for the elastic scattering and in the limit
of vanishing momentum transfer ($|\bm q| \to 0$),
the corresponding nuclear matrix elements become
proportional to the nucleon number $A$ since the scattering
with different nucleons is constructively coherent.
Consequently, the cross section of the SI interaction
scales as $A^2$. On the other hand, the SD operator
$\mathcal O_{4}$ in \geqn{eq:SD_Lg} is proportional
to nucleon spin density $\psi_N^\dagger \bm S_N \psi_N $
and hence the scattering with different nucleons is
still coherent with small momentum transfer but most
of them are destructive since the nucleon spins add
up destructively to give a not that large nuclear spin.
Finally, the scattering
amplitude is proportional to the expectation value
of the total nuclei spin operator $\sum_{j = 1}^A\bm S_{N_j}$.
For spinless nuclei, the scattering amplitude and
hence the cross section should vanish. Those nuclei
with non-zero spin typically have one unpaired nucleon,
which provides the dominant
contribution to the matrix element. For this reason,
the cross section of elastic SD scattering does not
scale with the nucleon number $A$. 

The simple picture described above assumes point-like
nuclei and is only applicable in the limit of $|\bm q| \to 0$.
However, the momentum transfer in the DM-nucleus
scattering can actually be quite large. According to
\geqn{eq:NR_scatter_exc_nucleus_q_sols}, a DM particle
with mass $m_\chi = 1\,$TeV can transfer a momentum
as large as $|\bm q| \sim 200\,$MeV to a xenon nucleus.
Such process is then sensitive to the distance scale
$1/|\bm q| \sim 1\,$fm that is already smaller than
the xenon nucleus radius $R_A \approx 6\,$fm.
This makes the scattering sensitive to the nuclear
structure and can be taken into account by
introducing nuclear response functions to describe
the nucleon number and spin density distributions. 

Moreover, as we show in \gsec{sec:EFT_operators},
the list of possible effective operators is much
broader than $\mathcal O_{1}$ and $\mathcal O_{4}$.
Operators depending on number densities, spins,
momentum transfer and relative velocity in general require different response functions.
Additionally, it is possible to separate the nuclear
responses in the squared amplitude from the particle-physics
contents such as coupling and the DM spin that can be
absorbed into the DM response functions.

In this subsection, we review the formalism of DM and nuclear response functions and derive a general parametrization of the squared amplitude of elastic and inelastic DM-nucleus scattering.
In \gsec{sec:CM_motion}, we separate the center-of-mass
and intrinsic nucleon motions. Based on the multipole
decomposition explored in \gsec{sec:multipole}, we
simplify the result with reduced matrix elements
and establish the selection rules in \gsec{sec:selection_rules}.
The final \gsec{sec:responses} summarizes the DM and
nuclear response functions to give the spin-averaged
squared amplitude.

\subsubsection{Overall Nuclear and Intrinsic Nucleon Motions}
\label{sec:CM_motion}

For a composite particle such as a nucleus, the center-of-mass
motion can be separated from the intrinsic (internal)
motion of nucleons, with the total momenta $\bm p_A$
and $\bm p'_A$ describing the nucleus as a whole,
\begin{align}
  |A\rangle
=
  |\bm p_A\rangle\otimes|A_{\mathrm{int}}\rangle,
\qquad 
  |A'\rangle
=
  |\bm p'_A\rangle\otimes|A'_{\mathrm{int}}\rangle.
\label{eq:CM_state_factorization}
\end{align}
While $|\bm p_A \rangle$ and $|\bm p'_A \rangle$
describe the motion of the initial and final nuclei as a whole,
$|A_{\mathrm{int}}\rangle$ and $|A'_{\mathrm{int}}\rangle$
are the intrinsic nuclear states describing the internal
nucleonic motion with respect to the nuclear center-of-mass
frame.

Originally we define the operators $\mathcal O_i$
as functions of the momentum transfer $\bm q$ and
the transverse relative DM-nucleon velocity $\bm v^\perp$.
Consequently, the nabla operators in
\geqn{eq:v_perp_definition} should become
\begin{align}
   \frac{-i}{m_N} \bm \nabla_{N}
&=
    \frac{-i}{m_N} \bm {\nabla}_{N,\, \text{int}}
    +
    \frac{\bm p_A}{m_A},
\qquad
    \frac{-i}{m_N} \overleftarrow {\bm \nabla}_{N}
=
    \frac{-i}{m_N}  \overleftarrow {\bm \nabla}_{N,\, \text{int}}
    -
    \frac{\bm p'_A}{m_A},
\label{eq:nabla}
\end{align}
where $\bm {\nabla}_{N,\, \text{int}}$ acts on the intrinsic
coordinates. In other words, the original nabla operators are replaced
by their intrinsic counterparts and the overall nucleus
momenta $\bm p_A$ and $\bm p'_A$. Then
the $T$ matrix can have an overall
energy-momentum conservation at the nuclear level,
\begin{subequations}
\begin{align}
  i T
& \equiv
    (2\pi)^4 \delta^{(4)} (p'_\chi + p'_A - p_\chi - p_A) \mathcal M,
\label{eq:amplitude_T}
\\
    \mathcal M
& \equiv
    \sum_{\tau = 0, 1}
    \sum_{i=1}^{15}
    c_i^\tau
    \int d^3 \bm x\, 
    e^{-i \bm q \cdot \bm x}
\notag
\\
&
\hspace{15mm}
    \me{\chi',\, A'_{\mathrm{int}}}{\psi_\chi^\dagger (0) \psi_N^\dagger (\bm x) \mathcal O_i \left( \bm q, \bm v^\perp, \overleftarrow{\nabla}_{N,\, \text{int}}, \overrightarrow \nabla_{N,\, \text{int}} \right)
    t^\tau \psi_N (\bm x) \psi_\chi (0)}{\chi,\, A_{\mathrm{int}}}.
\label{eq:amplitude_a}
\end{align}
\end{subequations}
Since both the DM and nuclear 
states have definite energies,
the energy conservation can be explicitly shown with the
delta function. Since the DM is in momentum eigenstate,
its plane-wave spatial dependence $e^{\pm i \bm p \cdot \bm x}$
can be factorized out, leaving $\psi_\chi(0)$ without
any spatial dependence.

To derive~\geqn{eq:amplitude_a}, one also needs to implement the factorization \geqn{eq:CM_state_factorization} by switching to the first-quantization picture, where a nucleus is described by a many-body wave function $\braket{\bm x_1 \dots \bm x_A}{A}$ depending on the coordinates of the $A$ nucleons in the laboratory frame \cite{Walecka:2001gs,DelNobile:2021wmp}. One then switches to the center-of-mass coordinate $\bm x_\text{CM} \equiv \frac{1}{A}\sum_{i=1}^A \bm x_i$ and the intrinsic coordinates $\tilde {\bm x}_i \equiv \bm x_i - \bm x_\text{CM}$ which are constrained by the condition $\sum_{i} \tilde {\bm x}_i = 0$.
In other words, the original $A$ coordinates of $\bm x_i$
are replaced by one $\bm x_{\rm CM}$ and $A-1$ coordinates
of $\tilde{\bm x}_i$ with some nontrivial Jacobian $A^3$.
The total wave function then separates into
the center-of-mass and intrinsic parts:
$\braket{\bm x_1, \dots, \bm x_A}{A}
=
\braket{\bm x_\text{CM}}{\bm p_A}
\braket{\tilde {\bm x}_1, \dots, \tilde {\bm x}_A}{A_{\text{int}}}$. The center-of-mass wave functions of the initial and final states, $\braket{\bm x_\text{CM}}{\bm p_A}$ and $\braket{\bm x_\text{CM}}{\bm p'_A}$, are plane waves describing the nucleus as a whole,
%at the nuclear level as a whole, 
leading to the momentum-conserving delta function in \geqn{eq:amplitude_T} after integration over $\bm x_\text{CM}$. Moreover, the standard choice of the wave function normalization cancels out with the Jacobian
$A^3$ of the coordinate transformation, ensuring that no extra coefficients arise in the amplitude \cite{Walecka:2001gs,DelNobile:2021wmp}.
The remaining integral involves only intrinsic nuclear coordinates and wave functions. In \geqn{eq:amplitude_a}, we have switched back to the second-quantization picture, which is more convenient for further evaluation.

Another consequence of the nabla replacements in \geqn{eq:nabla}
is that the perpendicular velocity operator in
\geqn{eq:v_perp_definition} splits into the center-of-mass
part $\bm v_T^\perp$ and the intrinsic nucleon part $\bm v_N$,
\begin{align}
    \bm{v}^\perp
&=
    \bm v_T^\perp
    -
    \bm v_N,
\quad
    \bm v_T^\perp
\equiv
    \frac{1}{2}
    \left(
    \frac{\bm p'_\chi + \bm p_\chi}{m_\chi} - \frac{\bm p'_A + \bm p_A}{m_A} \right),
\quad
    \bm v_N
\equiv
    \frac{i \overleftarrow{\bm \nabla}_{N,\, \text{int}} - i \overrightarrow{\bm \nabla}_{N,\, \text{int}}}{2m_N}.
\label{eq:v_perp_split}
\end{align}
To emphasize the separation of velocity operators,
we rewrote the argument of $\mathcal O_i$ in \geqn{eq:amplitude_a}.
From the conservation laws in \geqn{eq:E-mom-conservation}
and the above definition of $\bm v_T^\perp$, one can see that
$\bm v_T^\perp \cdot \bm q =-E^*$ since
$\bm q = \bm p'_\chi - \bm p_\chi = \bm p_A - \bm p'_A$
with a vanishing $\bm p_A$ in the lab frame.
So both $\bm v_T^\perp$ and $\bm v_N$ are orthogonal to
$\bm q$ only in the elastic regime, $E^* = 0$. 
For further evaluation, it is convenient to work with
velocities that are orthogonal to $\bm q$ in the inelastic
regime. They can be built as,
\begin{align}
    \bm{v}^\perp
\equiv
    \widetilde{\bm v}_T^\perp - \bm v_N^\perp,
\quad
    \widetilde{\bm v}_T^\perp
\equiv
    \bm v_T^\perp + \bm q \frac{E^*}{|\bm q|^2},
\quad
    \bm v_N^\perp
\equiv
    \bm v_N + \bm q
    \frac{E^*}{|\bm q|^2},
\end{align}
with $\widetilde{\bm v}^\perp_T \cdot \bm q = 0$.
Since $\bm v^\perp \cdot \bm q = 0$ according to
\geqn{eq:v_perp_definition}, $\bm v^\perp_N$ is also
perpendicular to $\bm q$ as $\bm v^\perp_N \cdot \bm q = 0$.
In the elastic limit with $E^* \rightarrow 0$,
$\widetilde{\bm v}_T^\perp = \bm v_T^\perp$,
allowing one to easily connect the elastic and
inelastic regimes. The lower limit on the DM
velocity in the lab frame as summarized by
\geqn{eq:vmin} implies that the magnitude of
the inelastic correction, $E^* /|\bm q| \leq v \sim 10^{-3}$,
is much smaller than the typical speed
($p_F/m_N \sim 0.1$) of a nucleon in the nucleus.
Such small corrections are typically negligible
in practical nuclear physics computations
\cite{Anand:2013yka}. For this reason, we
approximate $\bm v_N^\perp \approx \bm v_N$
in the following discussions.

As noted in \gsec{sec:NR_EFT_operators}, the
effective operators $\mathcal O_i$ are in general
not a direct product of DM and nucleonic operators.
However, it is possible to separate the DM and
hadronic parts of the amplitude in \geqn{eq:amplitude_a}
\cite{Fitzpatrick:2012ix,Anand:2013yka}.
According to \geqn{eq:NR_EFT_operators} and \geqn{eq:v_perp_split},
there are five types of nucleonic operators: vector
and axial-vector charges $1_N$ and $\bm S_N \cdot \bm v_N$,
spin current $\bm S_N$, convection current $\bm v_N$,
and spin-velocity current $\bm S_N \times \bm v_N$. 
So the amplitude can be generally written in the form as
\begin{align}
    \mathcal M
&=
    \sum_{\tau = 0, 1}
    \int d^3 \bm x\, 
    e^{-i \bm q \cdot \bm x}
    \bigg[
    \me{\chi'}{\psi_\chi^\dagger (0)\ell_0^\tau \psi_\chi (0)}{\chi}
    \me{A'_{\mathrm{int}}}{\psi_N^\dagger (\bm x) 1_N t^\tau \psi_N (\bm x)}{A_{\mathrm{int}}}
\notag
\\
&+
    \me{\chi'}{\psi_\chi^\dagger (0)\ell_0^{A\tau}\psi_\chi (0)}{\chi}
    \me{A'_{\mathrm{int}}}{\psi_N^\dagger (\bm x)
   (\bm \sigma_N \cdot \bm v_N)t^\tau
    \psi_N (\bm x)}{ A_{\mathrm{int}}}
\notag
\\
&+
    \me{\chi'}{\psi_\chi^\dagger (0)\bm \ell_5^{\tau}\psi_\chi (0)}{\chi} \cdot
    \me{A'_{\mathrm{int}}}{\psi_N^\dagger (\bm x)
    \bm \sigma_N t^\tau \psi_N (\bm x)}{A_{\mathrm{int}}}
\notag
\\
&+
    \me{\chi'}{\psi_\chi^\dagger (0)\bm\ell_M^{\tau}
    \psi_\chi (0)}{\chi} \cdot
    \me{A'_{\mathrm{int}}}{\psi_N^\dagger (\bm x)
    \bm v_N t^\tau \psi_N (\bm x)}{A_{\mathrm{int}}}
\notag
\\
&+
    \me{\chi'}{\psi_\chi^\dagger (0)\bm \ell_E^{\tau}\psi_\chi (0)}{\chi} \cdot
    \me{A'_{\mathrm{int}}}{\psi_N^\dagger (\bm x)
    i[\bm \sigma_N \times\bm v_N]t^\tau \psi_N (\bm x)}{A_{\mathrm{int}}}
    \bigg],
\label{eq:amplitude_current_factorization_a}
\end{align}
where the nucelon spin operator has been replaced by
$\bm S_N \equiv \bm \sigma / 2$.
On the other hand, the DM current operators, 
$\ell_0^\tau$, $\ell_0^{A\tau}$, $\bm{\ell}^\tau_5$,
$\bm{\ell}^\tau_M$, and $\bm{\ell}^\tau_E$, act only
on the DM states and are defined as \cite{Anand:2013yka},
\begin{subequations}
\begin{align}
    \ell_0^\tau
&\equiv
    c_1^\tau 
    + i c_5^\tau
    \left[\frac{\bm q}{m_N} \times \widetilde{\bm v}_T^{\perp} \right] \cdot \bm S_\chi
    +
    c_8^\tau (\widetilde{\bm v}_T^{\perp} \cdot \bm S_\chi) 
    + i c_{11}^\tau \left(\frac{\bm q}{m_N} \cdot \bm S_\chi \right),
\\
    \ell_0^{A\tau}
&\equiv
    -\frac{1}{2} \left[c_7^\tau + i c_{14}^\tau \left(\frac{\bm q}{m_N} \cdot \bm S_\chi \right) \right],
\\
    \bm \ell_5^\tau
&\equiv
    \frac{1}{2}
\bigg\{
    ic_3^\tau \left[\frac{\bm q}{m_N} \times \widetilde{\bm v}_T^{\perp} \right]
    +c_4^\tau \bm S_\chi
    + c_6^\tau \frac{\bm q}{m_N} \left(\frac{\bm q}{m_N} \cdot \bm S_\chi \right)
    +c_7^\tau \widetilde{\bm v}_T^{\perp}
\notag
\\
& + ic_9^\tau \left[\frac{\bm q}{m_N} \times \bm S_\chi \right]
  + ic_{10}^\tau \frac{\bm q}{m_N}
  + c_{12}^\tau
    \left[\widetilde{\bm v}_T^{\perp} \times \bm S_\chi \right]
+ ic_{13}^\tau \frac{\bm q}{m_N} \left(\widetilde{\bm v}_T^{\perp} \cdot \bm S_\chi \right)
\notag
\\
&
+ ic_{14}^\tau  \widetilde{\bm v}_T^{\perp}  \left(\frac{\bm q}{m_N} \cdot \bm S_\chi \right)
+ c_{15}^\tau \left[\frac{\bm q}{m_N} \times \widetilde{\bm v}_T^{\perp} \right] \left(\frac{\bm q}{m_N} \cdot \bm S_\chi \right)
\bigg\},
\\
  \bm \ell_M^\tau
& \equiv
    ic_{5}^\tau \left[\frac{\bm q}{m_N} \times \bm S_\chi \right]
    - c_8^\tau \bm S_\chi,
\\
  \bm \ell_E^\tau
& \equiv
    \frac{1}{2}
    \bigg\{
    c_3^\tau \frac{\bm q}{m_N}
    + i c_{12}^\tau \bm S_\chi
    - c_{13}^\tau \left[\frac{\bm q}{m_N} \times \bm S_\chi \right]
    - i c_{15}^\tau \frac{\bm q}{m_N} \left(\frac{\bm q}{m_N} \cdot \bm S_\chi \right)
    \bigg\}.
\end{align}
\label{eq:DM_currents_def}
\end{subequations}
These currents incorporate all the new physics contents of the problem. 
In the following, we treat their matrix elements
$\langle \ell \rangle \equiv \me{\chi'}{\psi_\chi^\dagger (0) \ell \psi_\chi(0)}{\chi}$
as coefficients of the nuclear transition operators,
\begin{align}
    \mathcal M
&=
    \sum_{\tau = 0, 1}
    \bigg\langle A'_{\mathrm{int}} \bigg|
    \int d^3 \bm x\, 
    \bigg[
    \langle\ell_0^\tau\rangle
    \psi_N^\dagger (\bm x)  e^{-i \bm q \cdot \bm x} t^\tau \psi_N (\bm x)
\notag
\\
&+
    \frac{\langle\ell_0^{A\tau}\rangle}{2 m_N}
    \psi_N^\dagger (\bm x)
    \left\{
    i \overleftarrow{\nabla}_{N,\, \text{int}} \cdot \bm \sigma_N\,e^{-i \bm q \cdot \bm x}
    -
    i e^{-i \bm q \cdot \bm x}\, \bm \sigma_N \cdot \overrightarrow \nabla_{N,\, \text{int}}
    \right\} t^\tau
    \psi_N (\bm x)
\notag
\\
&+
    \langle \bm \ell_5^{\tau}\rangle
    \cdot
    \psi_N^\dagger (\bm x)
    \bm \sigma_N  e^{-i \bm q \cdot \bm x} t^\tau \psi_N (\bm x)
\notag
\\
&+
    \frac{\langle \bm \ell_M^{\tau} \rangle}{2 m_N}
    \cdot
    \psi_N^\dagger (\bm x)
    \left\{
    i \overleftarrow{\nabla}_{N,\, \text{int}} \,e^{-i \bm q \cdot \bm x}
    -
    i e^{-i \bm q \cdot \bm x}\,\overrightarrow \nabla_{N,\, \text{int}}
    \right\} t^\tau
    \psi_N (\bm x)
\notag
\\
&+
    \frac{\langle \bm \ell_E^{\tau} \rangle}{2 m_N}
    \cdot
    \psi_N^\dagger (\bm x)
    \left\{
    [\overleftarrow{\nabla}_{N,\, \text{int}} \times \bm \sigma_N]\,e^{-i \bm q \cdot \bm x}
    +
    e^{-i \bm q \cdot \bm x}\, [\bm \sigma_N \times \overrightarrow \nabla_{N,\, \text{int}}]
    \right\}
    t^\tau \psi_N (\bm x)
    \bigg]
    \bigg| A_{\mathrm{int}} \bigg\rangle.
\label{eq:amplitude_current_factorization_b}
\end{align}
Note that the intrinsic nucleon velocity operator
$\bm v_N$ in \geqn{eq:v_perp_split} has been implemented
to make evaluation more explicit.

\subsubsection{Multipole Decomposition}
\label{sec:multipole}
The nuclear states are typically eigenstates of angular momentum, $\ket{A_{\mathrm{int}} } \equiv \ket{J_i\, M_i}$ and $\ket{A'_{\mathrm{int}} } = \ket{J_f\, M_f}$.
Therefore, to evaluate the nuclear transition matrix element $\mathcal{M}$ in \geqn{eq:amplitude_current_factorization_b}, it is convenient to perform the multipole expansion of the nuclear transition operators \cite{Fitzpatrick:2012ix,Anand:2013yka}.
Note that the nuclear states $\ket{A_{\mathrm{int}} }$
and $\ket{A'_{\mathrm{int}} }$ as well as the operators
$\bm{\sigma}_N$, $\overrightarrow \nabla_{N,\, \text{int}}$,
and $\overrightarrow \nabla_{N,\, \text{int}}$ in
\geqn{eq:amplitude_current_factorization_b} are
independent of the momentum transfer $\bm{q}$. The only place that the momentum transfer $\bm q$
enters is the phase $e^{- i \bm q \cdot \bm x}$
that originally comes from the DM plane waves.
With a dot product between $\bm q$ and $\bm x$,
the absolute direction of the momentum transfer
is not important. The value of the matrix
element $\mathcal{M}$ then becomes independent
of the orientation of $\bm{q}$
after integrating over the spatial coordinate $\bm x$.
We therefore choose the momentum transfer to be
aligned with the $z$ axis and
expand $e^{-i \bm q \cdot \bm x} = e^{-i |\bm q| |\bm x| \cos \theta_x} = 
\sum_{J = 0}^\infty (-i)^J \sqrt{4\pi (2J+1)} j_J (|\bm q| |\bm x|) Y_{J 0}
(\theta_x)%(\Omega_{\bm x})
$,
with $j_J$ being the spherical Bessel functions of
the first kind and $Y_{J M} (\Omega_{\bm x})$ being the spherical harmonics 
depending on the angular variables $\Omega_{\bm x} \equiv (\theta_x, \phi_x)$ of the position,
for the first two terms in \geqn{eq:amplitude_current_factorization_b} that involve scalar nuclear operators. Since its the direct dot
production $\bm q \cdot \bm x$ that appears as a function of
only $\cos \theta_x$, the exponential is
independent of the azimuthal angle $\phi_x$. Consequently,
only the $\phi_x$-independent spherical harmonics $Y_{J 0}
(\theta_x)$ survive in the expansion.
The remaining terms in \geqn{eq:amplitude_current_factorization_b}
have the form of dot products $\bm \ell \cdot \bm A$ of a DM current $\bm \ell$ and a nuclear operator $\bm A$. In this case, one can write the dot product in spherical basis, $\bm \ell \cdot \bm A = \sum_{\lambda = 0, \pm 1} \ell_\lambda A_\lambda^* = \sum_{\lambda = 0, \pm 1} \ell_\lambda (\bm A \cdot \bm e_\lambda^*)$ with $\bm e_\lambda$ being the unit vectors of the spherical basis. One then decomposes the products as \cite{Walecka:1995mi}
\begin{subequations}
\begin{align}
    e^{-i \bm q \cdot \bm x} \bm e^*_{\pm 1}
& =
    \sum_{J \geq 1}
    (-i)^J
    \sqrt{2\pi(2J + 1)}
    \left\{
    \mp 
    j_J (|\bm q| |\bm x|)
    \bm Y_{J, \mp 1}^J
    -
    \frac{\bm \nabla}{|\bm q|}
    \times \left[ j_J (|\bm q| |\bm x|) \bm Y_{J, \mp 1}^J \right]
    \right\},
\\
    e^{-i \bm q \cdot \bm x} \bm e_{0}^*
&=
    \frac{i}{|\bm q|}
    \sum_{J \geq 0}
    (-i)^J
    \sqrt{4\pi(2J + 1)}
    \,
    \bm \nabla  \left[j_J (|\bm q| |\bm x|) Y_{J 0}\right],
\end{align}
\label{eq:multipole_vector_Y}
\end{subequations}
where $\bm Y_{J M}^L$ are the vector spherical harmonics
and the nabla operator without overhead arrow acts
only on the expressions in the square brackets. 
The connection between $\bm e_\lambda$ and $\bm Y_{J M}^L$ arises from the inverted definition of the vector spherical harmonics, $Y_{Lm} \bm e_\lambda = \sum_{JM} \CG{Lm}{1 \lambda}{JM} \bm Y_{J M}^L$ \cite{Walecka:1995mi}. 
Each term in this decomposition behaves as an irreducible tensor operator and respects the angular momentum conservation. Notice that the decomposition in \geqn{eq:multipole_vector_Y} has different forms for the transverse ($\lambda = \pm 1$) and longitudinal ($\lambda = 0$) components. So the corresponding terms in the amplitude behave differently (e.g., satisfy different selection rules) and should be considered separately. For further convenience, we
introduce~\cite{Fitzpatrick:2012ix,Anand:2013yka}
\begin{align}
    M_{JM} ( \bm q \cdot \bm x)
& \equiv
    j_J (|\bm q| |\bm x|) Y_{J M} (\Omega_{\bm x}),
\qquad
    \bm M_{J M}^L ( \bm q \cdot \bm x)
\equiv
    j_L (|\bm q| |\bm x|) \bm Y_{J M}^L (\Omega_{\bm x}).
\label{eq:multipole_M_functions}
\end{align}

Then the amplitude in
\geqn{eq:amplitude_current_factorization_b}
can be written as,
\begin{align}
    \mathcal M
& =
    \sum_{\tau = 0,1}
    \bigg \langle J_f\, M_f \bigg|
% \Bigg\{
    %
    \sum_{J = 0}^\infty \sqrt{4\pi (2 J + 1)} (-i)^J
    \left[\langle\ell_0^\tau\rangle M_{J0;\tau} (|\bm q|) 
    - i \langle\ell_0^{A\tau}\rangle \frac{|\bm q|}{m_N} \widetilde \Omega_{J0;\tau} (|\bm q|)\right]
\label{eq:amplitude_general}
\\
&+\sum_{J = 1}^\infty \sqrt{2\pi (2 J + 1)} (-i)^J
    \sum_{\lambda = \pm 1} (-1)^\lambda
    \bigg\{\langle \ell_{5\lambda}^\tau \rangle\left[\lambda \Sigma_{J, -\lambda; \tau} (|\bm q|) + i \Sigma'_{J, -\lambda; \tau} (q) \right]
\notag
\\
&- 
    i \frac{|\bm q|}{m_N} \langle \ell_{M\lambda}^\tau \rangle
    \left[\lambda \Delta_{J, -\lambda; \tau} (|\bm q|) + i \Delta'_{J, -\lambda; \tau} (|\bm q|)  \right]
-
    i \frac{|\bm q|}{m_N} \langle \ell_{E\lambda}^\tau \rangle \left[\lambda \widetilde \Phi_{J, -\lambda; \tau} (|\bm q|) + i \widetilde \Phi'_{J, -\lambda; \tau} (|\bm q|)  \right] \bigg\}
\notag
\\
&+
    \sum_{J = 0}^\infty \sqrt{4\pi (2 J + 1)} (-i)^J
    \left[
    i \langle \ell_{50}^\tau \rangle \Sigma''_{J0;\tau} (|\bm q|)
    +\frac{|\bm q|}{m_N} \langle \ell_{M 0}^\tau\rangle \widetilde \Delta''_{J 0;\tau} (|\bm q|)
    +\frac{|\bm q|}{m_N} \langle \ell_{E 0}^\tau \rangle \Phi''_{J 0;\tau} (|\bm q|)
    \right]
    %
%    \Bigg\}
    \bigg| J_i\, M_i \bigg \rangle.
\nonumber
\end{align}
Here we use eleven nuclear multipole operators
\cite{Fitzpatrick:2012ix,Anand:2013yka}. The first
line corresponds to the charge-type operators
$M_{J0;\tau} (|\bm q|)$ and $\widetilde \Omega_{J0;\tau} (|\bm q|)$,
the second and third line to the two transverse
($\lambda = \pm$) components of each current (unprimed
$\Sigma_{J, -\lambda; \tau} (|\bm q|)$,
$\Delta_{J, -\lambda; \tau} (|\bm q|)$,
and $\widetilde \Phi_{J, -\lambda; \tau} (|\bm q|)$
for the ``magnetic'' together with primed
$\Sigma'_{J, -\lambda; \tau} (|\bm q|)$,
$\Delta'_{J, -\lambda; \tau} (|\bm q|)$,
and $\widetilde \Phi'_{J, -\lambda; \tau} (|\bm q|)$
for the ``electric'' multipole operators),
and the last line to the longitudinal ($\lambda = 0$)
components of the currents (with double-primed multipole
operators $\Sigma''_{J0;\tau} (|\bm q|)$,
$\widetilde \Delta''_{J 0;\tau} (|\bm q|)$, and
$\Phi''_{J 0;\tau} (|\bm q|)$).
For transverse operators, the summation starts from $J=1$,
reflecting the fact that these operators change the
projection of the nuclear spin by at least one unit.
These mentioned multipole operators can be explicitly written
as \cite{Fitzpatrick:2012ix,Anand:2013yka},
\begin{subequations}
\begin{align}
    M_{JM; \tau} (|\bm q|)
& \equiv
    \int d^3 \bm x\,
    \psi_N^\dagger (\bm x)
    M_{JM} ( \bm q \cdot \bm x) t^\tau
    \psi_N (\bm x),
\\
    \widetilde \Omega_{JM; \tau} (|\bm q|)
& \equiv
    \int d^3 \bm x\,
    \psi_N^\dagger (\bm x)
\left[
  M_{JM} (\bm q \cdot \bm x)
  \bm \sigma \cdot \frac{\overrightarrow{\bm \nabla}}{|\bm q|}
+
  \bm \sigma \cdot \frac{\bm \nabla}{|2 \bm q|} M_{JM} (\bm q \cdot \bm x)
\right] t^\tau \psi_N (\bm x),
\\
    \Sigma_{JM; \tau} (|\bm q|)
& \equiv
    \int d^3 \bm x\,
    \psi_N^\dagger (\bm x)
    \bm M_{J M}^J ( \bm q \cdot \bm x)
    \cdot
    \bm \sigma \, t^\tau
    \psi_N (\bm x),
\\
    \Sigma'_{JM; \tau} (|\bm q|)
& \equiv
    -i
    \int d^3 \bm x\,
    \psi_N^\dagger (\bm x)
    \left[ \frac {\bm \nabla}{|\bm q|}  \times \bm M_{J M}^J ( \bm q \cdot \bm x) \right]
    \cdot
    \bm \sigma \, t^\tau
    \psi_N (\bm x),
\label{eq:SigmaP_def}
\\
    \Sigma''_{JM; \tau} (|\bm q|)
&\equiv
    \int d^3 \bm x\,
    \psi_N^\dagger (\bm x)
%    \left\{\frac{1}{|\bm q|}
%    \bm \nabla  M_{JM} ( \bm q \bm x) \right\}
%    \cdot \bm \sigma \,  t^\tau
    \frac{\bm \sigma \cdot \bm \nabla}{|\bm q|}
    M_{JM} ( \bm q \cdot \bm x)
    t^\tau
    \psi_N (\bm x),
\label{eq:SigmaPP_def}
\\
    \Delta_{JM; \tau} (|\bm q|)
&\equiv
    \int d^3 \bm x\,
    \psi_N^\dagger (\bm x)
    \bm M_{J M}^J ( \bm q \cdot \bm x)
    \cdot
    \frac{\overrightarrow{\bm \nabla}}{|\bm q|} t^\tau \psi_N (\bm x),
\\
    \Delta'_{JM; \tau} (|\bm q|)
& \equiv
    -i\int d^3 \bm x\,
    \psi_N^\dagger (\bm x)
    \left[ \frac{\bm \nabla}{|\bm q|} \times \bm M_{J M}^J (\bm q \cdot \bm x) \right]
    \cdot
    \frac{\overrightarrow{\bm \nabla}}{|\bm q|} t^\tau
    \psi_N (\bm x),
\\
    \widetilde \Delta''_{JM; \tau} (|\bm q|)
& \equiv
    \int d^3 \bm x\,
    \psi_N^\dagger (\bm x)
    \left\{
    \left[ \frac{\bm \nabla}{|\bm q|} M_{JM} (\bm q \cdot \bm x)\right]
    \cdot
    \frac{\overrightarrow{\bm \nabla}}{|\bm q|}
    -
    \frac{1}{2} M_{JM} (\bm q \cdot \bm x)
    \right\} \, t^\tau \psi_N (\bm x),
\\
    \widetilde \Phi_{JM; \tau} (|\bm q|)
&\equiv
    i
    \int d^3 \bm x\,
    \psi_N^\dagger (\bm x)
\left\{
  \bm M_{JM}^J ( \bm q \cdot \bm x)
  \cdot \frac{\bm \sigma \times \overrightarrow{\bm \nabla}}{|\bm q|}
+ \left[ \frac{\bm \nabla}{2 |\bm q|} \times \bm M_{JM}^J ( \bm q \cdot \bm x) \right] \cdot \bm \sigma
\right\} t^\tau \psi_N (\bm x),
\\
    \widetilde \Phi'_{JM; \tau} (|\bm q|)
& \equiv
  \int d^3 \bm x\,
  \psi_N^\dagger (\bm x)
\left\{
    \left[ \frac{\bm \nabla}{|\bm q|} \times \bm M_{JM}^J ( \bm q \cdot \bm x)\right]
    \cdot \frac{\bm \sigma \times \overrightarrow{\bm \nabla}}{|\bm q|}
    + \bm M_{JM}^J ( \bm q \cdot \bm x) \cdot \frac {\bm \sigma} 2
\right\} \,  t^\tau \psi_N (\bm x),
\\
    \Phi''_{JM; \tau} (|\bm q|)
& \equiv
  i \int d^3 \bm x\,
  \psi_N^\dagger (\bm x)
  \frac{\bm \nabla}{|\bm q|} M_{JM} ( \bm q \cdot \bm x)
  \cdot \frac{\bm \sigma \times \overrightarrow{\bm \nabla}}{|\bm q|}
  t^\tau \psi_N (\bm x).
\end{align}
\label{eq:multipole_operators_def}
\end{subequations}
Note that the coordinates and nabla operators here are intrinsic,
with the index ``$N,\, \text{int}$'' on the nabla omitted for brevity.
To avoid ambiguity, the $\bm \nabla$ operator without
overhead arrow acts only on functions $M_{JM}$ and $\bm M_{JM}^J$,
while $\overrightarrow{\bm \nabla}$ applies on everything
to the right.

\subsubsection{Reduced Matrix Elements and Selection Rules}
\label{sec:selection_rules}

The multipole operators $\mathcal T_{JM}$ are irreducible
tensor operators of rank $J$, so for further evaluation
of nuclear matrix elements, we apply the Wigner-Eckart
theorem,
\begin{align}
  \bra{J_f,\, M_f} \mathcal T_{JM} \ket{J_i,\, M_i}
=
  \frac 1 {\sqrt{2 J_f + 1}} \CG{J_i M_i}{JM}{J_f M_f}
  \rme{J_f}{\mathcal T_J}{J_i},
\label{eq:mat-reduce-mat}
\end{align}
where $\CG{J_i M_i}{JM}{J_f M_f}$ are the Clebsch-Gordan
(CG) coefficients and $\rme{J_f}{\mathcal T_J}{J_i}$ is
the reduced matrix element of the operator $\mathcal T_{JM}$.
Therefore, the amplitude in \geqn{eq:amplitude_general}
can be rewritten as
\begin{align}
  \mathcal M
= &
    \frac 1 {\sqrt{2 J_f + 1}}
    \sum_{\tau = 0,1}
\Bigg\{
\nonumber
\\
&
  \sum_{J = 0}^\infty \sqrt{4\pi (2 J + 1)} (-i)^J \CG{J_i M_i}{J 0}{J_f M_f}
  \left[
    \langle\ell_0^\tau\rangle \rme{J_f}{M_{J;\tau}}{J_i}
  - i \langle\ell_0^{A\tau} \rangle \frac{|\bm q|}{m_N}
  \langle J_f || \widetilde \Omega_{J;\tau} || J_i \rangle
  \right]
\nonumber
\\
+ &
    \sum_{J = 1}^\infty 
    \sqrt{2\pi (2 J + 1)} (-i)^J 
    \sum_{\lambda = \pm 1} (-1)^\lambda
    \CG{J_i M_i}{J, -\lambda}{J_f M_f}
\bigg[
  \langle \ell_{5\lambda}^\tau\rangle
  \left(
    \lambda \rme{J_f}{\Sigma_{J;\tau}}{J_i}
  + i \rme{J_f}{\Sigma'_{J;\tau}}{J_i}
  \right)
\nonumber
\\
- & 
  i \frac{|\bm q|}{m_N} \langle \ell_{M\lambda}^\tau\rangle
  \left(
    \lambda \rme{J_f}{\Delta_{J;\tau}}{J_i}
  + i \langle J_f || \Delta'_{J;\tau} || J_i \rangle
  \right)
-
  i \frac{|\bm q|}{m_N} \langle\ell_{E\lambda}^\tau\rangle
  \left(
    \lambda \langle J_f || \widetilde \Phi_{J;\tau} || J_i \rangle
  + i \langle J_f || \widetilde \Phi'_{J;\tau} || J_i \rangle
  \right)
\bigg]
\nonumber
\\
+ &
    \sum_{J = 0}^\infty \sqrt{4\pi (2 J + 1)} (-i)^J \CG{J_i M_i}{J 0}{J_f M_f}
\nonumber
\\
&
    \left[
    i \langle \ell_{50}^\tau\rangle \rme{J_f}{\Sigma''_{J;\tau}}{J_i}
    + \frac{|\bm q|}{m_N} \langle\ell_{M 0}^\tau\rangle
      \langle J_f || \widetilde \Delta''_{J;\tau} || J_i \rangle
    +\frac{|\bm q|}{m_N} \langle \ell_{E 0}^\tau \rangle
    \langle J_f || \Phi''_{J;\tau} || J_i \rangle
    \right]
    \Bigg\}.
\label{eq:amplitude_general_rme}
\end{align}

After reducing the nuclear matrix elements, all terms
in the amplitude involve the CG coefficients
%$\CG{J_i M_i}{J \mu}{J_f M_f}$ 
$\CG{J_i M_i}{J M}{J_f M_f}$ 
and satisfy the
following selection rules,
\begin{align}
   |J_f - J_i| &\leq J \leq J_f + J_i,
\qquad
   M_i + M = M_f,
\label{eq:multipole_rank_limit}
\end{align}
due to angular momentum conservation. The former condition
limits the rank $J$ of the multipole operators in the summation 
while the latter balance the magnetic quantum
numbers. Since the charge and longitudinal multipoles
$M_{JM;\tau}$,
$\widetilde \Omega_{JM;\tau}$, $\Sigma''_{JM;\tau}$,
$\widetilde \Delta''_{JM;\tau}$, and $\Phi''_{JM;\tau}$
have intrinsic $M = 0$ while the transverse
multipoles $\Sigma_{JM;\tau}$, $\Sigma'_{JM;\tau}$,
$\Delta_{JM;\tau}$, $\Delta'_{JM;\tau}$, $\widetilde \Phi_{JM;\tau}$,
and $\widetilde \Phi'_{JM;\tau}$ take $M = \pm 1$,
these two groups of multipole operators cannot
survive simultaneously. Given a pair of $M_i$ and
$M_f$, either the charge and longitudinal operators
contribute with $M_i = M_f$ or their transverse
counterparts survives with $|M_i - M_f| = 1$.
No transition can happen with $|M_i - M_f| > 1$.
In addition, the operator multipole expansion
for the transverse operators starts from
$J = 1$, or equivalently $J \geq 1$, since
the magnetic quantum number is already $M = \pm 1$
as denoted by $\lambda$ in \geqn{eq:amplitude_general_rme}.

Both the nuclear states $|J_i \rangle$ and
$\langle J_f|$ as well as the multipole operators
$\mathcal T_{JM;\tau}$ have definite parities.
Those operators defined in \geqn{eq:multipole_operators_def}
have parities $\pi_{\mathcal T}$
\cite{Fitzpatrick:2012ix,Anand:2013yka},
\begin{subequations}
\begin{align}
  \pi_{\mathcal T} = (-1)^J,
& \quad
  \mathcal T_{JM;\tau} = M_{JM;\tau},
  \Sigma_{JM;\tau},
  \Delta'_{JM;\tau},
  \widetilde \Delta''_{JM;\tau},
  \widetilde \Phi'_{JM;\tau},
  \Phi''_{JM;\tau},
\\
  \pi_{\mathcal T} = (-1)^{J + 1},
& \quad
  \mathcal T_{JM;\tau} = \widetilde \Omega_{JM;\tau},
  \Sigma'_{JM;\tau},
  \Sigma''_{JM;\tau},
  \Delta_{JM;\tau},
  \widetilde \Phi_{JM;\tau}.
\end{align}
\label{eq:P_selection_rules}
\end{subequations}
Note that the scalar and vector spherical harmonics
in the definition \geqn{eq:multipole_M_functions} of
$M_{JM}$ and $\bm M^L_{JM}$ contribute a parity
of $(-1)^L$ while the nabla operator in
\geqn{eq:multipole_operators_def} has odd parity.
For given parities of the initial and final nuclear
states $\pi_i$ and $\pi_f$, the transition should satisfy
$\pi_i \pi_f \pi_{\mathcal T}= 1$ which would
limit the operator rank $J$ to only even or odd values.

In addition, the operators $\mathcal T_{JM ; \tau}$ also
have definite time-reversal ($T$ or equivalently CP) properties,
$T \mathcal T_{JM ; \tau}^\dagger T^{-1} = \eta_{\mathcal{T}} \mathcal T_{JM ; \tau}$
with \cite{Anand:2013yka},
\begin{subequations}
\begin{align}
  \eta_{\mathcal{T}} & = +1,
\quad
  \mathcal T_{JM; \tau}
= M_{JM;\tau}, \widetilde \Omega_{JM;\tau}, \widetilde \Phi_{JM;\tau}, \widetilde \Phi'_{JM;\tau}, \Phi''_{JM;\tau},
\label{eq:multipole_T_properties_a}
\\
  \eta_{\mathcal{T}} &= - 1,
\quad
  \mathcal T_{JM; \tau}
= \Sigma_{JM;\tau}, \Sigma'_{JM;\tau}, \Sigma''_{JM;\tau},
     \Delta_{JM;\tau}, \Delta'_{JM;\tau}, \widetilde \Delta''_{JM;\tau}.
\label{eq:multipole_T_properties_b}
\end{align}
\label{eq:multipole_T_properties}
\end{subequations}
Using the $T$-transformation property
$T \ket{jm} = (-1)^{j + m} \ket{j, -m}$ \cite{Edmonds_1957}
of the angular momentum eigenstate $|j m \rangle$,
one can first transform the matrix element as
$\me{J_f M_f}{\mathcal T_{JM}}{J_i M_i}
=
    (-1)^{J_i + M_i + J_f + M_f}
    \eta_{\mathcal{T}}
    \me{J_i, - M_i}
    {\mathcal T_{JM}}{J_f, - M_f}$.
Then applying the Wigner-Eckart theorem in 
\geqn{eq:mat-reduce-mat}, one may reduce the two
matrix elements. Note that the two CG coefficients
with opposite magnetic quantum numbers $\pm M_i$
and $\pm M_f$ as well as switching $i$ with $f$ 
are correlated as
$\CG{J_i M_i}{J M}{J_f M_f} = (-1)^{J + M}\sqrt{\frac{2 J_f + 1}{2 J_i + 1}}
\CG{J_f, -M_f}{J M}{J_i, -M_i}$
with a sign difference $(-1)^{J+M}$.
Then only those angular momenta $J_i$, $J_f$,
and $J$ can survive in the overall sign difference
\cite{DelNobile:2021wmp,Walecka:2001gs},
\begin{align}
    \rme{J_f}
    {\mathcal T_{J}}{J_i}
&=
    (-1)^{J_f - J_i - J}
    \eta_{\mathcal{T}}
    \rme{J_i}{\mathcal T_{J}}{J_f},
\label{eq:multipole_t_property_e}
\end{align}
with the prefactors $\sqrt{2 J_i + 1}$ and $\sqrt{2 J_f + 1}$
being canceled by the CG coefficients.

For the elastic scattering with $J_i = J_f$,
one then obtains a requirement on the operator
$T$-transformation property,
\begin{align}
  \eta_{\mathcal{T}} = (-1)^{J}.
\label{eq:T_selection_elastic}
\end{align}
Combining the values of $\eta_{\mathcal{T}}$
for specific operators as listed in \geqn{eq:multipole_T_properties},
this general requirement leads to the $T$ selection
rules to give the allowed multipole ranks $J$.
For convenience, one may take \geqn{eq:T_selection_elastic}
equivalently as the $T$ selection rule.
However, there is no such selection rule
for the inelastic scattering with $J_i \neq J_f$,
since the reduced matrix elements on the two
sides of \geqn{eq:multipole_t_property_e}
are in general different from each other.

Therefore, for an inelastic transition between the initial and final states with spins and parities $J_i^{\pi_i}$ and $J_f^{\pi_f}$, the allowed rank $J$ of a multipole operator $\mathcal T_{JM ; \tau}$ is simultaneously constrained by two requirements. The triangular condition of \geqn{eq:multipole_rank_limit} constrains its minimal and maximal values, while the parity selection rule limits $J$ to be odd or even. In general, these constraints still allow all the multipole operators in \geqn{eq:multipole_operators_def} to contribute to the amplitude of \geqn{eq:amplitude_general_rme}.
This can be seen from \gtab{tab:selection_rules}, where we explicitly show the ranks of operators contributing to the kinematically accessible transitions from the ground states of $^{129}_{54}$Xe and $^{131}_{54}$Xe.

In the special case of elastic scattering, an additional $T$-selection rule arises due to~\geqn{eq:T_selection_elastic}. Since both the parity and $T$ selection rules independently enforce $J$ to be even or odd, they end up to be incompatible for some operators. For example, for elastic scattering in $^{129}_{54}$Xe via the operator $\widetilde \Omega_{JM;\tau}$, the parity selection rule requires $J$ to be odd while the $T$ selection rule requires it to be even. As a result, such transition is forbidden as labeled with a cross in \gtab{tab:selection_rules}.
Overall, only six multipole operators,
$M_{JM;\tau}$, $\Sigma'_{JM;\tau}$, $\Sigma''_{JM;\tau}$, $\Delta_{JM;\tau}$, $\widetilde \Phi'_{JM;\tau}$, and $\Phi''_{JM;\tau}$ can survive in the elastic transitions \cite{Fitzpatrick:2012ix,Anand:2013yka}. Generally speaking, the inelastic scattering regime involves a much richer nuclear dynamics than its elastic counterpart.

Applying the selection rules established above,
we can elaborate on the statement we made earlier in
\gsec{sec:NR_EFT_operators} about different behavior
of ``spin-dependent'' operators
($\mathcal O_3$, $\mathcal O_4$, $\mathcal O_6$, $\mathcal O_7$, $\mathcal O_9$,
$\mathcal O_{10}$, and from $\mathcal O_{12}$ to $\mathcal O_{15}$)
in more detail. 
From the definition of multipole operators \geqn{eq:multipole_operators_def}, we see that the operators  $\widetilde \Omega,\, \Sigma,\, \Sigma',\, \Sigma'',\, \widetilde \Phi,\, \widetilde \Phi',$ and $\Phi''$ involve the spin of the nucleon. 
For elastic transitions on spinless nuclei targets (such as
$^{128}$Xe, $^{130}$Xe, $^{132}$Xe, $^{134}$Xe, and $^{136}$Xe
in \gtab{tab:xenon_isotopes}),
one has $J_i = J_f = 0$ and consequently only the multipole
rank $J = 0$ is allowed due to the angular momentum conservation
\geqn{eq:multipole_rank_limit}. This already excludes
transverse operators ($\Sigma_{JM;\tau}$,
$\Sigma'_{JM;\tau}$, 
$\widetilde \Phi_{JM;\tau}$, and $\widetilde \Phi'_{JM;\tau}$
with $M = \pm 1$) as they require $J \geq 1$.
For the remaining operators, $\widetilde \Omega,\, \Sigma'',$ and $\Phi''$, we then take the parity selection rules into account. Since $\pi_i = \pi_f$ in elastic scattering, then, according to \geqn{eq:P_selection_rules} the allowed multipole rank must be even for the operator $\Phi''$ and odd for
$\widetilde \Omega$ and $\Sigma''$.
Given the fact that $J = 0$ is already enforced,
only $\Phi''$ can survive.
From \geqn{eq:multipole_T_properties_a} and
\geqn{eq:T_selection_elastic}, one can see
that the time selection rule does not lead to extra
constraint for $\Phi''$. Since the matrix elements of
$\mathcal O_4$, $\mathcal O_6$, $\mathcal O_7$, $\mathcal O_9$,
$\mathcal O_{10}$, $\mathcal O_{13}$, and $\mathcal O_{14}$
do not even contain $\Phi''$, these operators should have vanishing
contribution due to selection rules even without
involving the spin cancellation argument.
Only $\mathcal O_3$, $\mathcal O_{12}$, and $\mathcal{O}_{15}$
can give rise to $\Phi''$ and hence may have
non-vanishing elastic scattering event rate
on spinless xenon isotopes.

%----------------------
%\setlength{\tabcolsep}{3pt}
\begin{table}[t]
\centering
{
\setlength{\tabcolsep}{5pt}
\begin{tabular}{lcl|c|cccccc|ccccc}
\multicolumn{4}{c|}{}
%&&&
& \multicolumn{6}{c|}{$^{129}_{54}$Xe} 
& \multicolumn{5}{c}{$^{131}_{54}$Xe} \\
\hline
\multicolumn{4}{c|}{$E^*$ (keV)}
%\multicolumn{3}{c|}{}
%&&
%& $E^*$ (keV)
& 0 & 39.58 & 236.1 & 274.3 & 318.2 & 321.7
& 0 & 80.19 & 163.9 & 341.1 & 364.5
\\
\multicolumn{4}{c|}{$J^{\pi_i}_i$ (ground only) or $J_f^{\pi_f}$}
%\multicolumn{3}{c|}{}
%&&
%&$J_f^{\pi_f}$
& $1/2^+$ & $3/2^+$ & $11/2^-$ & $9/2^-$ & $3/2^+$ & $5/2^+$
& $3/2^+$ & $1/2^+$ & $11/2^-$ & $9/2^-$ & $5/2^+$
\\
\hline
% \hline
% % \multicolumn{4}{c|}{\gred{$J$ range}}
% & \multicolumn{2}{c||}{$T$, $P$ factor} 
% & \gred{$J$ range}
% & \gred{$0,1$} & \gred{$1,2$} & \gred{$5,6$} & \gred{$4,5$} & \gred{$1,2$} & \gred{$2,3$}
% & \gred{$0-3$} & \gred{$1,2$} & \gred{$4-7$} & \gred{$3-6$} & \gred{$1-4$}
% % \multicolumn{2}{c|}{AM}
% % & $0,1$ & $1,2$ & $5,6$ & $4,5$ & $1,2$ & $2,3$
% % & $0, 1, 2, 3$ & $1,2$ & $4,5,6,7$ & $3,4,5,6$ & $1,2,3,4$
% \\
%\cline{2-15}
%\multicolumn{3}{c|}{\gblue{$\pi_i \pi_f$}}
&\gred{$\eta_{\mathcal T}$}& \gblue{$\pi_{\mathcal T}$} & \gblue{$\pi_i \pi_f$}
& \gblue{+} & \gblue{+} & \gblue{$-$} & \gblue{$-$} & \gblue{+} & \gblue{+}
& \gblue{+} & \gblue{+} & \gblue{$-$} & \gblue{$-$} & \gblue{+}
\\
\hline
$M$ 
& \gred{$+$}
 & \gblue{$(-)^J$}
& $J$
& 0 & 2 & 5 & 5 & 2 & 2
& 0, 2 & 2  & 5, 7 & 3, 5 & 2, 4
\\
\hline
$\widetilde \Omega$ 
& \gred{$+$}
& \gblue{$(-)^{J+1}$}
& $J$
& $\times$ & 1 & 6 & 4 & 1 & 3
& $\times$ & 1  & 4, 6 & 4, 6 & 1, 3
\\
\hline
$\Sigma$
& \gred{$-$}
& \gblue{$(-)^{J}$}
& $J$
& $\times$ & 2 & 5 & 5 & 2 & 2
& $\times$ & 2  & 5, 7 & 3, 5 & 2, 4
\\
\hline
$\Sigma'$
& \gred{$-$}
& \gblue{$(-)^{J+1}$}
& $J$
& 1 & 1 & 6 & 4 & 1 & 3
& 1, 3 & 1  & 4, 6 & 4, 6 & 1, 3
\\
\hline
$\Sigma''$
& \gred{$-$}
& \gblue{$(-)^{J+1}$}
& $J$
& 1 & 1 & 6 & 4 & 1 & 3
& 1, 3 & 1  & 4, 6 & 4, 6 & 1, 3
\\
\hline
$\Delta$
& \gred{$-$}
& \gblue{$(-)^{J + 1}$}
& $J$
& 1 & 1 & 6 & 4 & 1 & 3
& 1, 3 & 1  & 4, 6 & 4, 6 & 1, 3
\\
\hline
$\Delta'$
& \gred{$-$}
& \gblue{$(-)^{J}$}
& $J$
& $\times$ & 2 & 5 & 5 & 2 & 2
& $\times$ & 2  & 5, 7 & 3, 5 & 2, 4
\\
\hline
$\widetilde \Delta''$
& \gred{$-$}
& \gblue{$(-)^{J}$}
& $J$
& 1 & 1 & 6 & 4 & 1 & 3
& 1, 3 & 1  & 4, 6 & 4, 6 & 1, 3
\\
\hline
$\widetilde \Phi$
& \gred{$+$}
& \gblue{$(-)^{J +1}$}
& $J$
& $\times$ & 1 & 6 & 4 & 1 & 3
& $\times$ & 1  & 4, 6 & 4, 6 & 1, 3
\\
\hline
$\widetilde \Phi'$
& \gred{$+$}
& \gblue{$(-)^{J}$}
& $J$
& $\times$ & 2 & 5 & 5 & 2 & 2
& 2 & 2  & 5, 7 & 3, 5 & 2, 4
\\
\hline
$\Phi''$
& \gred{$+$}
& \gblue{$(-)^{J}$}
& $J$
& 0 & 2 & 5 & 5 & 2 & 2
& 0, 2 & 2  & 5, 7 & 3, 5 & 2, 4
\\
\hline
\end{tabular}
}
\caption{
Selection rules for multipole operators $\mathcal T_{JM;\tau}$
contributing to transitions from ground states with
spin and parity $J^{\pi_i}_i$ to excited states with
$J_f^{\pi_f}$ and energy $E^*$ for $^{129}$Xe and $^{131}$Xe. 
For each transition, we show  
the product of the initial and final parities $\pi_i \pi_f$
which can be used for the parity selection rule.
For the inelastic scattering, the operator angular momentum
rank $J$ needs to satisfy only 
the triangular condition of Eq.\,\eqref{eq:multipole_rank_limit} 
and the parity selection rule $\pi_i \pi_f \pi_{\mathcal T}= 1$
simultaneously. In addition, the elastic transition
to the ground state receives extra constraint from
the time-reversal selection rule
according to Eq.\,\eqref{eq:T_selection_elastic}, 
with the phase factors
$\eta_{\mathcal T}$ of Eq.\,\eqref{eq:multipole_T_properties} 
and the parities $\pi_{\mathcal T}$ of
Eq.\,\eqref{eq:P_selection_rules}.
Forbidden transitions are indicated
with a cross symbol $\times$.
For the transverse operators, $\Sigma$, $\Sigma'$,
$\Delta$, $\Delta'$, $\widetilde \Phi$,
$\widetilde \Phi'$, an additional requirement $J \geq 1$
needs to be satisfied.
}
\label{tab:selection_rules}
\end{table}

\subsubsection{Squared Amplitude and Response Functions}
\label{sec:responses}

For an unpolarized target nucleus, squaring the amplitude
\geqn{eq:amplitude_general_rme} and averaging over the
nuclear spins results in a common factor
$\sum_{M_{i} M_{f}}\CG{J_i M_{i}}{J M}{J_f M_{f}} \CG{J_i M_{i}}{J', M'}{J_f M_{f}} =
\frac{2 J_f + 1}{2 J + 1}\delta_{J J'}\delta_{M M'}$
for all terms in the squared amplitude. The factor
$\delta_{J J'}$ enforces all cross-terms in the squared
amplitude to have the same multipole rank $J$ while
$\delta_{M M'}$ ensures that the charge and longitudinal
multipoles do not interfere with those transverse multipoles.

The parity selection rules only  allow interference
among operators from the same group of \geqn{eq:P_selection_rules}.
Additionally, while the condition \geqn{eq:T_selection_elastic}
does not set strict limits on the contributions of operators
\geqn{eq:multipole_operators_def} to the amplitude,
it does restrict the interference between different
operators in the spin-averaged squared amplitude.
Specifically, for given angular momentum $J_i$,
$J_f$ and the multipole rank $J$, only those operators
with the same $\eta_{\mathcal{T}}$ may interfere.
For example, the operator $M_{JM;\tau}$ interferes
with $\Phi''_{JM;\tau}$ from the same group, \geqn{eq:multipole_T_properties_a}, but not with
$\Sigma''_{JM;\tau}$ which belongs to a different
group, \geqn{eq:multipole_T_properties_b}.
Overall, three pairs of operators can interfere in the inelastic scattering: $M_{JM;\tau}$ and
$\Phi''_{JM;\tau}$, $\Sigma'_{JM;\tau}$ and $ \Delta_{JM;\tau}$,
as well as $\Sigma_{JM;\tau}$ and $\Delta'_{JM;\tau}$.

Taking into account the %those
constraints on the interference
between different multipoles, the spin-averaged squared
amplitude for unpolarized DM and nucleus takes the form %as,
\begin{align}
    \overline{|\mathcal M|^2}
& =
    \frac{4 \pi}{2 J_i + 1}
    \sum_{\tau = 0,1;\,\tau' = 0,1}
%%%% L0 Elastic
\Bigg(
     R^{\tau \tau'}_M
     W_{M}^{\tau \tau'}
+
    R^{\tau \tau'}_{\Sigma''}
    W^{\tau \tau'}_{\Sigma''}
+
    \frac{\bm q^2}{m_N^2} 
    R^{\tau \tau'}_{\Phi''}
    W^{\tau \tau'}_{\Phi''}
+
    \frac{\bm q^2}{m_N^2}
    R^{\tau \tau'}_{\Phi'' M}
    W^{\tau \tau'}_{\Phi'' M}
\notag
\\
%%%% T Elastic
&
+ 
    R^{\tau \tau'}_{\Sigma'}
    W^{\tau \tau'}_{\Sigma'}
+ 
    \frac{\bm q^2}{m_N^2} 
    R^{\tau \tau'}_{\Delta}
    W^{\tau \tau'}_{\Delta}
+
    \frac{\bm q^2}{m_N^2} 
    R^{\tau \tau'}_{\Delta \Sigma'}
    W^{\tau \tau'}_{\Delta \Sigma'}
+
    \frac{\bm q^2}{m_N^2}  
    R^{\tau \tau'}_{\widetilde \Phi'}
    W^{\tau \tau'}_{\widetilde \Phi'}
\notag
\\
%%%% L0 Inelastic
&+ 
     \frac{\bm q^2}{m_N^2} R^{\tau \tau'}_{\widetilde \Omega}
     W^{\tau \tau'}_{\widetilde \Omega}
+
    \frac{\bm q^2}{m_N^2} R^{\tau \tau'}_{\widetilde \Delta''}
    W^{\tau \tau'}_{\widetilde \Delta''}
\notag
\\
%%%% T Inelastic
&+
    R^{\tau \tau'}_\Sigma 
    W^{\tau \tau'}_\Sigma 
+
    \frac{\bm q^2}{m_N^2} 
    R^{\tau \tau'}_{\Delta'}
    W^{\tau \tau'}_{\Delta'}
+
    \frac{\bm q^2}{m_N^2} 
    R^{\tau \tau'}_{\Delta' \Sigma}
    W^{\tau \tau'}_{\Delta' \Sigma}
+
    \frac{\bm q^2}{m_N^2} 
    R^{\tau \tau'}_{\widetilde \Phi}
    W^{\tau \tau'}_{\widetilde \Phi}
\Bigg).
\label{eq:amp_squared_total_w_nuc_resp}
\end{align}
For each term, the first factors $R^{\tau \tau'}_{\mathcal T}$
are the DM response functions and the second
$W^{\tau \tau'}_{\mathcal T}$ are the nuclear response functions.
The first two lines contain the responses that appear
in both the elastic and inelastic scattering while the
last two lines only in the inelastic scattering.
The DM response functions $R^{\tau'\tau}_{\mathcal T}$
arise from the spin average of the cross products of the
DM matrix elements $\langle \ell \rangle$,
\begin{subequations}
\label{eq:DM_responses}
\begin{align}
    R^{\tau\tau'}_M
& \equiv 
    c_1^{\tau} c_1^{\tau' } 
    + \frac{j_\chi(j_\chi + 1)}{3}
    \left[
    \frac{\bm q^2}{m_N^2} (\widetilde{v}_T^{\perp})^2 c_5^{\tau}c_5^{\tau' }
    + (\widetilde{v}_T^{\perp})^2 c_8^{\tau} c_8^{\tau' }
    + \frac{\bm q^2}{m_N^2} c_{11}^{\tau} c_{11}^{\tau' }
    \right], 
    \\
     R^{\tau\tau'}_{\Sigma''} & \equiv
     \frac{\bm q^2}{4m_N^2}  c_{10}^{\tau} c_{10}^{\tau' }
     + \frac{j_\chi(j_\chi + 1)}{12}
     \left[
     c_{4}^{\tau} c_{4}^{\tau' }
     + \frac{\bm q^2}{m_N^2} (c_{4}^{\tau}c_{6}^{\tau' } + c_{6}^{\tau}c_{4}^{\tau' })
     + \frac{\bm q^4}{m_N^4} c_{6}^{\tau}c_{6}^{\tau' } \right. \notag \\
     & \left. \hspace{50mm}
     + c_{12}^{\tau}c_{12}^{\tau' } (\widetilde{v}_T^{\perp})^2
     + \frac{\bm q^2}{m_N^2} 
     (\widetilde{v}_T^{\perp})^2 c_{13}^{\tau}c_{13}^{\tau' }
     \right], 
     \label{eq:R_SigmaPP_def}\\
    R^{\tau\tau'}_{\Phi''} & = 
    \frac{\bm q^2}{4m_N^2} c_3^{\tau} c_3^{\tau' }
    + \frac{j_\chi(j_\chi + 1)}{12}
    \left(c_{12}^{\tau} - \frac{\bm q^2}{m_N^2} c_{15}^{\tau}\right)
    \left(c_{12}^{\tau'  } - \frac{\bm q^2}{m_N^2} c_{15}^{\tau'  }\right), \\
    R^{\tau\tau'}_{\Phi'' M} & \equiv 
    c_3^{\tau} c_1^{\tau'  }
    + \frac{j_\chi(j_\chi + 1)}{3}
    \left(c_{12}^{\tau} - \frac{\bm q^2}{m_N^2} c_{15}^{\tau}\right)
    c_{11}^{\tau'  }, \\
    R^{\tau\tau'}_{\Sigma'} & \equiv 
    R^{\tau\tau'}_{\Sigma} = \frac{1}{8}
    \left[
    \frac{\bm q^2}{m_N^2} (\widetilde{v}_T^{\perp})^2
    c_{3}^{\tau}c_{3}^{\tau' }
    +(\widetilde{v}_T^{\perp})^2
    c_{7}^{\tau}c_{7}^{\tau' }
    \right]
    + \frac{j_\chi(j_\chi + 1)}{12}
    \left[
    c_{4}^{\tau}c_{4}^{\tau' }
    + c_{9}^{\tau}c_{9}^{\tau' } \frac{\bm q^2}{m_N^2} \right. \notag \\
    & \left. \hspace{20mm}
    + \frac{(\widetilde{v}_T^{\perp})^2}{2}
    \left(
    c_{12}^{\tau} - \frac{\bm q^2}{m_N^2} c_{15}^{\tau}
    \right)
    \left(
    c_{12}^{\tau' } - \frac{\bm q^2}{m_N^2} c_{15}^{\tau' }
    \right)
    + \frac{\bm q^2}{2m_N^2} (\widetilde{v}_T^{\perp})^2
    c_{14}^{\tau}c_{14}^{\tau' }
    \right], 
    \label{eq:R_Sigma_def}
    \\
    R^{\tau\tau'}_{\Delta} & = R^{\tau\tau'}_{\Delta'} \equiv
    \frac{j_\chi(j_\chi + 1)}{3}
    \left(
    \frac{\bm q^2}{m_N^2} c_{5}^{\tau}c_{5}^{\tau' }
    + c_{8}^{\tau}c_{8}^{\tau' }
    \right), \\
    R^{\tau\tau'}_{\Delta \Sigma'} & \equiv -R^{\tau\tau' }_{\Delta' \Sigma} =
    \frac{j_\chi(j_\chi + 1)}{3}
    (c_{5}^{\tau}c_{4}^{\tau' } - c_{8}^{\tau}c_{9}^{\tau' }), 
    \label{eq:R_DeltaSigmaP_def}
    \\
    R^{\tau\tau'}_{\widetilde{\Phi}'} & \equiv R^{\tau\tau'}_{\widetilde{\Phi}} =
     \frac{j_\chi(j_\chi + 1)}{12}
     \left(c_{12}^{\tau} c_{12}^{\tau' }
     + \frac{\bm q^2}{m_N^2} c_{13}^{\tau} c_{13}^{\tau' } \right), 
     \label{eq:R_PhiT_def}
     \\
    R^{\tau\tau'}_{\widetilde{\Omega}} & \equiv
    \frac{1}{4}
    \left(
    c_{7}^{\tau}c_{7}^{\tau' }
    + \frac{j_\chi(j_\chi + 1)}{3}
    \frac{\bm q^2}{m_N^2} 
    c_{14}^{\tau}c_{14}^{\tau' }
    \right), 
    \label{eq:R_OmegaT_def}\\
    R^{\tau\tau'}_{\Delta''} & \equiv
    \frac{j_\chi(j_\chi + 1)}{3} c_{8}^{\tau}c_{8}^{\tau' }.
\end{align}
\end{subequations}
For convenience, we have assumed all couplings to be real.

On the other hand, the nuclear response functions are
constructed from all allowed combinations of the reduced
nuclear matrix elements.
Eight elastic nuclear response functions are defined as
\cite{Anand:2013yka},
\begin{subequations}
\begin{align}
\hspace{-5mm}
    W_{M}^{\tau \tau'}
&\equiv
    \sum_{J = 0}^\infty
    \rme{J_f}{M_{J;\tau}}{J_i}  
    \rme{J_f}{M_{{J};{\tau'}}}{J_i}^*,
\quad
    W^{\tau \tau'}_{\Sigma''}
\equiv
    \sum_{J = 0}^\infty
    \rme{J_f}{\Sigma''_{J;\tau}}{J_i}
     \rme{J_f}{\Sigma''_{{J};{\tau'}}}{J_i}^*,
\\
\hspace{-5mm}
    W^{\tau \tau'}_{\Phi''}
&\equiv
    \sum_{J = 0}^\infty
    \rme{J_f}{\Phi''_{J;\tau}}{J_i}
     \rme{J_f}{\Phi''_{{J};{\tau'}}}{J_i}^*,
\quad
    W^{\tau \tau'}_{\Phi'' M}
\equiv
    \sum_{J = 0}^\infty
    \rme{J_f}{M_{J;\tau}}{J_i}
     \rme{J_f}{\Phi''_{{J};{\tau'}}}{J_i}^*,
\\
\hspace{-5mm}
    W^{\tau \tau'}_{\Sigma'}
&\equiv
    \sum_{J = 1}^\infty
    \rme{J_f}{\Sigma'_{J;\tau}}{J_i}
     \rme{J_f}{\Sigma'_{{J};{\tau'}}}{J_i}^*,
\quad
    W^{\tau \tau'}_{\Delta}
\equiv
    \sum_{J = 1}^\infty
        \rme{J_f}{\Delta_{J;\tau}}{J_i}
     \rme{J_f}{\Delta_{{J};{\tau'}}}{J_i}^*,
\\
\hspace{-5mm}
    W^{\tau \tau'}_{\Delta \Sigma'}
&\equiv
    \sum_{J = 1}^\infty
     \rme{J_f}{\Delta_{J;\tau}}{J_i}
     \rme{J_f}{\Sigma'_{{J};{\tau'}}}{J_i}^*,
\quad
    W^{\tau \tau'}_{\widetilde \Phi'}
\equiv
  \sum_{J = 1}^\infty
  \langle J_f || \widetilde \Phi'_{J;\tau}  || J_i \rangle
  \langle J_f || \widetilde \Phi'_{J;\tau'} || J_i \rangle^*.
\end{align}
\label{eq:W_elastic}
\end{subequations}
There are also six functions that only appear for
inelastic scattering,
\begin{subequations}
\begin{align}
\hspace{-5mm}
    W^{\tau \tau'}_{\widetilde \Omega}
&\equiv
    \sum_{J = 0}^\infty
    \langle J_f || \widetilde \Omega_{J;\tau}  || J_i \rangle
    \langle J_f || \widetilde \Omega_{J;\tau'} || J_i \rangle^*,
\quad
    W^{\tau \tau'}_{\widetilde \Delta''}
\equiv
  \sum_{J = 0}^\infty
  \langle J_f || \widetilde \Delta''_{J;\tau}  || J_i \rangle
  \langle J_f || \widetilde \Delta''_{J;\tau'} || J_i \rangle^*,
\\
\hspace{-5mm}
    W^{\tau \tau'}_\Sigma
&\equiv
    \sum_{J = 1}^\infty
    \rme{J_f}{\Sigma_{J;\tau}}{J_i}
    \rme{J_f}{\Sigma_{{J};{\tau'}}}{J_i}^*,
\quad
    W^{\tau \tau'}_{\Delta'}
\equiv
    \sum_{J = 1}^\infty
    \rme{J_f}{\Delta'_{J;\tau}}{J_i}\rme{J_f}{\Delta'_{{J};{\tau'}}}{J_i}^*,
\\
\hspace{-5mm}
    W^{\tau \tau'}_{\Delta' \Sigma}
&\equiv
    \sum_{J = 1}^\infty
    \rme{J_f}{\Sigma_{J;\tau}}{J_i}
    \rme{J_f}{\Delta'_{{J};{\tau'}}}{J_i}^*,
\quad
    W^{\tau \tau'}_{\widetilde \Phi}
\equiv
  \sum_{J = 1}^\infty
  \langle J_f || \widetilde \Phi_{J;\tau}  || J_i \rangle 
  \langle J_f || \widetilde \Phi_{J;\tau'} || J_i \rangle^*.
\end{align}
\label{eq:W_inelastic}
\end{subequations}
For easy reference, the correspondence between the NR EFT operators, DM currents, nuclear multipole operators, and the response functions is summarized in~\gtab{tab:operators_responses}.
\begin{table}[h!]
\centering
\begin{tabular}{cccccc}
\toprule
$\mathcal O_i$ &DM current &Multipole& Multipole (Elastic) & Response (Inelastic) & Response (Elastic)\\
\midrule
$\mathcal O_1$ & $\ell_0$& $M$ & $M$
& $W_M$, $W_{\Phi'' M} $ &  $W_M$, $W_{\Phi'' M} $
\\
$\mathcal O_3$ & $\bm \ell_5 $ &$\Sigma$,  $\Sigma'$
&$\Sigma'$
& $W_\Sigma$, $W_{\Sigma'}$ & $W_{\Sigma'}$
\\
& $\bm \ell_E $ & $
\Phi''$
&$\Phi''$
& $W_{\Phi''}$, $W_{\Phi'' M}$ & $W_{\Phi''}$, $W_{\Phi'' M}$
\\
$\mathcal O_4$& $\bm \ell_5 $ & $\Sigma$,  $\Sigma'$, $\Sigma''$
&$\Sigma'$, $\Sigma''$
& $W_\Sigma$,  $W_{\Sigma'}$, $W_{\Sigma''}$, $W_{\Delta \Sigma'}$, $W_{\Delta' \Sigma}$
& $W_{\Sigma'}$, $W_{\Sigma''}$, $W_{\Delta \Sigma'}$
\\
$\mathcal O_5$& $\ell_0$& $M$ & $M$
& $W_M$ & $W_M$
\\
& $\bm \ell_M$& $\Delta$, $\Delta'$& $\Delta$
& $W_\Delta$, $W_{\Delta'}$, $W_{\Delta \Sigma'}$, $W_{\Delta' \Sigma}$
& $W_\Delta$, $W_{\Delta \Sigma'}$
\\
$\mathcal O_6$& $\bm \ell_5$&$\Sigma''$
&$\Sigma''$
& $W_{\Sigma''}$ & $W_{\Sigma''}$
\\
$\mathcal O_7$& $\ell_0^A$& $\widetilde \Omega$ & -
& $W_{\widetilde \Omega}$ & -
\\
& $\bm \ell_5$& $\Sigma$,  $\Sigma'$
&$\Sigma'$
& $W_\Sigma$,  $W_{\Sigma'}$
& $W_{\Sigma'}$
\\
$\mathcal O_8$& $\ell_0$& $M$ & $M$
& $W_M$ & $W_M$
\\
     & $\bm \ell_M$ & $\Delta$, $\Delta'$, $\widetilde \Delta''$& $\Delta$
& $W_\Delta$, $W_{\Delta'}$, $W_{\widetilde \Delta''}$, $W_{\Delta \Sigma'}$, $W_{\Delta' \Sigma}$
& $W_\Delta$, $W_{\Delta \Sigma'}$
\\
$\mathcal O_9$& $\bm \ell_5$& $\Sigma$,  $\Sigma'$
&$\Sigma'$
& $W_\Sigma$,  $W_{\Sigma'}$, $W_{\Delta \Sigma'}$, $W_{\Delta' \Sigma}$& $W_{\Sigma'}$, $W_{\Delta \Sigma'}$
\\
$\mathcal O_{10}$& $\bm \ell_5$& $\Sigma''$
&$\Sigma''$
& $W_{\Sigma''}$ & $W_{\Sigma''}$
\\
$\mathcal O_{11}$& $\ell_0$ & $M$ & $M$
& $W_M$, $W_{\Phi'' M} $ &  $W_M$, $W_{\Phi'' M} $
\\
$\mathcal O_{12}$& $\bm \ell_5$ & $\Sigma$,  $\Sigma'$, $\Sigma''$
&$\Sigma'$, $\Sigma''$
& $W_\Sigma$,  $W_{\Sigma'}$, $W_{\Sigma''}$
& $W_{\Sigma'}$, $W_{\Sigma''}$
\\
& $\bm \ell_E$ & $\widetilde \Phi$, $\widetilde \Phi'$, $ 
\Phi''$
&$\widetilde \Phi'$, $ 
\Phi''$
& $W_{\widetilde \Phi}$, $W_{\widetilde \Phi'}$, $ 
W_{\Phi''}$ & $W_{\widetilde \Phi'}$, $ 
W_{\Phi''}$
\\
$\mathcal O_{13}$& $\bm \ell_5$ & $\Sigma''$
&$\Sigma''$
& $W_{\Sigma''}$ & $W_{\Sigma''}$
\\
& $\bm \ell_E$ & $\widetilde \Phi$, $\widetilde \Phi'$
&$\widetilde \Phi'$
& $W_{\widetilde \Phi}$, $W_{\widetilde \Phi'}$ & $W_{\widetilde \Phi'}$
\\
$\mathcal O_{14}$& $\ell_0^A$ & $\widetilde \Omega$ & - & $W_{\widetilde \Omega}$ & -\\
& $\bm \ell_5$ & $\Sigma$,  $\Sigma'$
&$\Sigma'$
&$W_\Sigma$,  $W_{\Sigma'}$ &$W_{\Sigma'}$
\\
$\mathcal O_{15}$& $\bm \ell_5$ & $\Sigma$,  $\Sigma'$
&$\Sigma'$
&$W_\Sigma$,  $W_{\Sigma'}$&$W_{\Sigma'}$\\
& $\bm \ell_E$ &$\Phi''$
&$\Phi''$
& $W_{\Phi''}$ & $W_{\Phi''}$\\
\bottomrule
\end{tabular}
\caption{
Correspondence between the NR effective operators
$\mathcal O_i$ in Eq.\,\eqref{eq:NR_EFT_operators},
the DM currents in Eq.\,\eqref{eq:DM_currents_def},
the nuclear multipole operators in
Eq.\,\eqref{eq:multipole_operators_def} as well as the
nuclear response functions in Eq.\,\eqref{eq:W_elastic}
and Eq.\,\eqref{eq:W_inelastic}. For the nuclear
multipole operators and response functions, we show
separately the general contributions and the elastic ones.
}
\label{tab:operators_responses}
\end{table}

For elastic scattering, our analytical formula for the averaged squared amplitude~\geqn{eq:amp_squared_total_w_nuc_resp} reduced to \cite[Eq.\,(40)]{Anand:2013yka}, with the same definitions for the response functions.
However, our result has some differences from
Ref.\,\cite{Arcadi:2019hrw}, which considers
the general case of inelastic and elastic scatterings.

For evaluating the squared amplitudes
\geqn{eq:amp_squared_total_w_nuc_resp} and event
rates \geqn{eq:diff_event_rate}, we modify 
the Mathematica
package \textbf{DMFormFactor} \cite{Anand:2013yka} to
take into account the inelastic responses and the limits
on DM velocity \geqn{eq:vmin}. 
The code for computation of inelastic nuclear response
functions \geqn{eq:W_inelastic} was taken from the
\textbf{Mu2E} Mathematica package \cite{Haxton:2022piv},
which employs the same multipole formalism for nuclear
$\mu \to e$ conversion.
The crucial part for evaluation of nuclear response functions
is the description of nuclear states, whose contributions
are encoded in nuclear density matrices, as we elaborate
in the next section.

\section{Relativistic configuration-interaction density functional theory for elastic and inelastic processes}
\label{sec:nuclear}

As mentioned above, the DM with sufficient kinetic
energy can induce inelastic scattering.
In other words, it can excite the final-state nuclei to their low-lying excited states.
Since the inelastic scattering occurs concurrently with
its elastic counterpart in a specific kinematic region,
it is thus essential to explore its contribution to the
expected spectrum and the scattering rates for a complete
experimental direc detection of DM.
To achieve this goal, one first needs to establish a unified description for both the ground states and the low-lying states of the target nuclei.

In the present work, the nuclear ground and excited states are calculated using the state-of-the-art Relativistic Configuration-interaction Density functional (ReCD) theory, which can be considered as a configuration-interaction approach based on the microscopic relativistic density functional theory.
For completeness, we provide a pedagogical introduction to the ReCD theory.
To achieve this, we first expound on the basic ideas of the nuclear relativistic density functional theory (RDFT) as basis of the ReCD theory in \gsec{sec:denFunc}.
Then, we show how to incorporate the nuclear many-body
correlations into the RDFT via angular momentum projection
in \gsec{sec:rotation} and configuration mixing in
\gsec{sec:configuration}.
Based on this, the concepts of ReCD theory can be clearly elucidated
in \gsec{sec:NME}.
Finally, we demonstrate the calculations of the nuclear
density matrix, an crucial nuclear structure input for
DM-nucleus scattering, within the framework of ReCD theory
in \gsec{sec:comparison}.

\subsection{Basic ideas of nuclear relativistic density functional theory}
\label{sec:denFunc}

The nuclear density functional theory (DFT) is a widely used microscopic approach in the field of nuclear physics.
It is founded upon the fundamental theorem of Hohenberg and Kohn \cite{Hohenberg1964PR}, which demonstrates that the ground-state energy of the many-body fermionic system can be expressed as a universal functional of the particle density $\rho(\bm x)$.
Therefore, in the framework of DFT, the intractable nuclear many-body problem is reduced to a tractable one-body problem because $\rho(\bm x)$ depends only on three spatial coordinates $\bm x$.
Once a universal density functional is built, the ``correct" density that minimizes the total energy of the system can be determined by the so-called {\it Hohenberg-Kohn variational principle}.
With the obtained ``correct" density, one can in general realize a universal description for the nuclear ground-state energy.

The practical usefulness of nuclear DFT hinges entirely on whether an accurate energy density functional can be found.
The Kohn–Sham DFT \cite{Kohn1965PR} offers a highly useful approach for constructing the energy density functional. 
It achieves this by introducing a fictitious non-interacting system, the density of which is assumed to be the same as the ``correct" ground-state density. 
With the assistance of this fictitious non-interacting system, the application of Hohenberg-Kohn variational principle results in the self-consistent {\it Kohn–Sham equation}.
Note that the Kohn–Sham equation is a single-particle equation and it formally resembles the {\it Hartree–Fock equation}. 
Nevertheless, in the Kohn–Sham equation, one only needs to deal with the local densities and potentials while the complicated non-local terms that appear in the Hartree–Fock equation do not appear in the Kohn–Sham equation.

In the past few decades, the nonrelativistic Gogny and Skyrme DFTs \cite{Bender2003RMP} as well as the relativistic DFTs \cite{Meng2016book} have been developed to study  nuclear properties across the entire nuclear chart.
Among these, the relativistic DFTs leverage the Lorentz symmetry and offer an efficient description of nuclei with underlying large scalar and vector fields on the order of a few hundred MeV.
The most significant advantages of relativistic DFT are as follows: 1) The large spin-orbit splitting, which is crucial  for the description of nuclear shell effects, is naturally incorporated; 2) The Lorentz symmetry guarantees that the couplings of time-odd components (currents) of the potentials are identical to those of the time-even potentials. Consequently, no new parameters are required for the time-odd fields.

The starting point of the relativistic DFT is the following Lagrangian density~\cite{Burvenich2002PRC}
\begin{align}\label{eq:Lagrangian-density}
     \mathcal{L} &= \bar{\psi}(i\gamma_\mu\partial^\mu - m_N)\psi
     \notag \\
     &-\frac{1}{2}\alpha_S(\bar{\psi}\psi)(\bar{\psi}\psi) 
     - \frac{1}{2}\alpha_V(\bar{\psi}\gamma_\mu\psi)(\bar{\psi}\gamma^\mu\psi) 
     -\frac{1}{2}\alpha_{TV}(\bar{\psi}\vec{\tau}\gamma_\mu\psi)(\bar{\psi}\vec{\tau}\gamma^\mu\psi)
     \notag \\
     & -\frac{1}{3}\beta_S(\bar{\psi}\psi)^3 
     - \frac{1}{4}\gamma_S(\bar{\psi}\psi)^4 
     - \frac{1}{4}\gamma_V[(\bar{\psi}\gamma_\mu\psi)(\bar{\psi}\gamma^\mu\psi)]^2
     \notag \\
     &-\frac{1}{2}\delta_S\partial_\nu(\bar{\psi}\psi)\partial^\nu(\bar{\psi}\psi) 
     - \frac{1}{2}\delta_V\partial_\nu(\bar{\psi}\gamma_\mu\psi)\partial^\nu(\bar{\psi}\gamma^\mu\psi)
     -\frac{1}{2}\delta_{TV}\partial_\nu(\bar{\psi}\vec{\tau}\gamma_\mu\psi)\partial^\nu(\bar{\psi}\vec{\tau}\gamma^\mu\psi)
     \notag\\
    &-\frac{1}{4}F^{\mu\nu}F_{\mu\nu} - e\bar{\psi}\gamma^\mu \frac{1-\tau_3}{2} \psi A_\mu,
\end{align}
where $\psi$ denotes the nucleonic field and $m_N$ represents the mass of the nucleon.
In addition, $\alpha_S$, $\alpha_V$, $\alpha_{TV}$, $\beta_S$, $\gamma_S$, $\gamma_V$, $\delta_S$, $\delta_V$, and $\delta_{TV}$ are defined as coupling constants.
The subscripts $S$, $V$, and $TV$ represent the scalar, vector, and isovector, respectively.
The Greek indices $\mu$ and $\nu$ run over the Minkowski indices $0$, $1$, $2$, and $3$.
The last two terms in the Lagrangian density signify the electromagnetic interaction of protons.

Based on the Lagrangian density in \geqn{eq:Lagrangian-density}, one can construct the density functional within the framework of Kohn–Sham DFT.
Firstly, we treat the nucleus as a non-interacting system such that its ground state $|\Phi \rangle$ at the meanfield level can be expressed as a Slater determinant,
\begin{align}\label{eq:Slater}
  |\Phi\rangle
\equiv
  \prod_{i\leq A} c^\dagger_i|0\rangle,
\qquad \mbox{with} \qquad
  \langle\Phi|\Phi\rangle
= 1,
\end{align}
where $c^\dagger_i$ is the creation operator of the $i$-th nucleon.
With $A$ being the nuclear mass number, $i \leq A$
indicates that the above product state is confined
to the occupied states below the Fermi surface.
In addition, for a time-independent nuclear Hamiltonian, the nucleon field operator can be expanded in terms of the stationary single-particle wavefunctions as
\begin{align}
\label{eq:field-expansion}
  \psi(x)
=
  \sum_i \psi_i(\bm{x}) e^{-i\varepsilon_i t} c_i,
\qquad
  \psi^\dagger(x)
=
  \sum_i \psi^\dagger_i(\bm{x}) e^{i\varepsilon_i t} c_i^\dagger.
\end{align}
The nucleon creation and annihilation operators $c_i^\dagger$
and $c_i$ should satisfy the anticommutation rules,
$\{c_i, c^\dagger_{i'}\} = \delta_{ii'}$,
$\{c_i, c_{i'}\} = 0$, and
$\{c_i^\dagger, c_{i'}^\dagger\} = 0$.

The nuclear ground-state energy, namely, the Hamiltonian
expectation value $E_{\mathrm{DFT}} \equiv \langle\Phi|H|\Phi\rangle$
can then be expressed as 
\begin{align}
\label{eq:E-density-functional}
  E_{\mathrm{DFT}}
& =
  \int d\bm{x}
\bigg[
  \sum_{i=1}^A\psi^\dagger_k(\bm{\alpha}\cdotp\bm{p}+\beta m)\psi_k 
        + \frac{1}{2} \alpha_S\rho_S^2 
        + \frac{1}{2} \alpha_V j_\mu j^\mu 
        + \frac{1}{2} \alpha_{TV} (\vec{j}_{TV})_\mu\vec{j}^\mu_{TV}
\notag \\
& \hspace{15mm}
+ \frac{1}{3}\beta_S\rho_S^3 
+ \frac{1}{4}\gamma_S\rho^4_S 
+ \frac{1}{4}\gamma_V(j_\mu j^\mu)^2
\notag \\
& \hspace{15mm}
- \frac{1}{2}\delta_S\bm{\nabla}\rho_S\cdotp\bm{\nabla}\rho_S 
- \frac{1}{2}\delta_V\bm{\nabla}j_\mu\cdotp\bm{\nabla}j^\mu 
- \frac{1}{2}\delta_{TV}\bm{\nabla}(\vec{j}_{TV})_\mu\cdotp\bm{\nabla}\vec{j}^\mu_{TV} 
+\frac{1}{2}eA_\mu j_p^\mu
\bigg].
\end{align}
It is evident that $E_{\mathrm{DFT}}$ is a functional of nuclear density and currents,
\begin{subequations}
\begin{align}
  \rho_S(\bm{x})
\equiv
  \langle\Phi|\bar{\psi}\psi|\Phi\rangle
& =
  \sum_{i=1}^A \bar{\psi}_i(\bm{x)} \psi_i(\bm{x}),
\\
  j^\mu(\bm{x})
\equiv
  \langle\Phi|\bar{\psi}\gamma^\mu\psi|\Phi\rangle
& = \sum_{i=1}^A \bar{\psi}_i(\bm{x)}\gamma^\mu\psi_i(\bm{x}),
\\
  \vec{j}^\mu_{TV}(\bm{x})
\equiv
  \langle\Phi|\bar{\psi}\gamma^\mu\tau\psi|\Phi\rangle
& =
  \sum_{i=1}^A\bar{\psi}_i(\bm{x)}\gamma^\mu\tau\psi_i(\bm{x}),
\\
  j^\mu_p(\bm{x})
\equiv
  \langle\Phi|\bar{\psi}\gamma^\mu\frac{1-\tau_3}{2}\psi|\Phi\rangle
& =
  \sum_{i=1}^A\bar{\psi}_i(\bm{x)}\gamma^\mu\frac{1-\tau_3}{2}\psi_i(\bm{x}).
\end{align}
\label{eq:dens-curr}
\end{subequations}

As mentioned before, the ``correct"
density and currents
that minimize the nuclear total energy can be determined
by the variational principle, 
\begin{align}
\label{eq:variation}
  \delta
\left[
  E_{\mathrm{DFT}}(\rho_S, j_\mu, \vec j^\mu_{TV})
- \varepsilon_i \int d\bm{x} \psi^\dagger_i(\bm{x})\psi_i(\bm{x})
\right]
= 0,
\end{align}
where $\varepsilon_i$ is a Lagrange multiplier.
Its value should be adjusted to guarantee the
renormalization condition of the single-particle wavefunctions.
Regarding $\psi^\dagger_i(\bm{x})$ as the variational
parameter, \geqn{eq:variation} leads to the following
relativistic Kohn-Sham equation,
\begin{align}
\left[
  \bm{\alpha} \cdot (- i \bm \nabla - \bm V)
+ V^0
+ \beta(m+S)
\right]
  \psi_k(\bm{x})
=
  \varepsilon_k\psi_k(\bm{x}).
\label{eq:R-Kohn-Sham}
\end{align}
The scalar single-particle potential is defined as
\begin{align}
  S(\bm{x})
\equiv
  \alpha_S\rho_S
+ \beta_S\rho_S^2
+ \gamma_S\rho_S^3
+ \delta_S\Delta\rho_S,
\end{align}
with $\Delta \equiv \bm \nabla \cdot \bm \nabla$.
The vector potential $V^\mu = (V^0, \bm V)$ is defined as
\begin{align}
  V^\mu(\bm{x})
\equiv
  V^{\mu}_{V}(\bm{x})
+ \tau \cdotp \vec{V}^\mu_{TV}(\bm{x}),
\end{align}
with
\begin{subequations}
\begin{align}
  V^\mu_V(\bm x)
& \equiv
  \alpha_V j^\mu(\bm{x})
+ \gamma_V j^\mu(\bm{x}) j_\nu(\bm{x}) j^\nu(\bm{x})
+ \delta_V\Delta j^\mu
+ e\frac{1-\tau_3}{2}A^\mu,
\\
  \vec V^\mu_{TV}(\bm{x})
& \equiv
  \alpha_{TV} \vec j^\mu_{TV}(\bm{x})
+ \delta_{TV}\Delta\vec{j}^\mu_{TV}(\bm{x}).
\end{align}
\end{subequations}
Both scalar and vector single-particle potentials
depend on the nuclear density and currents which
are connected with the single-particle wavefunction
$\psi_i(\bm{x})$ as shown in \geqn{eq:dens-curr}. 
Therefore, the relativistic Kohn-Sham equation needs
to be solved iteratively in practical calculations.

For open-shell nuclei, in which the neutron and/or
proton numbers are not the magic numbers,
pairing correlations are known to play crucial
roles in nuclear structural properties. It can
be readily verified that the Slater determinant
$|\Phi\rangle$ in \geqn{eq:Slater} does not
incorporate two-body pairing correlations,
\begin{align}
  \langle\Phi|c_{l'}^\dagger c_l^\dagger|\Phi\rangle
=
  \langle \Phi|c_{l'} c_{l}|\Phi\rangle
= 0.
\end{align}
To take the pairing correlations into account, one needs to extend $|\Phi\rangle$ in \geqn{eq:Slater} to a more general product state by introducing the so-called quasiparticle concept~\cite{Ring2004Manybody} as
\begin{equation}
  \beta^\dagger_k
\equiv
  \sum_l U_{lk} c_l^\dagger + V_{lk} c_l,
\qquad
  \beta_k
\equiv
  \sum_l U_{lk}^\ast c_l + V_{lk}^\ast c_l^\dagger.
\label{eq:Bogoliubov-transformation}
\end{equation}
Here $\beta^\dagger_k$ and $\beta_k$ are quasiparticle
creation and annihilation operators while
$U$ and $V$ are the quasiparticle wave functions.
Based on these operators, the meanfield-level nuclear
ground state in \geqn{eq:Slater} can be reformulated as,
\begin{align}
\label{eq:Slater-qp}
  |\Phi_0\rangle
\equiv
  \prod_k \beta_k| 0 \rangle,
\end{align}
which satisfies $\beta_k|\Phi_0\rangle = 0$, i.e., $|\Phi_0\rangle$ is the quasiparticle vacuum with respect to the operators $\beta_k$.
Moreover, unlike the Slater determinant \geqn{eq:Slater}, it is straightforward to verify that the expectation values 
$\langle\Phi_0|c^\dagger_{l'}c^\dagger_l|\Phi_0\rangle$
and
$\langle\Phi_0|c_lc_{l'}|\Phi_0\rangle$
are in general nonzero.
This implies that $|\Phi_0\rangle$ contains components with different particle numbers connected by pair creation and annihilation operators.
Consequently, $|\Phi_0\rangle$ provides an appropriate trial state for describing the two-body pairing correlations.

With the above general product state, one
can define the {\it normal density} $\rho_{ll'}$
and the {\it abnormal density} ({\it pairing density})
$\kappa_{ll'}$ as,
\begin{align}\label{eq:abnormal-den}
  \rho_{ll'}
\equiv
  \langle\Phi_0|c_{l'}^\dagger c_l|\Phi_0\rangle
= [V^\ast V^T]_{ll'},
\qquad
  \kappa_{ll'}
\equiv
  \langle \Phi_0|c_{l'} c_l|\Phi_0\rangle
= [V^\ast U^T]_{ll'}.
\end{align}
Since now one has two different densities,
it is necessary to define a {\it generalized density
matrix} \cite{Ring2004Manybody} as,
\begin{align}
\label{eq:R-matrix}
  \mathcal R
\equiv 
\left\lgroup
\begin{array}{cc}
  \rho & \kappa \\
- \kappa^\ast & 1 - \rho^\ast
\end{array}
\right\rgroup.
\end{align}

Besides the extension of nuclear product state
$|\Phi\rangle$, one needs also introduce a
{\it pairing Hamiltonian}, 
\begin{align}
\label{eq:Hamiltonian-pair}
  \hat{H}_{\text{pair}}
\equiv
  \frac 1 2
  \int d\bm{x}d\bm{x}' \psi^\dagger(\bm{x})\psi^\dagger(\bm{x}')V^{pp}(\bm{x},\bm{x}')\psi(\bm{x})\psi(\bm{x}'),
\end{align}
with $V^{pp}(\bm{x},\bm{x}')$ being the two-body pairing interaction.
Combining \geqn{eq:field-expansion}, \geqn{eq:Slater-qp}, and
\geqn{eq:abnormal-den}, one obtains the so-called {\it pairing energy},
\begin{align}
\label{eq:pair-E}
  E_{\text{pair}}
\equiv
  \frac 1 2
  \int d\bm{x}d\bm{x}' \kappa^\ast(\bm{x},\bm{x}') V^{pp}(\bm{x},\bm{x}')\kappa(\bm{x},\bm{x}'),
\end{align}
which is a functional of the pairing density $\kappa$. 
The original energy density functional in \geqn{eq:E-density-functional}
then becomes
%\begin{align}
$ \mathcal{E}_{DFT}
\equiv
  E_{\mathrm{DFT}} + E_{\mathrm{pair}}$.
% \end{align}
%
The first term $E_{\mathrm{DFT}}$ is a functional of normal density $\rho$, while the second term $E_{\mathrm{pair}}$ is a functional of pairing density $\kappa$.

A simultaneous variation on both $\rho$ and $\kappa$,
namely the general density matrix $\mathcal{R}$ in
\geqn{eq:R-matrix}, leads to the following
relativistic {\it Hartree-Bogoliubov equation}
\cite{Meng2016book},
\begin{align}
\label{eq:RHB}
\left\lgroup
\begin{array}{cc}
  \hat{h}_D-\lambda & \hat{\Delta} \\
- \hat{\Delta}^\ast & -\hat{h}_D^\ast + \lambda
\end{array}
\right\rgroup
\left\lgroup
\begin{array}{c}
    U_k\\
    V_k
\end{array}
\right\rgroup
=
  E_k
\left\lgroup
\begin{array}{c}
    U_k\\
    V_k
\end{array}
\right\rgroup,
\end{align}
with $U_k$ and $V_k$ being the quasiparticle wave
functions as defined in \geqn{eq:Bogoliubov-transformation}
together with the quasi-particle energies $E_k$.
On the other hand, $\hat h_D \equiv \bm \alpha \cdot (\bm p -\bm V) + \beta (m + S) + V^0$
inherited from the Kohn-Sham equation in \geqn{eq:R-Kohn-Sham}
is the single-particle Dirac Hamiltonian.
The Fermi surface $\lambda$ acts as a Lagrange multiplier and is
introduced to ensure that the average particle
number equals the real particle number
\cite{Ring2004Manybody}.

Note that within the framework of quasiparticle
picture, the scalar field $S$ and the vector field
$V^\mu$ are still connected with the nuclear density
and currents defined in \geqn{eq:dens-curr}. 
Nevertheless, the single-particle wave functions
$\psi_k(\bm{x})$ and $\psi_k^\dagger(\bm{x})$
appeared in \geqn{eq:dens-curr} need to be
replaced by the quasiparticle wave functions
$V_k(\bm{x})$ and $V^\dagger_k(\bm{x})$, respectively. 
More details about the RDFT and relativistic
Hartree-Bogoliubov equation can be found in
\cite{Meng2016book,Niksic2014CPC}.

\subsection{Rotational Symmetry Restoration}
\label{sec:rotation}

As previously mentioned, within the framework of Kohn-Sham
DFT, the nuclear system with complex many-body correlations
is substituted by a fictitious non-interacting system,
namely a system with nucleons moving independently in a meanfield potential.
Consequently, correlated nucleons within the nucleus are modeled as independent nucleons moving in a self-consistent meanfield potential and the nuclear ground state is represented as a product state, namely, a Slater determinant as shown in Eqs.\,\eqref{eq:Slater} and \eqref{eq:Slater-qp}. 
The introduction of the product state and the meanfield potential significantly facilitates the nuclear structure calculations. 
The product state enables a straightforward implementation of the symmetrization principle of quantum mechanics for identical nucleons and allows the application of Wick's theorem, which is frequently employed in field theory. 
Moreover, based on the mean-field approximation, numerous sophisticated nuclear phenomena can be comprehended in an intuitive manner. 
For instance, the occurrence of nuclear rotational bands, in which the nuclear excited energies scale with the nuclear angular momentum $J$ as $J(J+1)$, similar to the spectra of a deformed rotor in quantum mechanics, can be well explained by breaking the rotational symmetry of the meanfield potential.
That is, it can be explained by introducing a deformed nuclear meanfield potential and the concept of nuclear rotation, as in done in molecular physics.

Nevertheless, while the meanfield picture offers advantages and simplicity, it is also accompanied by significant drawbacks. The meanfield potentials with symmetry breaking lead to the obtained nuclear states not being invariant with respect to the symmetry operations. 
As a result, the nuclear states at the meanfield level cannot be characterized by the symmetry quantum numbers, such as angular momentum. 
This impedes a quantitative description of nuclear transitions and scattering, where the initial and final nuclear states are labeled with angular momentum quantum numbers, and the transition rates strongly depend on the selection rules of these quantum numbers.

To overcome the drawbacks of symmetry-breaking meanfield theory, we need to implement symmetry restoration techniques. 
In the following, we demonstrate how to obtain a rotational invariant state with good angular momentum $J$.
For the restoration of the rotational symmetry broken by the deformed meanfield, the relevant symmetry group is SO(3). 
The corresponding group element is
% \begin{align}
$ \hat{R}(\Omega)
\equiv
  e^{-i \alpha \hat{J}_z} e^{-i \beta \hat{J}_y} e^{-i \gamma \hat{J}_z}$
% \end{align}
%
where $\hat{J}_y$ and $\hat{J}_z$ are the angular
momentum operators along the $y$ and $z$ axes.
The rotation is parameterized in terms of the
Euler angles $\Omega \equiv (\alpha, \beta, \gamma)$
with $\alpha \in [0, 2\pi]$, $\beta \in [0, \pi]$ and $\gamma \in [0, 2\pi]$.
Since the nuclear Hamiltonian $\hat H$ is invariant
under the operation $\hat{R}(\Omega)$, its expectation
value should also be rotationally invariant,
\begin{align}\label{eq:E-Phi}
    \frac{\langle \Phi_0|\hat{R}^\dagger(\Omega) \hat{H} \hat{R}(\Omega)|\Phi_0\rangle}{\langle \Phi_0|\hat{R}^\dagger(\Omega) \hat{R}(\Omega)|\Phi_0\rangle}
    =
    \frac{\langle \Phi_0|\hat{H}|\Phi_0\rangle}{\langle\Phi_0|\Phi_0\rangle}.
\end{align}
In other words, those states $\hat{R}(\Omega)|\Phi_0\rangle$
with different orientations are degenerate in energy.
We then construct a generalized state in the form of linear combination,
\begin{align}\label{eq:Psi}
  |\Psi\rangle
\equiv
  \int d\Omega F(\Omega) \hat{R} (\Omega) |\Phi_0\rangle.
\end{align}
The variational parameters $F$ can be determined
by minimizing the energy expectation value
\begin{align}
  E
\equiv
  \frac {\langle \Psi|\hat{H}|\Psi\rangle}
        {\langle \Psi|\Psi\rangle}.
\label{eq:E-Psi}
\end{align}
Note that the energy expectation obtained in \geqn{eq:E-Psi}
is typically lower than that in \geqn{eq:E-Phi} since
the Slater determinant
$|\Phi_0\rangle = \hat{R}(0)|\Phi_0\rangle$ is just
a subset of the more general state $|\Psi\rangle$.
Refer to Ref.\,\cite{Sheikh2021JPG} for more details.

It is easy to verify that the constructed trial state $|\Psi\rangle$ in \geqn{eq:E-Psi} is rotationally invariant. 
Nevertheless, the angular momentum quantum number
is not explicitly present in $|\Psi\rangle$ at this stage.
To obtain a state with well-defined angular momentum,
we expand the weight $F(\Omega)$ in terms of the
continuous single-valued representations of the
SO(3) group,
\begin{align}
  F(\Omega)
\equiv
  \sum_{JMK}
  \frac {2J+1}{8\pi^2}
  F^J_{MK}
  D^{J\ast}_{MK} (\Omega),
\label{eq:FOmega}
\end{align}
where $D^J_{MK}$ is the Wigner-$D$ function and
$F^J_{MK}$ the variational variables.
Then $|\Psi\rangle$ in \geqn{eq:Psi} can be expressed as
\begin{align}\label{eq:Psi-sum-JMK}
  |\Psi\rangle
=
  \sum_{JMK} F^J_{MK} \frac{2J + 1}{8 \pi^2}
  \int d \Omega D^{J\ast}_{MK} (\Omega) \hat{R}(\Omega) |\Phi_0\rangle 
\equiv
  \sum_{JMK} F^J_{MK} \hat{P}^{J}_{MK} |\Phi_0\rangle,
\end{align}
where 
% \begin{align}
$\hat{P}^J_{MK}
\equiv
  \frac{2J + 1}{8 \pi^2}
  \int d \Omega D^{J\ast}_{MK} (\Omega) \hat{R}(\Omega)$
%\end{align}
%
is the so-called {\it angular momentum projection operator}.

Suppose that $|\mu JM\rangle$ represents a state
with a well-defined angular momentum $J$, magnetic
quantum number $M$, and $\mu$ is the remaining
quantum number that uniquely specifies the state. 
Then, we have the closure relation,
\begin{align}\label{eq:closure-rule}
  \sum_{\mu JM} |\mu JM\rangle \langle \mu JM|
= 1,
\end{align}
and the multiplet relation
\begin{align}\label{eq:multiplet-relation}
  \hat{R}(\Omega) |\mu J K\rangle
  \equiv
  \sum_M D^J_{MK} (\Omega) |\mu JM\rangle.
\end{align}
The Wigner-$D$ function acts like the
representation matrix of the rotational
operation $\hat R(\Omega)$ on the states
$|\mu J M \rangle$.
Based on \geqn{eq:multiplet-relation} and the
orthogonality of the Wigner-$D$ function,
\begin{align}
  \int d \Omega D^{J\ast}_{MK} (\Omega) D^{J'}_{M'K'}(\Omega) 
=
  \frac{8\pi^2}{2J + 1} \delta_{JJ'} \delta_{MM'} \delta_{KK'},
\end{align}
we have
\begin{align}\label{eq:Projector-act}
  \hat{P}^J_{MK} |\mu J' K'\rangle 
= \frac{2J + 1}{8\pi^2}
  \int d\Omega D^{J\ast}_{MK}(\Omega) \hat{R}(\Omega) |\mu J' K'\rangle
= \delta_{JJ'} \delta_{KK'} |\mu J M\rangle.
\end{align}
In other words, $\hat P^J_{MK}$ projects
out those states with $J' = J$ and $K' = K$.

From \geqn{eq:Projector-act} and \geqn{eq:closure-rule},
we are able to obtain the spectral representation
and the sum rule of the angular momentum projector,
\begin{align}
\label{eq:spectral-sum-rule}
  \hat{P}^J_{MK}
\equiv
  \sum_\mu |\mu JM\rangle \langle \mu JK\rangle, 
\quad
  \sum_{JM} \hat{P}^J_{MM}
= 1.
\end{align}
Moreover, by using the spectral representation,
one can readily verify the following properties
for $\hat{P}^J_{MK}$,
\begin{align}
  \hat P^{J\dagger}_{MK}
= \hat{P}^J_{KM},
\quad
  \hat{P}^J_{KM} \hat{P}^{J'}_{M'K'}
= \delta_{JJ'} \delta_{MM'} \delta_{KK'}.
\end{align}
With the spectral representation of $\hat{P}^J_{MK}$ as shown in \geqn{eq:spectral-sum-rule}, we obtain
\begin{align}\label{eq:PJ-act-Phi}
    \hat{P}^J_{MK} |\Phi_0\rangle = \sum_\mu |\mu JM\rangle \langle \mu JK|\Phi_0\rangle = \sum_\mu C_{\mu J K} |\mu J M\rangle,
\end{align}
where the coefficients $C_{\mu J K} \equiv \langle \mu JK|\Phi_0\rangle$ are merely $c$ numbers.
Since $|\mu J M\rangle$ is an eigenstate of the angular momentum operator $\hat{J}$, one can conclude that $\hat{P}^J_{MK} |\Phi_0\rangle$ is also an eigenstate of $\hat{J}$ with eigenvalues $J$ and $M$. 
Therefore, from \geqn{eq:PJ-act-Phi}, one can now recognize that the angular momentum operator $\hat{P}^J_{MK}$ projects out an eigenstate $|\mu JM\rangle$ from the rotational symmetry-breaking meanfield state $|\Phi_0\rangle$.

Based on these clarifications, the physical meaning of the generalized state $|\Psi_0\rangle$ in \geqn{eq:Psi-sum-JMK} becomes clear: $|\Psi\rangle$ is actually a symmetry-restored state expanded in terms of the projected basis $\hat{P}^J_{MK} |\Phi_0\rangle$ with well-defined quantum numbers $J$ and $M$. 
The weights $F^J_{MK}$ are determined by the variational principle, namely, by minimizing the energy as shown in \geqn{eq:E-Psi}. 
Note that the angular momentum eigenstates with different eigenvalues $J$ and $M$ are orthogonal to each other. 
Therefore, in practical calculations, one can perform the variational procedure by eliminating the summation over $J$ and $M$, that is, perform the variational procedure in a subspace spanned by given $J$ and $M$ quantum numbers. 
Then it is sufficient to carry out the variational calculations with
\begin{align}\label{eq:Psi-JM}
  |\Psi^{JM}\rangle
\equiv
  \sum_K F^J_K \hat{P}^J_{MK} |\Phi_0\rangle,
\end{align}
by removing the $J$ and $M$ indices from the
variational parameters $F^J_{JMK}$ originally
defined in \geqn{eq:FOmega}.

\subsection{Configuration Mixing}
\label{sec:configuration}

In the previous subsection, we introduced the method of restoring the broken rotational symmetry by introducing the concept of angular momentum projection. 
In this approach, the nuclear state with well-defined angular momentum $J$ and magnetic quantum number $M$ is represented as a linear combination of the projected basis $\hat{P}^J_{MK}|\Phi_0\rangle$, as shown in \geqn{eq:Psi-JM}. 
It should be noted that the projected state is generated from a single mean-field state $|\Phi_0\rangle$, which is referred to as the reference state. 
As previously mentioned, such a reference state is obtained by assuming that nucleons move independently in a self-consistent mean-field potential, and the nucleons fill the single-particle orbits from low to high energy.

Nevertheless, the use of a single reference state
fails to describe the nuclear ground-state properties
such as the spin and parity of some odd mass nuclei,
as well as the excited properties, including the
nuclear isomeric states, the rotational and
vibrational excitations, etc.
This indicates that some fundamental ingredients
are missing in the mean-field state.
In fact, from a pure physical perspective,
the actual nuclear structure should not be
described by a simple mean-field state. 
Nucleons interact with each other via highly complex two-body and even three-body nuclear forces. 
This leads to a non-negligible residual interaction beyond the one-body mean-field potential, which consequently alters the single-particle orbits obtained from the mean-field calculations. 
Such processes, which are beyond the independent single-particle motion, yield an additional amount of binding energy, referred to as {\it correlation energies}.
One then has to evaluate and include the correlation energies induced by the beyond mean-field correlations into the theoretical calculations.

\begin{figure}[t]
\centering
\includegraphics[width=0.8\textwidth]{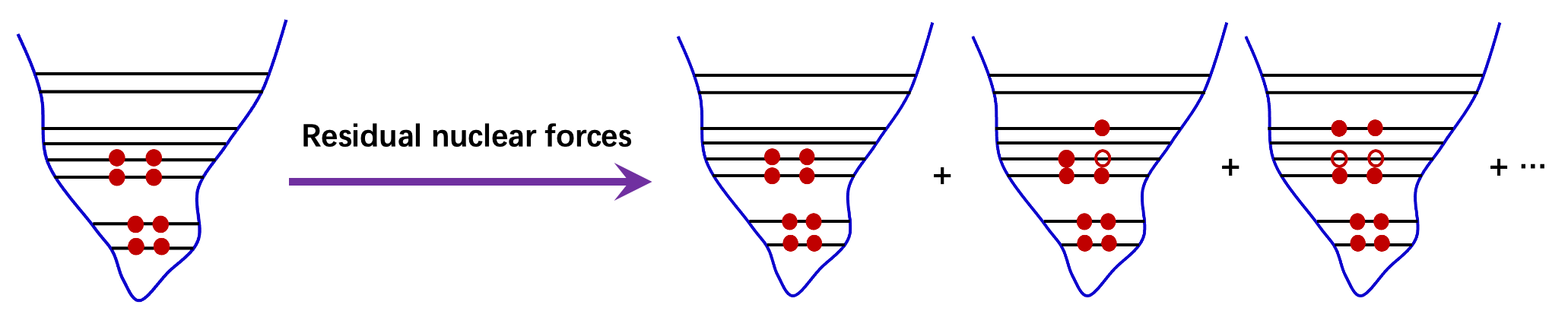}
\caption{Intuitive pictures for the nucleon occupation
in a mean-field potential (left) and the rearranged occupation
of nucleons excited by the non-negligible residual two-body
and three-body nuclear forces (right).}
\label{fig:particle-hole-exc}
\end{figure}
The basic idea of incorporating the correlations
that is missing in the mean-field state is to
enlarge the variational space, which is realized
again by generalizing the single $|\Phi_0\rangle$
to a state with linear combination. 
The physical principle guiding us on how to construct such a linear combination form is the fact that the non-negligible residual two-body and three-body nuclear forces, which cannot be fully captured by the mean-field potential, can excite nucleons from occupied orbits below the Fermi surface to the unoccupied orbits above the Fermi surface, as intuitively shown in \gfig{fig:particle-hole-exc}.

The excited states, sometimes referred to as configurations, describe the way nucleons are populated in the single-particle orbits as shown in the right panel of
\gfig{fig:particle-hole-exc}.
There configurations can be generated for even-even nuclei as
\begin{subequations}
\label{eq:configurations-even}
\begin{align}
  \hat{\beta}^\dagger_{\pi_i} \hat{\beta}^\dagger_{\pi_j} |\Phi_0\rangle,\quad
& \mbox{two~quasiproton~configurations},\\
  \hat{\beta}^\dagger_{\nu_i} \hat{\beta}^\dagger_{\nu_j} |\Phi_0\rangle,\quad
& \mbox{two~quasineutron~configurations},\\
  \hat{\beta}^\dagger_{\pi_i} \hat{\beta}^\dagger_{\pi_j} \hat{\beta}^\dagger_{\nu_k} \hat{\beta}^\dagger_{\nu_l} |\Phi_0\rangle  , \quad
& \mbox{four~quasiparticle~configurations},\\
\cdots, \quad
&
\end{align}
\end{subequations}
and for odd-$N$ nuclei as
\begin{subequations}
\label{eq:configurations-odd}
\begin{align}
  \hat{\beta}^\dagger_{\nu_i} |\Phi_0\rangle,\quad
& \mbox{one quasineutron configurations},\\
  \hat{\beta}^\dagger_{\nu_i} \hat{\beta}^\dagger_{\pi_j} \hat{\beta}^\dagger_{\nu_k} |\Phi_0\rangle  , \quad
& \mbox{three quasiparticle configurations},\\
  \cdots, \quad
&
\end{align}
\end{subequations}
where $\pi$ and $\nu$ are used to label the protons and neutrons, $\hat{\beta}^\dagger_{\pi_i(\nu_i)}$ represent the quasiparticle creation operators as
defined in \geqn{eq:Slater-qp}, and ``$\cdots$" is used to illustrate the higher quasi-particle configurations.

%\subsection{Relativistic configuration-interaction density functional theory}
Based on the constructed configurations, a more
general trial nuclear wave function, which not
only has good angular momentum $J$ but also
incorporates beyond the mean-field correlations,
can be extended from \geqn{eq:Psi-JM} as,
\begin{align}
\label{eq:wavefunction-ReCD}
  |\Psi^{JM}\rangle \equiv |JM\rangle
=
  \sum_{K\kappa} F^{J }_{K\kappa } \hat{P}^J_{MK} |\Phi_\kappa\rangle.
\end{align}
Here, $|\Phi_\kappa\rangle$ represents the
mean-field ground state $|\Phi_0\rangle$ as
defined in \geqn{eq:Slater-qp} or a certain
configuration in \geqn{eq:configurations-even}
and \geqn{eq:configurations-odd}.
The additional summation index $\kappa$ reflects the so-called {\it configuration mixing}.
The expansion coefficients $F^{J}_{K \kappa}$ are determined by minimizing the energy in \geqn{eq:E-Psi}, or, equivalently, by solving the following {\it Hill-Wheeler equation},
\begin{align}\label{eq:Hill-Wheeler}
  \sum_{K'\kappa'}
  \{ H^J_{K\kappa,K'\kappa'}  -E^J N^J_{K\kappa,K'\kappa'} \}
  F^J_{K'\kappa'}
=
  0,
\end{align}
where the nuclear {\it Hamiltonian kernel} reads
\begin{align}
  H^J_{K\kappa,K'\kappa'}
\equiv 
  \me{\Phi_\kappa }{\hat{H}\hat{P}^J_{KK'}}{\Phi_{\kappa'}},
\end{align}
and the norm kernel is
\begin{align}
  N^J_{K\kappa,K'\kappa'}
\equiv 
  \me{\Phi_\kappa }{\hat{P}^J_{KK'}}{\Phi_{\kappa'}}.
\end{align}
Both the Hamiltonian and the norm kernels
are evaluated using the Pfaffian algorithm
\cite{Carlsson2021PRL,Hu2014PLB}.
With the optimal coefficients $F^J_{K\kappa}$,
one is able to obtain the nuclear state $|JM\rangle$
and subsequently calculate the nuclear density matrix
relevant to the DM-nucleus elastic and inelastic scatterings.

\subsection{The ReCD-based Nuclear Density Matrix}
\label{sec:NME}

As shown in \geqn{eq:amp_squared_total_w_nuc_resp},
the nuclear structure effects on the squared amplitude
are encapsulated in the response functions
$W^{\tau\tau'}_\mathcal{T}$.
These response functions, in turn, are determined by
the reduced matrix elements
$\langle J_f||\hat{\mathcal{T}}_{J;\tau}||J_i\rangle$
as given in \geqn{eq:W_elastic} and \geqn{eq:W_inelastic}
for the elastic and inelastic scatterings, respectively.
It is easy to verify that $\langle J_f||\hat{\mathcal{T}}_{J;\tau}||J_i\rangle$
can be further decomposed as follows \cite{Anand:2013yka}:
\begin{align}
\label{eq:reduced-mat-D-T}
  \langle J_f || \hat{\mathcal{T}}_{J;\tau} || J_i \rangle
= 
    \sum_{|\alpha|,|\beta|} \mathcal{D}^{J;\tau}_{|\alpha|,|\beta|}
    \langle|\alpha|::\mathcal{T}_{J;\tau}(q\bm{x})::|\beta|\rangle.
\end{align}
While $|\alpha\rangle$ and $|\beta\rangle$ denote
the single-particle spherical harmonic oscillator
bases with quantum numbers
$\alpha \equiv \{n_\alpha, l_\alpha, j_\alpha, m_\alpha\}$
and $\beta \equiv \{n_\beta, l_\beta, j_\beta, m_\beta\}$,
the nonmagnetic quantum numbers are denoted by
$|\alpha| \equiv \{n_\alpha, l_\alpha, j_\alpha\}$
and $|\beta| \equiv \{n_\beta, l_\beta, j_\beta\}$.
Note that $\langle|\alpha|::\mathcal{T}_{J;\tau}(q\bm{x})::|\beta|\rangle$ is a reduced single-particle harmonic oscillator matrix element, with $::$ indicating reduction in both spin and isospin.
For a given $\mathcal{T}_{J;\tau}(q\bm{x})$, it
possesses an analytical expression \cite{Donnelly1979ADNDT}
and carries no nuclear structure information which
is actually carried by the one-body nuclear density
matrix \cite{Anand:2013yka}
\begin{align}
\label{eq:density-matrix-general}
    \mathcal{D}^{J,\tau}_{|\alpha|,|\beta|} \equiv
    \frac{1}{[J][\tau]}
    \rme{J_f}{[c^\dagger_{|\alpha|}\tilde{c}_{|\beta|}]_{J;\tau}}{J_i},
\end{align}
with $[J] = \sqrt{2J+1}$, and $[\tau] = \sqrt{2\tau + 1}$.
While $c^\dagger_\alpha \equiv c^\dagger_{n_\alpha l_\alpha j_\alpha m_\alpha}$
represents the creation operator of the spherical
harmonic oscillator basis,
$\tilde{c}_\beta \equiv (-1)^{j_\beta - m_\beta} c_{n_\beta l_\beta j_\beta -m_\beta}$.
The nuclear density matrix encodes the nuclear
structure information purely from the initial
and final nuclear states as calculated by the ReCD theory.

As indicated in \geqn{eq:mat-reduce-mat}, to obtain $\mathcal{D}^{J,\tau}_{|\alpha|,|\beta|}$, one has to calculate $\me{J_f M_f}{[c^\dagger_{|\alpha|}\tilde{c}_{|\beta|}]_{LM,\tau}}{J_i M_i}$, where $|J_i M_i\rangle$ and $|J_f M_f\rangle$ are the initial and final nuclear states provided by the ReCD theory.
Based on \geqn{eq:wavefunction-ReCD}, we have
\begin{align}
\hspace{-3mm}
  \me{J_f M_f}{ [c^\dagger_{|\alpha|} \tilde{c}_{|\beta|}]_{JM;\tau} }{J_i M_i}
= \sum_{K_f \kappa_f}
  \sum_{K_i \kappa_i}
  F^{J_f\dagger}_{K_f \kappa_f}
  F^{J_i}_{K_i \kappa_i}
  \me{ \Phi_{\kappa_f}}
  {\hat{P}^{J_f\dagger}_{M_f K_f} [c^\dagger_{|\alpha|} \tilde{c}_{|\beta|}]_{JM;\tau} \hat{P}^{J_i}_{M_i K_i}}
  {\Phi_{\kappa_i}}.
\label{eq:me_JfMf_LM_JiMi}
\end{align}
Since $[c^\dagger_{|\alpha|} \tilde{c_{|\beta|}}]_{JM;\tau}$
is a tensor operator of rank
$J$ and $\hat{P}^{J\dagger}_{MK} = \hat{P}^J_{KM}$,
the operator combination sandwiched between
the initial and final states becomes,
\begin{align}
%     & \hat{P}^{J_f\dagger}_{M_f K_f}
%     [ c^\dagger_{|\alpha|} \tilde{c}_{|\beta|} ]_{JM;\tau}
%     \hat{P}^{J_i}_{M_i K_i} = 
%     \hat{P}^{J_f}_{K_f M_f}
%     [ c^\dagger_{|\alpha|} \tilde{c}_{|\beta|} ]_{JM;\tau}
%     \hat{P}^{J_i\dagger}_{K_i M_i}
%     \notag \\
% & =
  \sum_{K_i'} \sum_{M'}
    \frac{2J_f + 1}{8\pi^2}
    \int d\Omega D^{J_f\ast}_{K_f M_f}(\Omega) D^{J_i}_{K_i' M_i}(\Omega) D^J_{M' M}(\Omega)
    [c^\dagger_{|\alpha|} \tilde{c}_{|\beta|}]_{JM';\tau} \hat{P}^{J_i}_{K_i' K_i}.
\end{align}
With the help of the integral formula for the product of three $D$-functions~\cite{Varshalovich1988book}, 
\begin{align}
    \int d\Omega D^{J_f\ast}_{K_f M_f}(\Omega) D^{J_i}_{K_i' M_i}(\Omega) D^J_{M' M}(\Omega) = 
    \frac{8\pi^2}{2J_f + 1}
    \CG{J_i K_i'}{LM'}{J_f K_f}
    \CG{J_i M_i}{LM}{J_f M_f},
\end{align}
one gets that 
\begin{align}\label{eq:PJ_cc_LM_PJ}
    \hat{P}^{J_f\dagger}_{M_f K_f}
    [ c^\dagger_{|\alpha|} \tilde{c}_{|\beta|} ]_{JM;\tau}
    \hat{P}^{J_i}_{M_i K_i} =
    \CG{J_i M_i}{JM}{J_f M_f}
    \sum_{K_i' M'}
    \CG{J_i K_i'}{JM'}{J_f K_f}
    [c^\dagger_{|\alpha|} \tilde{c}_{|\beta|}]_{JM';\tau}
    \hat{P}^{J_i}_{K_i' K_i}.
\end{align}
Substituting \geqn{eq:PJ_cc_LM_PJ} into \eqref{eq:me_JfMf_LM_JiMi}
and extracting the reduced matrix element
$\rme{J_f}{ [c^\dagger_{|\alpha|} \tilde{c}_{|\beta|}]_{J;\tau} }{J_i}$
according to \geqn{eq:mat-reduce-mat},
\begin{align}
\label{eq:rme_Jf_cc_L_Ji_v1}
    & \rme{J_f}{ [c^\dagger_{|\alpha|} \tilde{c}_{|\beta|}]_{J;\tau} }{J_i}
    \notag\\
    & = \sqrt{2J_f + 1}
    \sum_{K_f \kappa_f}
    \sum_{K_i \kappa_i}
    F^{J_f\dagger}_{K_f \kappa_f}
    F^{J_i}_{K_i \kappa_i}
    \sum_{K_i' M'}
    \CG{J_i K_i'}{JM'}{J_f M_f}
    \me{ \Phi_{\kappa_f}}
    { [c^\dagger_{|\alpha|} \tilde{c}_{|\beta|}]_{JM';\tau} \hat{P}^{J_i}_{K_i' K_i}}
    {\Phi_{\kappa_i}}.
\end{align}
This provides us the final expression of the ReCD-based nuclear density matrix as
\begin{align}\label{eq:density-matrix-v1}
    \mathcal{D}^{J,\tau}_{|\alpha|,|\beta|} &
=
%     \frac{1}{[J][\tau]}
%     \rme{J_f}{[c^\dagger_{|\alpha|}\tilde{c}_{|\beta|}]_{J;\tau}}{J_i}
% =
\frac{1}{[J]{[\tau]}}
    \sqrt{2J_f + 1}
    \sum_{K_f \kappa_f}
    \sum_{K_i \kappa_i}
    F^{J_f\dagger}_{K_f \kappa_f}
    F^{J_i}_{K_i \kappa_i}
    \notag \\
    & \times 
    \left[
    \sum_{M'} \CG{J_i K_f - M'}{JM'}{J_f K_f}
    \frac{2J_i + 1}{8\pi^2}
    \int d\Omega D^{J_i\ast}_{K_f - M' K_i}(\Omega)
    \me{\Phi_{\kappa_f}}
    { [c^\dagger_{|\alpha|} \tilde{c}_{|\beta|}]_{JM';\tau} \hat{R}(\Omega)}
    {\Phi_{\kappa_i}}
    \right].
\end{align}

\subsection{Comparison with Nuclear Shell Model}
\label{sec:comparison}

As previously noticed, the nuclear response functions
$W^{\tau \tau'}_{\mathcal T}$ or more precisely the
nuclear density matrices $\mathcal D^{J, \tau}_{|\alpha|, |\beta|}$,
are typically obtained via the NSM calculations.
The fundamental assumption of the NSM is that the low-lying
nuclear properties are predominantly governed by those
\textit{valence nucleons} near the Fermi surface \cite{Engel2017RPP}.
Thus the full nuclear Hamiltonian that should be solved
in a full model space is replaced by an effective
Hamiltonian solved in a limited model space composed
of the nucleon orbits around the Fermi surface.

Nevertheless, the dimension of the configuration space
generated by all possible particle-hole excitations
gradually becomes too large to handle even when limited
to such a restricted model space as the proton and
neutron numbers deviate from the magic numbers.
Take the Xe isotopes as an example, the NSM model
space consists of spherical single-particle orbits
$0g_{7/2}, 1d_{5/2}, 1d_{3/2}, 2s_{1/2}$ and $0h_{11/2}$
\cite{Fitzpatrick:2012ix,Menendez2012PRD}.
Those orbits that below and above this model space
are frozen and do not contribute to the properties
of Xe isotopes.
Moreover, the dimensions of configuration spaces are further truncated for odd-even isotopes $^{129,131}$Xe and sometimes also for even-even isotopes $^{128,130,132}$Xe.
For example, Ref.\,\cite{Fitzpatrick:2012ix} truncates
by fixing the $0h_{11/2}$ occupation numbers for $^{128,130,132}$Xe
to 4, 6, and 8, respectively, which are the
minimum allowed nucleon numbers in each isotope.
The configuration space for $^{129,131}$Xe is further restricted by limiting valence protons to the energetically favored $1d_{5/2}$ and $0g_{7/2}$ shells and by requiring neutrons to fully occupy the same shells.
In Ref.\cite{Menendez2012PRD}, the number of nucleon excitations into the $1d_{3/2}, 2s_{1/2}$ and $0h_{11/2}$ orbits is limited to three for $^{129}$Xe.

In comparison, the ReCD theory (a NSM approach built
on top of the nuclear relativistic density functional
theory) allows all nucleons to be involved in the practical calculations. No truncation is made to the numbers of active nucleons and single-particle orbits.
Furthermore, the self-consistent RDFT calculations can consider the important nuclear correlations by taking into account the nuclear deformation and pairing correlations. 
Thus, the single quasiparticle wave functions given by \geqn{eq:RHB} and the configurations in \geqn{eq:configurations-even} and \geqn{eq:configurations-odd} incorporate far more physical effects than the spherical harmonic oscillator orbits used in the NSM calculations.
In other words, much better bases are adopted to
construct the configuration space in the ReCD theory.
Since the ReCD method has much more physical effects
intrinsically incorporated, the dimension of the
configuration space in the ReCD theory is much smaller than that in the NSM, and one does not need to introduce any nonphysical truncation that may influence the accuracy of theoretical calculations into the configuration space.

\begin{figure}[t]
\centering
\includegraphics[width=0.7\textwidth]{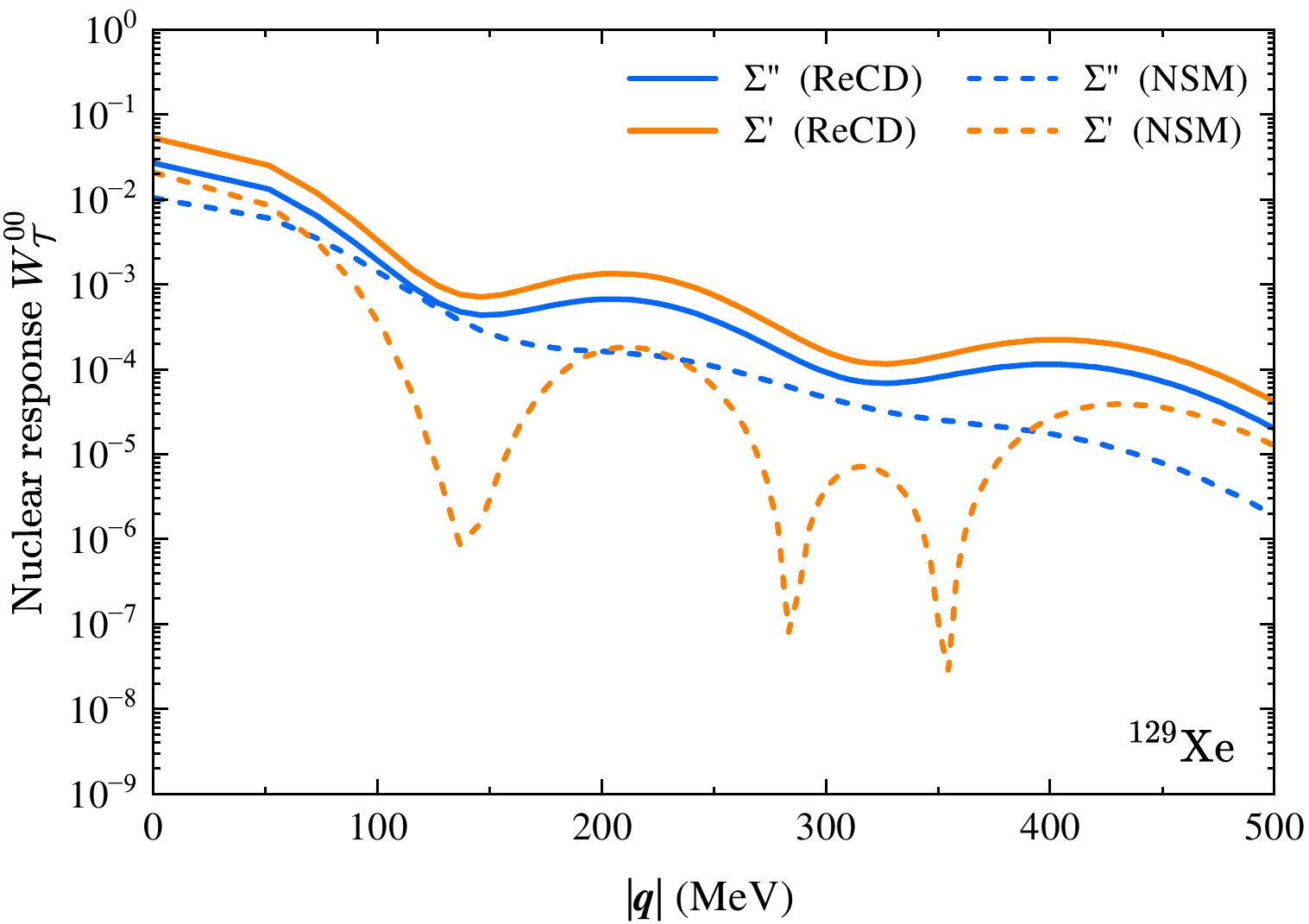}
\caption{The nuclear response $W^{00}_{\mathcal{T}}$ for $\mathcal{T} = \Sigma''$ and $\Sigma'$ as functions of momentum transfer amplitude $|\bm{q}|$ calculated by the ReCD theory and their comparison to the NSM results \cite{Anand:2013yka}.}
\label{fig:129Xe_Response_SM_ReCD}
\end{figure}

The NSM and the ReCD are two widely used many-body approaches in the field of nuclear structures. Since some approximations are indeed employed in the NSM when calculating the nuclear response functions, it is necessary to determine to what extent the nuclear response functions given by the ReCD theory and the NSM differ from each other.
To this end, we perform calculations for the nuclear response functions based on our ReCD and compare our results with those provided by the NSM.

In the ReCD calculations, the Hamiltonian $\hat{H}$ is provided by the well-known PC-PK1 effective interaction \cite{Zhao2010PRC} and the pairing Hamiltonian $\hat{H}_{\text{pair}}$ in \geqn{eq:Hamiltonian-pair} is provided by the finite-range separable force \cite{Tian2009PLB}.
The relativistic Hartree-Bogoliubov equation in \geqn{eq:RHB} is solved in terms of the three-dimensional harmonic oscillator basis with 8 major shells.
Its solution provides the nuclear ground state at the mean-field level, namely $|\Phi\rangle$ in \geqn{eq:Slater-qp}, based on which the ReCD configuration spaces \geqn{eq:configurations-even} and \geqn{eq:configurations-odd} can be constructed.
By solving the Hill-Wheeler equation \geqn{eq:Hill-Wheeler} in the Hilbert space spanned by the configuration space, one can obtain the nuclear initial and final states $|JM\rangle$ of \geqn{eq:wavefunction-ReCD} and consequently the density matrices as well as the nuclear response functions.

The nuclear response functions $W^{00}_{\mathcal{T}}$
for elastic DM scattering off $^{129}$Xe, calculated
by the ReCD theory for multiple operators $\Sigma''$
and $\Sigma'$ with $\tau = \tau' = 0$, are shown in
\gfig{fig:129Xe_Response_SM_ReCD} along with the NSM
results. At low momentum transfer $|\bm{q}|$, the
ReCD and NSM calculations yield similar responses.
Nevertheless, notable differences occur for higher
momentum transfer. For $W^{00}_{\Sigma''}$, the ReCD
results are approximately five times larger than its
NSM counterpart. The discrepancy for $W^{00}_{\Sigma'}$
is even more prominent. In the region
$98.5\,\mbox{MeV} < |\bm q| < 394.0$\,MeV, the ReCD-based response exhibits
smooth oscillation while the NSM results evolve
dramatically and show three sharp dips around
$|\bm q| = 136.8$\,MeV, $283.3$\,MeV, and $354.6$\,MeV.

\begin{table}[htbp]
\centering
\begin{tabular}{ccc}
\toprule
Orbits ($|\alpha|,|\beta|$) &   ReCD      &     NSM  \\
\midrule
$(0g_{7/2},0g_{7/2})$   &  0.025        &  0.114  \\
$(1d_{5/2},1d_{5/2})$   &  0.115        &  0.138  \\
$(1d_{3/2},1d_{3/2})$   &  $-0.163$     & $-0.755$  \\
$(2s_{1/2},2s_{1/2})$   &  2.070        &  0.904  \\
$(0h_{11/2},0h_{11/2})$ &  0.027        &  0.173  \\
\bottomrule
\end{tabular}
\caption{Nuclear density matrices
$\mathcal{D}^{J,\tau}_{|\alpha|,|\beta|}$ calculated
by the ReCD and NSM methods for the elastic DM scattering
off $^{129}$Xe with $J=1$ and $\tau=0$.
For clarity, the table shows only the results induced
by single-particle orbits around the Fermi surface, namely 
$0g_{7/2}$, $1d_{5/2}$, $1d_{3/2}$, $2s_{1/2}$, and $0h_{11/2}$,
which play most important roles in the nuclear responses.}
\label{Tab:density-compare}
\end{table}

As indicated by \geqn{eq:reduced-mat-D-T}, the discrepancies
in the nuclear responses arise from the differences in the
nuclear density matrices given by the ReCD and the NSM.
\gtab{Tab:density-compare} shows the reduced
nuclear density matrices
$\mathcal{D}^{J,\tau}_{|\alpha|,|\beta|}$ with $J=1$ and $\tau = 0$
that originates from the single-particle orbits
$0g_{7/2}$ ($|\alpha| = |\beta| = \{0,4,7/2\}$),
$1d_{5/2}$ ($|\alpha| = |\beta| = \{1,2,5/2\}$),
$1d_{3/2}$ ($|\alpha| = |\beta| = \{1,2,3/2\}$),
$2s_{1/2}$ ($|\alpha| = |\beta| = \{2,0,1/2\}$),
and $0h_{11/2}$ ($|\alpha| = |\beta| = \{0,5,11/2\}$).
Although the values of
$\mathcal{D}^{J,\tau}_{|\alpha|,|\beta|}$ with $J=0$ and
$\tau = 0$ are nonzero, they do not contribute to the nuclear
response functions of $\mathcal{T} = \Sigma'$ and $\Sigma''$.
For $\Sigma'$, this is simply because the parity selection rule requires $J \geq 1$, as shown in \geqn{eq:W_elastic}.
On the other hand, the reduced single-particle
harmonic oscillator matrix elements
$\langle|\alpha|::\mathcal{T}_{J;\tau}(q\bm{x})::|\beta|\rangle$
in \geqn{eq:reduced-mat-D-T} with rank $J=0$
become zero for $\Sigma''$ \cite{Donnelly1979ADNDT}.

As shown in \gtab{Tab:density-compare}, significant differences between the ReCD and the NSM results are observed when $|\alpha| = |\beta| = 1d_{3/2}$ and $|\alpha|=|\beta| = 2s_{1/2}$.
More specifically, in the ReCD calculations, $\mathcal{D}^{1,0}_{|\alpha|,|\beta|}$ values from $1d_{3/2}$ and $2 s_{1/2}$ have opposite signs, and the corresponding amplitude from $1d_{3/2}$ is significantly smaller than that from $2s_{1/2}$.
In the NSM calculations, $\mathcal{D}^{1,0}_{|\alpha|,|\beta|}$ values from $1d_{3/2}$ and $2 s_{1/2}$ also have opposite signs but their amplitudes are comparable with each other.
Further evaluations demonstrate that the nuclear response functions, particularly their oscillation behaviors as functions of the momentum transfer $|\bm q|$, depend sensitively on the relative amplitudes of $\mathcal{D}^{1,0}_{|\alpha|,|\beta|}$ from $1d_{3/2}$ and $2 s_{1/2}$.
In fact, simply replacing the NSM-based $\mathcal{D}^{1,0}_{|\alpha|,|\beta|}$ from $1 d_{3/2}$ and $2 s_{1/2}$ with the ReCD results, the response functions obtained from these two approaches, as plotted in \gfig{fig:129Xe_Response_SM_ReCD}, agree with each other.
The comparison between our ReCD results and the NSM ones highlights the nuclear structure effects on the DM-nucleus scattering.

\section{Event Rate Enhancement due to Inelastic Scattering}
\label{sec:results}

\subsection{First Excited State}

Using the general form of the squared amplitude
in \geqn{eq:amp_squared_total_w_nuc_resp} with
the density matrices obtained within the ReCD
theory, we evaluate the expected event rates
for a natural xenon target. We pay particular
attention to the contribution arising from the
inelastic scattering. 
As we show in \gsec{sec:kinematics}, this contribution can appear only for $^{129}$Xe and
$^{131}$Xe, since the excited states of other xenon isotopes are kinematically inaccessible.
For concreteness, we
consider PandaX-4T detector with a target mass
of $3.7$\,tonne\cite{PandaX-4T:2021bab}.
\gtab{tab:operators_ratios} shows
the ratios $R_{\text{inel}}^i/R_{\text{el}}^i$
of the inelastic event rates
to the first excited states of $^{129}$Xe and
$^{131}$Xe to the total elastic event rates
from all natural xenon isotopes, with parameters
presented in \gtab{tab:xenon_isotopes}, for
all NR EFT operators $\mathcal O_i$ in \geqn{eq:NR_EFT_operators}.

We focus on the case of isoscalar couplings
($c_i^p = c_i^n$) and consider two benchmark
scenarios with DM masses $m_\chi = 300\,$GeV
and $m_\chi = 10\,$TeV, respectively.
The former case was considered in the
literature before and it was shown that the effects of inelastic scattering are significant for some operators~\cite{Arcadi:2019hrw}. 
In the latter case, the reduced DM-nucleus mass $\mu$ is close to its asymptotic value $\mu \approx m_A$, which corresponds to the maximal possible ranges of momentum transfers and recoil energies for fixed target mass and excitation energy, as can be seen from \geqn{eq:NR_scatter_exc_nucleus_q_sols} and \geqn{eq:inel_recoil_energy_range}.

\begin{table}[t]
\centering
\begin{tabular}{ccccc}
\toprule
\multirow{3}{*}{$\mathcal O_i$} &\multicolumn{4}{c}{$R_{\text{inel}}^i/R_{\text{el}}^i $}\\
 \cline{2-5} & \multicolumn{2}{c}{$m_\chi = 300$\,GeV} & \multicolumn{2}{c}{$m_\chi = 10$\,TeV}\\
\cline{2-3} \cline{4-5} & ReCD &  NSM & ReCD &  NSM
\\
 \hline
 $\mathcal O_1$ & $1.1 \times 10^{-5}$ & $2.6 \times 10^{-5}$
 & $2.2 \times 10^{-5}$ & $3.7 \times 10^{-5}$\\ 
 $\mathcal O_3$ & $5.4 \times 10^{-4}$& $2.9 \times 10^{-4}$
 & $1.4 \times 10^{-3}$ & $3.5 \times 10^{-4}$\\ 
 $\mathcal O_4$ & $9.8 \times 10^{-2}$ & $0.15$
 &$0.26$ & $0.19$\\ 
 $\mathcal O_5$ & $8.5 \times 10^{-2}$ & $3.1 \times 10^{-2}$
 & $0.29$ & $0.10$\\ 
 $\mathcal O_6$ & $0.13$ & $0.14$
 & $0.24$ & $0.13$\\ 
 $\mathcal O_7$ & $67$ & $74$
 & $340$ & $170$\\ 
 $\mathcal O_8$ & $1.3 \times 10^{-2}$ & $4.4 \times 10^{-3}$
 & $3.2 \times 10^{-2}$ & $1.1 \times 10^{-2}$\\ 
 $\mathcal O_9$ & $0.21$ & $0.28$
 & $0.39$ & $0.35$\\ 
 $\mathcal O_{10}$ & $0.19$ & $0.21$
 &$0.37$ & $0.21$\\
 $\mathcal O_{11}$ & $4.4 \times 10^{-5}$ & $1.0 \times 10^{-4}$
 &  $1.0 \times 10^{-4}$ & $1.4 \times 10^{-4}$\\ 
 $\mathcal O_{12}$ & $2.3 \times 10^{-4}$ & $5.0 \times 10^{-4}$
 & $5.1 \times 10^{-4}$ & $6.1 \times 10^{-4}$\\ 
 $\mathcal O_{13}$ & $1.2$ &  $1.7$
 & $3.3$ & $1.9$\\ 
 $\mathcal O_{14}$ & $4.0 \times 10^2$ & $5.7 \times 10^2$
 & $1.9 \times 10^3$ & $1.2 \times 10^3$\\ 
 $\mathcal O_{15}$& $1.2 \times 10^{-3}$ & $3.3 \times 10^{-4}$
 & $3.3 \times 10^{-3}$ & $4.0 \times 10^{-4}$\\
 \bottomrule
\end{tabular}
\caption{The ratio $R_{\text{inel}}^i/R_{\text{el}}^i $ between
the inelastic event rate due to transition to the first
excited states of $^{129}$Xe plus $^{131}$Xe and the elastic
event rate with natural xenon via operators $\mathcal O_i$ defined in
Eq.\,\eqref{eq:NR_EFT_operators}. Two DM masses $m_\chi = 300$\,GeV
and $m_\chi = 10$\,TeV have been used for illustration.}
\label{tab:operators_ratios}
\end{table}

The second and the fourth columns of \gtab{tab:operators_ratios}
show our ReCD-based results for $m_\chi = 300\,$GeV and
$m_\chi = 10\,$TeV, respectively. As a consistency check,
we perform an additional computation with our
\geqn{eq:amp_squared_total_w_nuc_resp} using the
shell-model density matrices from \cite{Arcadi:2019hrw}
and \cite{Anand:2013yka} for the inelastic and elastic
event rates, respectively, as inputs. The corresponding
ratios, denoted as ``NSM'', are shown in the third
and the last columns of \gtab{tab:operators_ratios}.
It should be emphasized that the NSM 
%shell model 
calculations
\cite{Arcadi:2019hrw} and \cite{Anand:2013yka} adopted
different effective nucleon-nucleon interactions and model spaces.
Consequently, the NSM results in
\gtab{tab:operators_ratios} cannot directly compare 
with each other. More specifically, the elastic and inelastic
density matrices are not obtained in a unified manner.
Roughly speaking, the ReCD and NSM predictions
are mostly consistent with each other, being within the
same order of magnitude.

As one would naively expect, larger DM mass would lead
to larger momentum transfer such that the inelastic
event rate for $m_\chi = 10\,$TeV is larger than
the one for $m_\chi = 300\,$GeV. We can clearly
see such feature in the ReCD results and it is
basically true for most cases for the NSM methods.
The only two exceptions, for $\mathcal O_6$ and
$\mathcal O_{10}$, should come from the fact that
the two NSM calculations for the inelastic and elastic
density matrices are actually from two different references,
\cite{Arcadi:2019hrw} and \cite{Anand:2013yka}, respectively.

\gtab{tab:operators_ratios} shows that the relative size
of the elastic and inelastic scattering event rates vary
a lot among different effective operators. The concrete
values distribute in the range from $10^{-5}$ to $10^3$.
For $\mathcal O_1$, $\mathcal O_3$, $\mathcal O_8$,
$\mathcal O_{11}$, $\mathcal O_{12}$, and $\mathcal O_{15}$,
inelastic scattering contributes to the total event rate
about $1\%$ or less and is unlikely to affect the experimental searches. 
For operators $\mathcal O_4$, $\mathcal O_5$, $\mathcal O_6$,
$\mathcal O_9$, and $\mathcal O_{10}$, the inelastic
contributions are already at the noticeable order of
$10\%$ and can become relevant for the experimental
search. The remaining operators $\mathcal O_7$,
$\mathcal O_{13}$, and $\mathcal O_{14}$ are
particularly important with even larger inelastic
scattering rates than their corresponding
elastic counterparts. We will try to elaborate
the details below. Note that $^{129}$Xe (26.4\%)
and $^{131}$Xe (21.2\%) have roughly the same
natural abundances as summarized in
\gtab{tab:xenon_isotopes}. For simplicity,
we assume equal weight for these two
isotopes in the detailed discussions below.
For concreteness, we focus on the $m_\chi = 300\,$GeV
case.

For operator $\mathcal O_{13}$, we predict the inelastic
signal to have slightly larger size than the elastic one.
This can be explained with a simple analysis of the squared
amplitude in \geqn{eq:amp_squared_total_w_nuc_resp}.
The inelastic scattering for this operator proceeds via 
the responses $W_{\Sigma''}$, $W_{\widetilde \Phi}$,
and $W_{\widetilde \Phi'}$, while only $W_{\Sigma''}$
and $W_{\widetilde \Phi'}$ survive for elastic scattering,
as can be seen from \gtab{tab:operators_responses}.
Since only the $^{129}$Xe and $^{131}$Xe isotopes contribute,
the ratio between the inelastic and elastic scattering
event rates is approximately
\begin{align}
    \frac{R_{13\,\text{inel}}}{R_{13\,\text{el}}}
\sim
    \frac{\sum \limits_{A = 129, 131}
    \frac{\bm q^2_\text{inel}}{m_N^2}
    \left\{
     (\widetilde{v}_T^{\perp})^2 \,
     W_{\Sigma''}^\text{(inel)} (A)
     +
     \frac{\bm q^2_\text{inel}}{m_N^2}
     \left[W_{\widetilde \Phi}^\text{(inel)} (A) + W_{\widetilde \Phi'}^\text{(inel)}(A)\right]
     \right\}
    }{\sum \limits_{A = 129, 131}
    \frac{\bm q^2_\text{el}}{m_N^2}
    \left\{
     (\widetilde{v}_T^{\perp})^2 \,
     W_{\Sigma''}^\text{(el)} (A)
     +
     \frac{\bm q^2_\text{el}}{m_N^2}
     W_{\widetilde \Phi'}^\text{(el)} (A)
     \right\}
    }.
\end{align}
Here we emphasize that the typical momentum transfers in elastic and inelastic scattering regimes may differ. However, in the case of $\mathcal O_{13}$, the numerical difference between them turns out to be small, $\bm q^2_\text{inel}/m_N^2 \approx \bm q^2_\text{el}/m_N^2 \approx 2 \times 10^{-2}$ for both isotopes. 
Note that the elastic response $W_{\widetilde \Phi'}^\text{(el)}$
for $^{129}$Xe vanishes due to selection rules as shown
in \gtab{tab:selection_rules}.
All involved nuclear responses, $W_{\Sigma''}$, $W_{\widetilde \Phi}$, and $ W_{\widetilde \Phi'}$, are accompanied by
a common factor of $\bm q^2/m_N^2 \sim 10^{-2}$
arising from the DM response functions in
\geqn{eq:R_SigmaPP_def} and \geqn{eq:R_PhiT_def},
with similar typical values
for elastic and inelastic scattering.
However,
$W_{\Sigma''}$ has an extra small coefficient
$(\widetilde{v}_T^{\perp})^2 \sim 10^{-6}$ from
the DM response~\geqn{eq:R_SigmaPP_def},
while the functions $W_{\widetilde \Phi}$
and $W_{\widetilde \Phi'}$ are multiplied by a
much larger extra factor $\bm q^2/m_N^2$
appearing in \geqn{eq:amp_squared_total_w_nuc_resp}. 
Numerically, $W_{\widetilde \Phi}$ is strongly suppressed
while all the other surviving responses for both isotopes
have comparable typical values $\sim 10^{-3}$.
As a result, one finds that the total inelastic
signal is mainly due to the sum of responses
$W_{\widetilde \Phi'}^\text{(inel)}$ of $^{129}$Xe
and $^{131}$Xe, and the elastic signal is dominated
by the $W_{\widetilde \Phi'}^{\rm (el)}$ response of $^{131}$Xe.
Therefore, the ratio of inelastic and elastic rates can be estimated as
\begin{align}
  \frac{R_{13\,\text{inel}}}{R_{13\,\text{el}}}
\approx
  \frac {W_{\widetilde \Phi}^\text{(inel)}(^{129}\text{Xe})
       + W_{\widetilde \Phi'}^\text{(inel)}(^{131}\text{Xe})}
        {W_{\widetilde \Phi'}^\text{(el)}(^{131}\text{Xe})}
\sim
  1,
\label{eq:O13_ratios}
\end{align}
which is in agreement with \gtab{tab:operators_ratios}.

\begin{figure}[t]
\centering
\includegraphics[width=0.99\linewidth]{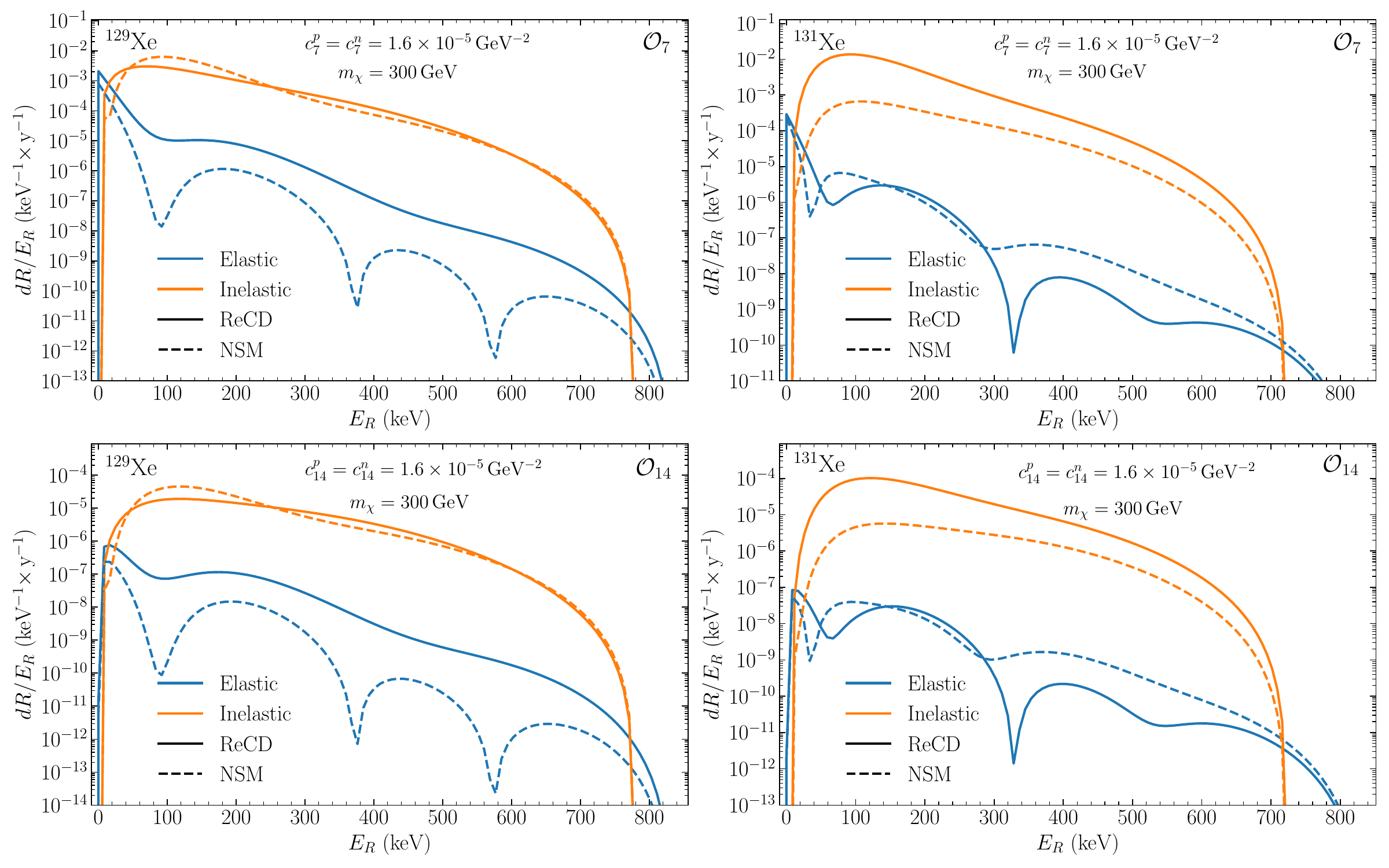}
\caption{Differential event rates due to the WIMP-nucleus
elastic (blue) and inelastic (orange) scatterings to the
first excited nuclear states on $^{129}$Xe (left panels)
and $^{131}$Xe (right panels) nuclei via operators
$\mathcal O_{7}$ (upper panels) and $\mathcal O_{14}$
(lower panels). Solid and dashed lines correspond to
ReCD and NSM calculations.}
\label{fig:O7_O14}
\end{figure}

For $\mathcal O_7$, we obtain a large inelastic-to-elastic
ratio at the level of $\mathcal O(10 \sim 10^{2})$.
This can be seen better from the upper panel of
\gfig{fig:O7_O14} which shows the individual elastic
(blue) and inelastic (orange) differential event
rates for $^{129}$Xe and $^{131}$Xe, calculated
with the ReCD (solid) and NSM (dashed),
respectively. For both isotopes,
the inelastic signal is strongly dominating.
We can make a similar analysis as for
$\mathcal O_{13}$. The elastic scattering for
$\mathcal O_7$ proceeds via the 
nuclear response $W_{\Sigma'}$ with a small
factor $(\widetilde{v}_T^{\perp})^2 \sim 10^{-6}$
from the corresponding DM response in \geqn{eq:DM_responses}.
The inelastic event rate gets contributions from the 
responses $W_{\Sigma}$ and
$W_{\Sigma'}$ that are also multiplied by $(\widetilde{v}_T^{\perp})^2$,
as well as from 
$W_{\widetilde \Omega}$
with a coefficient $\bm q^2_\text{inel}/m_N^2 \sim 4 \times 10^{-2}$. 
Here the numerical difference between the typical momentum transfers for elastic and inelastic regimes becomes noticeable, $\bm q^2_\text{el}/m_N^2:\bm q^2_\text{inel}/m_N^2 \approx 1:6$.
For the typical momentum transfers involved in the
inelastic process, $W_{\widetilde \Omega}$ is about
$1 \sim 2$ orders of magnitude smaller than the sum
$W_{\Sigma} + W_{\Sigma'}$. As a result, the inelastic
scattering is dominated by the $W_{\widetilde \Omega}$ 
%axial charge 
contribution,
and the total inelastic-to-elastic ratio is kinematically
enhanced by a factor of
$\bm q^2_\text{inel}/(m_N \widetilde v_T^\perp)^2 \sim A^2 \sim 10^4$
for both isotopes $^{129}$Xe and $^{131}$Xe. Numerically,
the total elastic signal is dominated by $^{129}$Xe with
$W_{\Sigma'}^{(\text{el})} \sim 10^{-2}$ while the inelastic
is primarily due to $^{131}$Xe with
$W_{\widetilde \Omega}^{(\text{inel})}\sim 10^{-4}$.
Combining everything together, one obtains a ratio
\begin{align}
    \frac{R_{7\,\text{inel}}}{R_{7\,\text{el}}}
&\sim
    \frac{\bm q^2_\text{inel}}{m_N^2 (\widetilde{v}_T^{\perp})^2}
    \frac{W_{\widetilde \Omega}^\text{(inel)}(^{131}\text{Xe})}{W_{\Sigma'}^\text{(el)}(^{129}\text{Xe})}
\sim
    10^2,
\label{eq:O7_ratios}
\end{align}
that is consistent with the exact result.

The most pronounced contribution from inelastic
scattering arises for the operator $\mathcal O_{14}$.
Our prediction gives a very large ratio $10^{2} \sim 10^{3}$.
The differential event rates shown in the lower panel
of \gfig{fig:O7_O14} illustrate this in more detail.
From \gtab{tab:operators_responses}, one can see that
the operators $\mathcal O_{7}$ and $\mathcal O_{14}$
share exactly the same nuclear responses $W_{\widetilde \Omega}$,
$W_\Sigma$, and $W_{\Sigma'}$.
The difference is that both the elastic and
inelastic signals for $\mathcal O_{14}$ have an
extra factor of $\bm q^2/m_N^2$ in the DM response
functions \geqn{eq:R_Sigma_def} and \geqn{eq:R_OmegaT_def}.
The ratio between elastic and inelastic momentum transfers is less dramatic than for $\mathcal O_7$, with $\bm q^2_\text{el}/m_N^2:\bm q^2_\text{inel}/m_N^2 \approx 1:2$, but the absolute values are slightly larger, $\bm q^2_\text{inel}/m_N^2 \sim 5 \times 10^{-2}$.
Overall, the kinematical enhancement ends up to be larger than for $\mathcal O_7$ by about one order of magnitude:
the inelastic contribution has a factor of
$\bm q^4_\text{inel}/m_N^4 \sim 10^{-3}$
while elastic contribution is accompanied by
$(\widetilde{v}_T^{\perp})^2 \times \bm q^2_\text{el}/m_N^2 \sim 10^{-8}$.
On the nuclear response side, the elastic signal is dominated by the $^{129}$Xe 
%transverse spin-current 
response $W_{\Sigma'} \sim 10^{-3}$
and the inelastic by the $^{131}$Xe %axial-charge
response $W_{\widetilde \Omega}\sim 10^{-5}$. 
Overall, we get an estimate
\begin{align}
  \frac{R_{14\,\text{inel}}}{R_{14\,\text{el}}}
\sim
  \frac {\bm q^4_\text{inel}}
        {m_N^4}
  \frac {m_N^2}{\bm q^2_\text{el} (\widetilde{v}_T^{\perp})^2}
  \frac {W_{\widetilde \Omega}^\text{(inel)}(^{131}\text{Xe})}{W_{\Sigma'}^\text{(el)}(^{131}\text{Xe})}
\sim
  10^{3},
\label{eq:O14_ratios}
\end{align}
agreeing with the exact ratios given in \gtab{tab:operators_ratios}.

\subsection{Higher Excited States}

As shown in \gtab{tab:xenon_isotopes}, not just the first
excited energy levels,
but also their higher counterparts
can be reached if the DM mass is large enough. While the
first excited level only requires the DM mass
to be larger than 14\,GeV and 31\,GeV for the two isotopes
$^{129}$Xe and $^{131}$Xe, the minimal DM mass required
for the second excited states is not terribly high.
With an
excitation energy at 236\,keV (164\,keV) for the second
excitation, $^{129}$Xe ($^{131}$Xe) requires the DM mass
to be $m_\chi \gtrsim 182$\,GeV (86\,GeV), respectively.
For a TeV scale DM, even the fifth (fourth) energy
level of $^{129}$Xe ($^{131}$Xe) can be  excited.
For other xenon isotopes, the excitation energies are
too large to allow physical transition.

On both $^{129}$Xe and $^{131}$Xe targets, the transitions to
the aforementioned higher excited states are highly suppressed
comparing with
the transition to the first excited state for most operators.
However, $^{131}$Xe can have several exceptions as summarized
in \gtab{tab:operators_ratios_higher_exc}.
At $m_\chi = 10\,$TeV, the contributions of the first and second excited states for $\mathcal{O}_6$ are roughly equal, both being about $13\%$ of the elastic event rate. An even more dramatic effect takes place for $\mathcal{O}_9$: the second excited state contributes almost $70\%$ of elastic event rate, more than the contribution of the first excited state which is about $50\%$ elastic.
Therefore, taking the second excited state into account, one can increase the predicted event rate for $^{131}$Xe by a factor of $2$. 
The other two operators $\mathcal O_4$ and $\mathcal O_{10}$
can also have sizable contributions to the second excitation
energy levels.

\begin{figure}[t]
\centering
\includegraphics[width=0.49\linewidth]{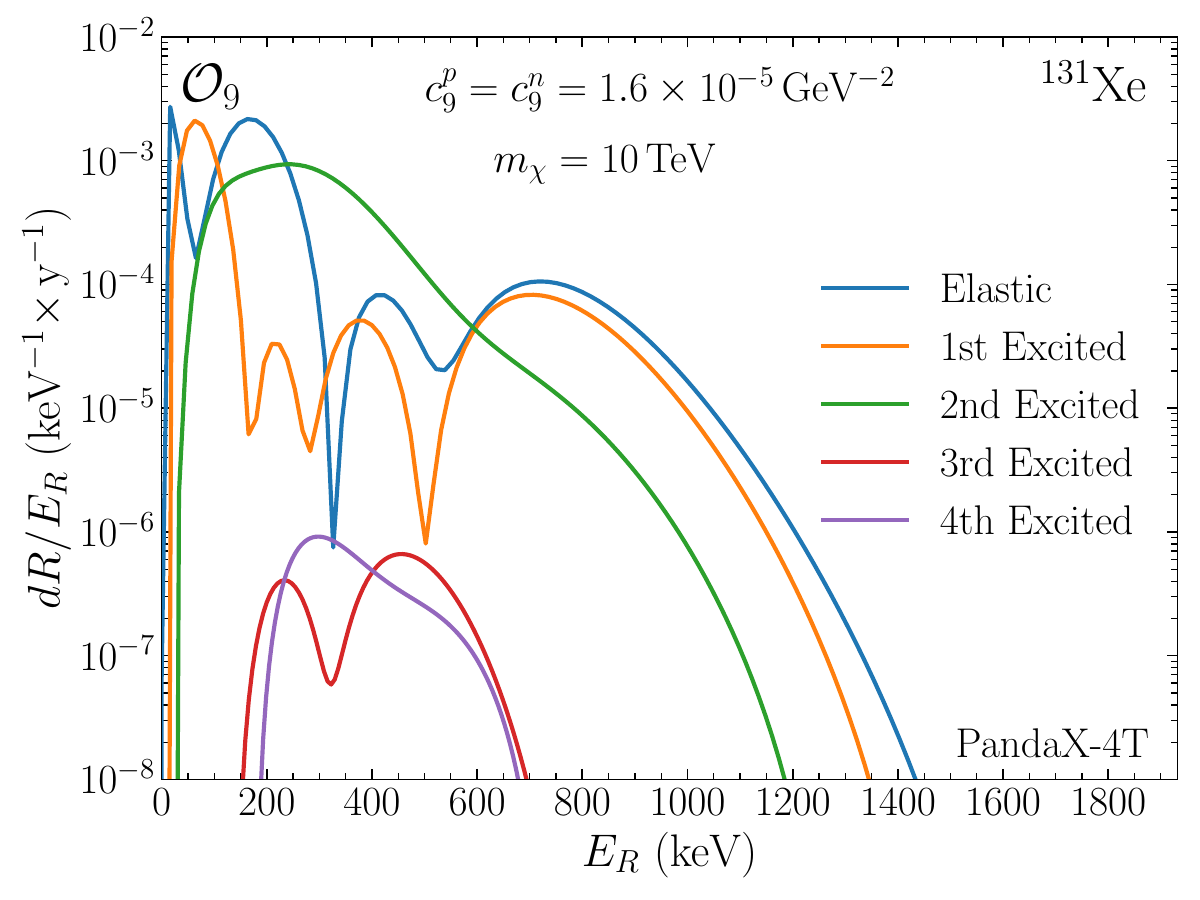}
\includegraphics[width=0.49\linewidth]{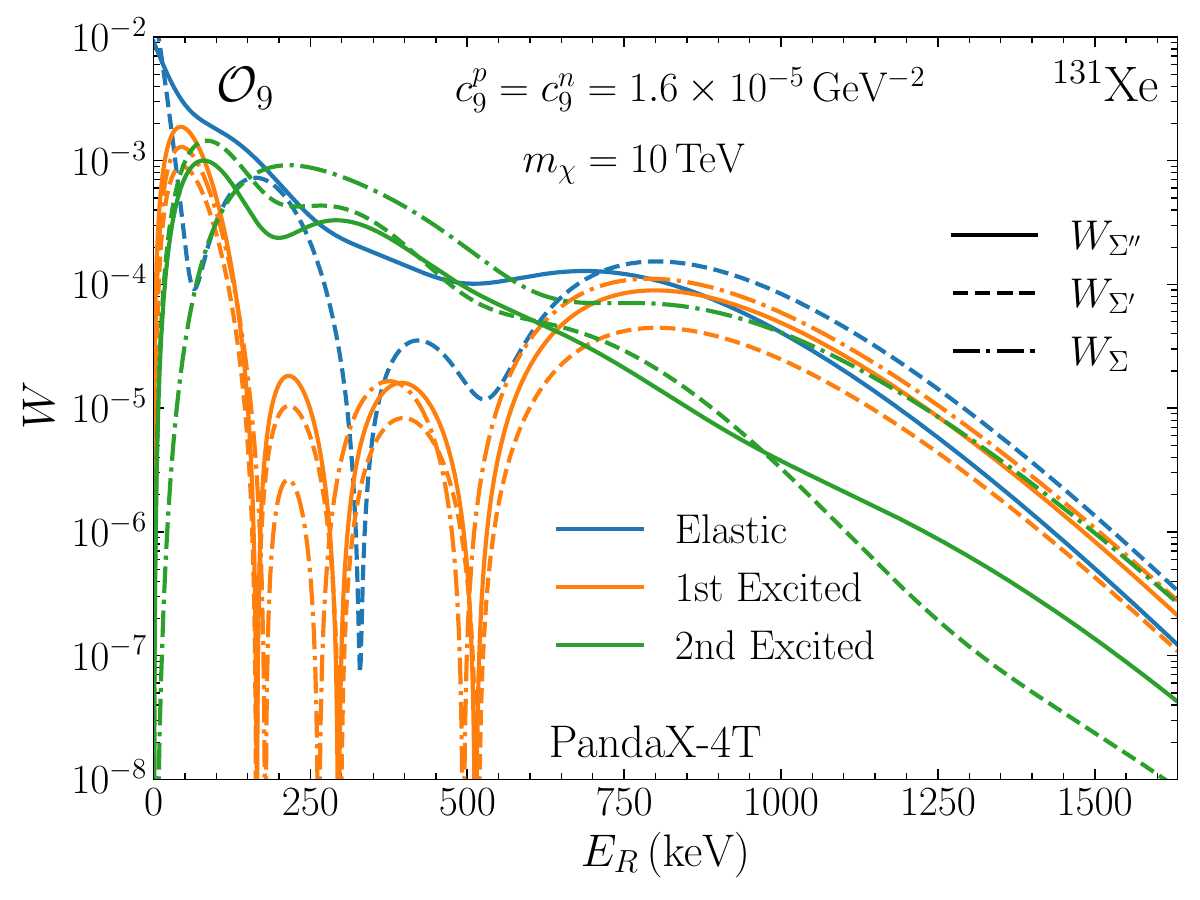}
\caption{{\bf Left:} Expected differential event rates of a WIMP
DM with mass $m_\chi = 10$\,TeV scattering on $^{131}$Xe
via operator $\mathcal O_9$ at PandaX-4T. We show the
recoil spectrum for elastic scattering (blue) and
inelastic contributions of all four kinematically
accessible excited nuclear states (orange, green,
red, and purple).
{\bf Right:} Elastic and inelastic spin nuclear
responses of $^{131}$Xe as functions of the recoil
energy $E_R$.}
\label{fig:131Xe_O9}
\end{figure}

\begin{table}[h]
\centering
\begin{tabular}{ccc}
\toprule
$^{131}$Xe &  $R_{\text{inel},1}^i/R_{\text{el}}^i $ & $R_{\text{inel},2}^i/R_{\text{el}}^i $\\
 \midrule
 $\mathcal O_6$ & $0.13$ & $0.13$\\
 $\mathcal O_9$ & $0.45$ & $0.69$\\ 
 \midrule
 $\mathcal O_4$ & $0.19$ & $0.07$\\
 $\mathcal O_{10}$ & $0.15$ & $0.08$\\
 \bottomrule
\end{tabular}
\caption{
The inelastic-to-elastic event rate ratios for
transitions to the first
$R_{\text{inel},1}^i/R_{\text{el}}^i$
and second $R_{\text{inel},2}^i/R_{\text{el}}^i$
excited states of $^{131}$Xe via operators $\mathcal O_i$
with DM mass $m_\chi = 10$\,TeV.
}
\label{tab:operators_ratios_higher_exc}
\end{table}

To understand this pattern, let us consider the case
of $\mathcal{O}_9$ more closely. The left panel of
\gfig{fig:131Xe_O9} shows the differential event
rates of the elastic and inelastic scatterings to
different final nuclear states. In the recoil energy
range from about $100$\,keV to $600$\,keV, the
contribution of the second excited state (green)
exceeds the first excited state one (orange).
From $250$\,keV to $600$\,keV, it even exceeds
the elastic channel (blue) with the peak values
comparable with the absolute maxima of the elastic spectrum.
Notice also that the contributions of the third
(red) and fourth (purple) excited states are
strongly suppressed and can be neglected in practice.

The spectrum shape for $\mathcal{O}_9$ and its difference
from the $\mathcal{O}_6$ case can be understood with the
underlying nuclear responses. All the four operators
($\mathcal O_4$, $\mathcal O_6$, $\mathcal O_9$, and $\mathcal O_{10}$)
in \gtab{tab:operators_ratios_higher_exc} all couple to
the nucleus through the spin-current multipoles
$\Sigma$, $\Sigma'$ and $\Sigma''$ as summarized in
\gtab{tab:operators_responses}.
Note that we assume $c_5 = c_8 = 0$ for this analysis, so both the $c_{5}c_{4}$ and $c_{8}c_{9}$
interference terms in the DM responses \geqn{eq:R_DeltaSigmaP_def} vanish, and the nuclear responses $W_{\Delta \Sigma'}$ and $W_{\Delta' \Sigma}$ do not contribute.
The primary difference between the operators arises at the level of nuclear response functions: while the scattering via $\mathcal O_{6}$ and $\mathcal O_{10}$ only involves the longitudinal response $W_{\Sigma''}$, $\mathcal O_{9}$ requies only the transverse responses $W_\Sigma$ and $W_{\Sigma'}$, and $\mathcal O_4$ couples to all three of these responses.
As shown in the right panel of \gfig{fig:131Xe_O9},
one can see that between approximately $150$\,keV and $550$\,keV,
all the responses of the first excited state (orange) are strongly
suppressed and experience multiple dips.
On the other hand, their counterparts for the second excited states (green) have large values and evolve slowly. The elastic response $W_{\Sigma'}$ (dashed blue) is also suppressed in this range while $W_{\Sigma''}$ (solid blue) behaves similarly to the responses of the second excited state. Since only $W_{\Sigma'}$ contributes to the elastic signal from the operator $\mathcal O_9$,
the relative contribution of inelastic scattering becomes enhanced in this region. 
Meanwhile, for the other three operators, the elastic signal involves $W_{\Sigma''}$. So there is no dramatic decrease in the elastic signals and the relative inelastic contributions are smaller than for $\mathcal O_9$. 
The additional difference between the operators $\mathcal O_{4}$, $\mathcal O_{6}$ and $\mathcal O_{10}$ comes from the DM responses in \geqn{eq:DM_responses}: for $\mathcal O_{4}$, the nuclear responses are accompanied by constant coefficients, while for $\mathcal O_{10}$
the contribution of higher momentum transfers is enhanced by a factor of $\bm q^2/m_N^2$, and $\mathcal O_{6}$ receives an even stronger enhancement with a factor of $\bm q^4/m_N^4$.

\subsection{Comment on Recent LZ Event}
\label{sec:LZ}

The LUX-ZEPLIN (LZ) collaboration has recently reported an observation of one event consistent with a nuclear recoil $E_R = 248 \pm 23\,\text{(stat)}\,\pm 23\,\text{(sys)}$\,keV in a region with a low expected background \cite{Akerib:2026jyz}.
The authors of Ref.~\cite{Akerib:2026jyz} tested signal hypotheses for NR EFT operators $\mathcal O_i$ and corresponding relativistic Lagrangians $\mathcal L_j$ of~\cite{Anand:2013yka}, obtaining a maximum local significance up to $3.4\,\sigma$. 
The analysis assumed only elastic scattering on the nuclear side and was based on NSM predictions
\cite{LZ:2023lvz}.

Observation of a DM-nucleus scattering event at this relatively high energy and non-observation of events at lower energies implies that the recoil spectrum must have a specific spectral shape, with a maximum of the spectrum achieved around $E_R \approx 250\,$keV and suppression in the low-energy region. 

However, two factors may affect the spectral shape prediction as we have already demonstrated above. First, the corresponding differential rate is sensitive to the nuclear model used as the input even if one considers only the elastic scattering regime. Second, the inelastic channel may give a comparable contribution and have a different shape from the elastic one. In other words, the total contribution may have a different shape from the ``purely elastic'' prediction.

\begin{figure}[t]
\centering
\includegraphics[width=0.8\linewidth]{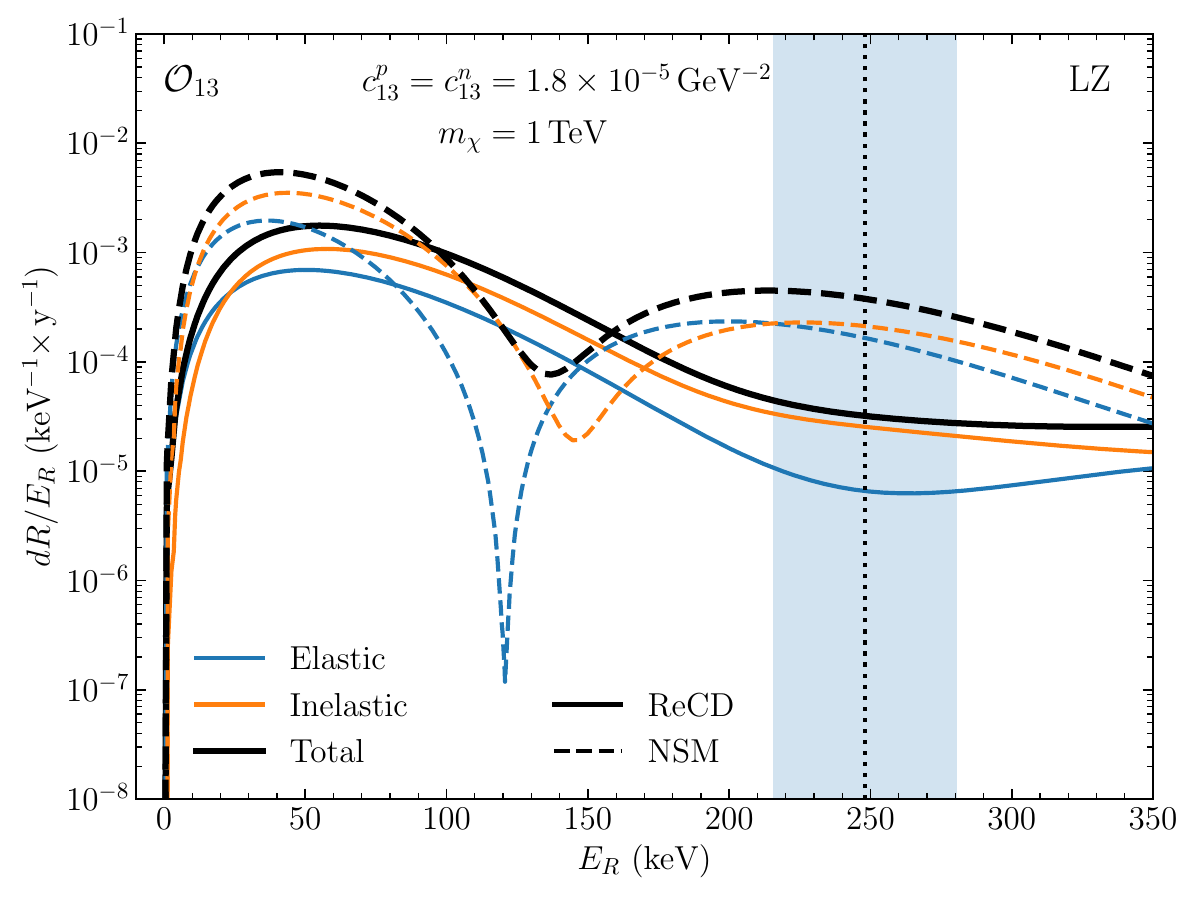}
\caption{Differential event rates due to the WIMP-nucleus
elastic (blue) and inelastic (orange) scatterings via operator $\mathcal O_{13}$ and their sum (black) for a natural xenon target. Solid and dashed curves correspond to
ReCD and NSM predictions, respectively. The dotted vertical line and the shaded region correspond to the recoil energy $E_R = 248 \pm 23\,\text{(stat)}\,\pm 23\,\text{(sys)}$\,keV reconstructed by LZ~\cite{Akerib:2026jyz}.}
\label{fig:O13_comparison}
\end{figure}

Let us consider a specific example.
A $3.4\,\sigma$ significance was obtained in
the LZ analysis \cite{Akerib:2026jyz} for the
DM mass $m_\chi = 1\,$TeV and isoscalar interaction
Lagrangian $\mathcal L_{16}^s$ which is simply
proportional to the effective operator $\mathcal O_{13}$
\cite{Anand:2013yka}. As pointed out earlier in
\gsec{sec:kinematics} and \gsec{sec:selection_rules},
only two isotopes $^{129}$Xe and
$^{131}$Xe can contribute to both elastic and
inelastic scatterings via $\mathcal O_{13}$.
With contributions summed over these two isotopes,
\gfig{fig:O13_comparison} compares the expected event
rates obtained with the ReCD and NSM methods.
Note that we use the density matrices of
Refs.\,\cite{Arcadi:2019hrw} and \cite{Anand:2013yka}
for the NSM predictions, which may differ from those
used in LZ analysis \cite{Akerib:2026jyz}.

The NSM elastic contribution (dashed blue) indeed has
a maximum (albeit local) around $E_R \approx 200\,$keV
which could provide an explanation of the observed
event as the LZ collaboration has tried \cite{Akerib:2026jyz}.
For comparison, the elastic signal obtained by the ReCD
method (solid blue) decreases slowly without a sharp
dip in this region. Contrast to the NSM method,
the ReCD result reaches a minimum at $E_R \approx 250\,$keV.
The difference between the NSM and ReCD results can
be even larger than one order. To be exact, the
ReCD prediction of an elastic scattering is
25 times smaller than its NSM counterpart at $E_R = 250\,$keV.

Besides the elastic channel, the inelastic scattering
can also have sizable contribution. As shown with the
orange curves in \gfig{fig:O13_comparison}, the NSM
method predicts a comparable inelastic signal as the
elastic one while ReCD tells that the inelastic one
can actually dominate for the region $E_R \approx 250\,$keV.
Between the elastic and inelastic channels, the spectra
take roughly the same overall features. For NSM,
both channels have a dip around $E_R \approx 150\,$keV
and rises up again to reach a second maximum around
$E_R \approx 200\,$keV. Neither the dip or the second
maximum can appear in the ReCD results whose elastic
and inelastic curves decreases slowly with the recoil
energy $E_R$ to reach a minimum around
$E_R \approx 250\,$keV. So the sum of elastic and
inelastic spectra (black) has similar features as the
individual ones.

Overall, the event rate at $E_R \approx 250\,$keV predicted by ReCD is one order of magnitude smaller than the NSM prediction, making the observation of a signal at this energy much less favorable.
Note also that according to our computations, both models predict a global maximum at around $50\,$keV, making the non-observation of low-energy events difficult to explain.
Thus, we expect that the results of the analysis
\cite{Akerib:2026jyz} performed by the LZ collaboration
may change significantly depending on the nuclear inputs.

\section{Conclusions}
\label{sec:conclusion}

In this work, we thoroughly explore the nuclear responses
in the DM direct detection, with particular emphasis on the
inelastic scattering to excited energy levels. Taking the
tonne scale liquid xenon detectors (such as PandaX, XENONnT,
and LZ) for illustration, two isotopes $^{129}$Xe and $^{131}$Xe
with nonzero nuclear spin can experience inelastic scattering
while all the spinless isotopes ($^{128}$Xe, $^{130}$Xe,
$^{132}$Xe, $^{134}$Xe, $^{136}$Xe) have too high excitation
energies ($E^* \gtrsim 400$\,keV) to be accessible as summarized
in \gtab{tab:xenon_isotopes}.
Putting together, $^{129}$Xe (26.4\%) and $^{131}$Xe (21.2\%)
contribute almost $50\%$ of the natural xenon abundance, which makes
the inelastic channel quite promising for xenon-based experiments.

Based on the multipole expansion of various effective operators,
we detail their angular momentum, parity, and time-reversal
properties to establish the selection rules.
Even without invoking the spin cancellation
argument, these selection rules can already efficiently show
which operators should give vanishing contribution to the
elastic and inelastic transitions. We point out that not just
the nucleon spin but also the nucleon velocity can get involved.
Consequently, the spin cancellation argument cannot simply
apply. In particular, non-zero spin allows probing
more non-trivial interaction operators and nuclear
responses with the inelastic scattering as a prominent
example. With less stringent selection rules, the inelastic
scattering has richer dynamics and involves more nuclear
responses than its elastic counterpart.

We employ the state-of-the-art ReCD theory to evaluate
both the elastic and inelastic nuclear response functions.
Comparing with the traditional NSM method, the ReCD
calculation is a configuration-interaction approach
rooted in the relativistic nuclear density functional
theory and has the advantage of handling all nucleons
without imposing severe model space truncations.
This is particularly suitable for medium-mass and
heavy nuclei such as the xenon isotopes.

Our results show that, the inelastic channel can
significantly increase the expected event rate for
a DM with mass $m_\chi \gtrsim 100\,$GeV.
For the effective NR operator $O_{13}$, the elastic
and inelastic contributions turn out to be comparable.
Taking the inelastic contribution into consideration
can increase the expected total event rate by at
least a factor of two. For $\mathcal O_{7}$ and
$\mathcal O_{14}$, the enhancement is even much
more significant by factors up to $10^2$ and $10^3$,
respectively.
For the first time, we quantitatively study the role
of higher excited nuclear states. Our results show
that for spin-dependent operators such as $\mathcal O_6$
and $\mathcal O_9$, the contribution of the second
excited states can be comparable or larger than
that of the first excited state.
Taking into account the inelastic scatterings
can significantly improve the sensitivity of the
DM direct detection experiments and set much more
stringent constraints on the DM-nucleon interactions.

The differential event spectrum typically has dips and peaks, and their positions and magnitudes differ between NSM and ReCD results.
This can lead to
quite different interpretation of the recently
published LZ event with a recoil energy around
250\,keV. While the NSM predicts a local maximum,
the ReCD result reaches a local minimum instead.
The difference can be as large as a factor of 25.
Especially, the ReCD prediction has only a prominent
peak at low recoil energy. It would be very difficult
to interpret the LZ event by a DM signal if the
ReCD calculation of the nuclear responses is adopted.

\section*{Acknowledgements}

The authors would like to thank Jianglai Liu, Lei Wu, Wen-Na Yang, and Ning Zhou for useful discussions.
We thank Giorgio Arcadi and Stefan Vogl for carefully inspecting our calculations and for useful correspondence concerning the results of Ref.\,\cite{Arcadi:2019hrw}.
We are grateful to Wick Haxton for helpful correspondence regarding the DMFormFactor code.
The authors are supported by the National Natural Science
Foundation of China (12375101, 12425506, and 12105004) and the
State Key Laboratory of Dark Matter Physics.
SFG is also an affiliate member of Kavli IPMU, University of Tokyo.

%\bibliographystyle{utphysGe}
%\bibliography{dmInelastic}
%\nocite{*}

\end{document}